\documentclass{aastex701}
\graphicspath{{./}{figures/}}
\usepackage{graphicx}
\usepackage{gensymb}
\usepackage{footnote}

\begin{document}

\title{Constraints on the Hot Circumgalactic Medium around Nearby $L^*$ Galaxies from SRG/eROSITA All Sky Survey}
\author[0000-0002-7875-9733]{Lin He}
\email[show]{helin@smail.nju.edu.cn}
\affiliation{School of Astronomy and Space Science, Nanjing University, Nanjing 210023, China}
\affiliation{Key Laboratory of Modern Astronomy and Astrophysics, Nanjing University, Nanjing 210023, China}

\author[0000-0003-0355-6437]{Zhiyuan Li}
\email[show]{lizy@nju.edu.cn}
\affiliation{School of Astronomy and Space Science, Nanjing University, Nanjing 210023, China}
\affiliation{Key Laboratory of Modern Astronomy and Astrophysics, Nanjing University, Nanjing 210023, China}
\affiliation{Institute of Science and Technology for Deep Space Exploration, Suzhou Campus, Nanjing University, Suzhou 215163, China}

\begin{abstract}
The circumgalactic medium (CGM), a multi-phase gas acting as the dynamic interface between the main body of a normal galaxy and the intergalactic medium, is a key ingredient of the galactic ecosystem and provides crucial diagnostics to a wide array of physical processes affecting galaxy evolution.
However, direct evidence for a predominantly hot (at a characteristic temperature of a few million Kelvin) CGM around present-day $L^*$ galaxies remains elusive, despite extensive observational searches and strong physical reasons that it should exist. Here, we present a systematic search for the hot CGM around a representative sample of nearby $L^*$ galaxies selected from the 50 Mpc Galaxy Catalog (50MGC), by stacking their X-ray images and spectra from the SRG/eROSITA all-sky survey. Significant ($7.4\sigma$) diffuse X-ray emission is detected out to a galactocentric radius $\gtrsim$ 50 kpc, the cumulative spectrum of which requires a substantial thermal component, consistent with a hot plasma with log-normal temperature distribution, and thus argues against a predominantly non-thermal origin. The diffuse emission exhibits a relatively steep intensity profile ($\beta=0.52_{-0.06}^{+0.08}$), inferring a moderate amount ($2\times10^{10}~M_{\odot}$) of hot gas within 10--200 kpc.
The radial distribution and total amount of the hot CGM gas are broadly in agreement with prediction by IllustrisTNG simulations for present-day Milky Way-like galaxies, while also showing dependencies on both star formation and nuclear activity.
The constraints on the hot CGM derived in this study hold promise for calibrating key physical processes, such as stellar feedback and active galactic nucleus feedback, in next-generation cosmological simulations.

\end{abstract}

\section{Introduction} \label{sec:intro}
The circumgalactic medium (CGM), broadly referred to an extended, dynamic, and multi-phase gas surrounding the main stellar content of a normal galaxy, is increasingly recognized as a key ingredient of the so-called galactic ecosystem \citep{Putman_2012,Tumlinson_2017,FG_2023,Fumagalli_2024,Chen_2024}, in which the host galaxy's evolution is governed by both internal processes such as stellar feedback and active galactic nucleus (AGN) feedback \citep{Mathews_1971,Chevalier_1985,Choi_2020}, and external processes such as accretion of the inter-galactic medium or ram pressure exerted by an intra-cluster medium \citep{White_1978,White_1991,Benson2010,Hou_2024}.   
While extensive theoretical and numerical studies posited the prevalence of a CGM around present-day galaxies, in particular a hot CGM with a characteristic temperature of a few million degrees around Milky Way[MW]-sized galaxies (i.e., with a total stellar mass $\sim 10^{10-11}\rm~M_\odot$ and a dark matter halo mass $\sim 10^{12}\rm~M_\odot$), they differ substantially in the predicted amount, spatial extent and physical state of the CGM, often owing to different physical assumptions adopted, and recipes implemented, for the feedback processes \citep{Silich_2025,Zhang_2025ApJ...991..170Z}.  
Direct detection of the hot CGM, best with spectroscopic imaging observations in soft ($\lesssim$2 keV) X-ray band, is notoriously difficult, due to its tenuous and extended nature, which requires sufficient sensitivity and sky coverage simultaneously.  

Nearby galaxies offer the advantage of detection sensitivity for the diffuse X-ray emission from the hot CGM. 
Existing pointed X-ray observations, mainly conducted by the {\it Chandra} X-ray Observatory and the X-ray Multi-Mirror Mission-{\it Newton}, have led to firm detections of diffuse hot gas in and around dozens of nearby galaxies \citep{Strickland_2004,Tullmann2006A&A...448...43T,Li_2007b,Boroson2011ApJ...729...12B,LJT_2013a,Bogdan_2013,Anderson_2016,LJT_2016,Bogdan_2017}, but these observations were often biased towards actively star-forming galaxies and/or massive early-type galaxies, and their field-of-view of was typically limited to the inner region of the target galaxy, rarely reaching beyond a few tens of kpc from the galactic center, where the bulk of the hot CGM of a MW-sized galaxy should reside.  Thanks to the recent advent of the Spectrum-Roentgen-Gamma/eROSITA all-sky survey \citep{Sunyaev_2021}, a highly uniform and sensitive soft-X-ray census of the hot CGM around nearby galaxies, assisted with the technique of imaging stacking, is now within reach. 

Indeed, efforts have been made to detect the diffuse and faint X-ray emission from hot baryons in circumgalactic and intragroup media, primarily by stacking wide-field X-ray survey data from ROSAT and eROSITA (\citealp{Anderson_2013ApJ...762..106A,Li_2024,Chadayammuri_2022,Comparat_2022}). 
The latest constraints on the hot CGM around MW-mass and more massive galaxies have been established by \cite{Zhang_2024}, who accumulated four SRG/eROSITA All-Sky Surveys (eRASS:4) data of nearly 30,000 MW-mass galaxies and revealed the stacked signal extending up to the virial radius, with an intensity profile  conforming to a $\beta$-model with $\beta\sim 0.43$. 
Meanwhile, \cite{Sacchi_2025} investigated the near-galactic hot medium and uncovered anisotropy associated with galactic outflows/bubbles in the halo of a considerable sample of highly inclined galaxies, utilizing archival data from {\it Chandra}. 
These stacking studies have predominantly focused on more distant objects (e.g., $D > 100$ Mpc), relying on wide-field galaxy surveys such as Two Micron All Sky Survey (2MASS), Sloan Digital Sky Survey (SDSS) and Galaxy And Mass Assembly (GAMA). Given that the signal-to-noise ratio scales with $N_{\rm gal}^{1/2}$, where $N_{\rm gal}$ is the number of stacked galaxies, a large sample is often necessary to achieve a robust detection. However, the accuracy of stacking results hinges upon the completeness and purity of defined sample. As discussed in \cite{Zhang_2025} and \cite{Shreeram_2025b}, satellite boost bias (i.e., satellite galaxies misclassified as centrals) will enhance the stacked X-ray signals in the halo of central galaxies, and the contamination effect becomes more severe toward lower stellar masses. Also shown in \cite{Shreeram_2025a}, contribution from satellites with massive host halos is predicted to dominate the stacked X-ray emission beyond $\sim$ 40 kpc. Therefore, precisely identifying central and satellite galaxies, as well as examining their large-scale environments, are essential to obtain reliable stacking results, yet both become more challenging at greater distances than in the nearby Universe.


We have constructed a sample of nearby galaxies that can be broadly considered as MW-like hosts, which have the following advantages: i) Their proximity ensures a good sensitivity for detecting the hot CGM; ii) Similarly, a good sensitivity for point source detection helps to minimize potential contamination by unresolved discrete sources associated with the host galaxies and their satellites, such as X-ray binaries (XRBs) and/or active galactic nuclei (AGNs); iii) The large-scale environment of these galaxies are relatively well constrained, helping to minimize potential contamination by the ambient, such as a hot intra-cluster medium; iv) The galactic halo, loosely defined by a spherical volume enclosed within the virial radius, which is practically taken to be 200 kpc for a MW-sized galaxy here, can be well resolved even under the moderate angular resolution of eROSITA, facilitating a spatially resolved analysis of the hot CGM. 
Specifically, we selected a representative sample of 474 $L^*$ galaxies from the 50 Mpc Galaxy Catalog (50MGC; \citealp{Ohlson_2024}), which are all located within a distance of 50 Mpc and with integrated properties (in particular stellar mass and star formation rate) broadly similar to the MW, as illustrated in  Figure~\ref{fig:histo} (see Section~\ref{sec:sample} for details about the sample construction).

This paper is organized as follows: We describe our sample selection flow in Section~\ref{sec:sample}, and introduce the data preparation and stacking method in Section~\ref{sec:method}. Our major results, including the stacked surface brightness profile and spectrum, are presented in Section~\ref{sec:results}. We discuss the origin of the observed diffuse emission and compare our results with previous studies and cosmological simulation predictions in Section~\ref{sec:discussion}. Our main conclusions are summarized in Section \ref{sec:summary}. 
All uncertainties are quoted at $1\sigma$ confidence level unless stated otherwise.

\section{Nearby $L^*$ galaxy sample selection} \label{sec:sample}

We construct a sample of nearby galaxies that can be broadly considered as MW-like hosts. The 50 Mpc Galaxy Catalog (50MGC; \citealp{Ohlson_2024}), a catalog of 15424 galaxies located within a distance of 50 Mpc, compiled from HyperLeda, the NASA-Sloan Atlas, and the Catalog of Local Volume Galaxies, is taken as the parent sample. 
The 50MGC, with consistent and homogenized magnitude, color and distance measurements provided, allows for the extraction of a galaxy sample with both a sufficiently large size and a sufficient proximity, ensuring good statistics for detecting and quantifying the halo X-ray emission.
Instead of defining a MW-like galaxy sample in terms of the total stellar mass or halo mass, which is commonly adopted and easily constructed in studies based on simulated galaxies or large-scale extragalactic surveys, we construct a sample of $L^*$ galaxies from the 50MGC, according to the optical ($B$-band) galaxy luminosity function \citep{Tortorelli_2020}. Compared to the stellar mass or halo mass, which is not a direct observable, the optical luminosity is the more readily available, uniformly sourced parameter in the 50MGC. Besides this practical consideration, we stress that the MW is generally considered a $L^*$ galaxy, and hence a $L^*$ galaxy sample may well reflect the properties of the hot CGM around the MW and MW-like galaxies. 

\begin{figure*}[htbp]
\centering
\includegraphics[width=0.8\textwidth]{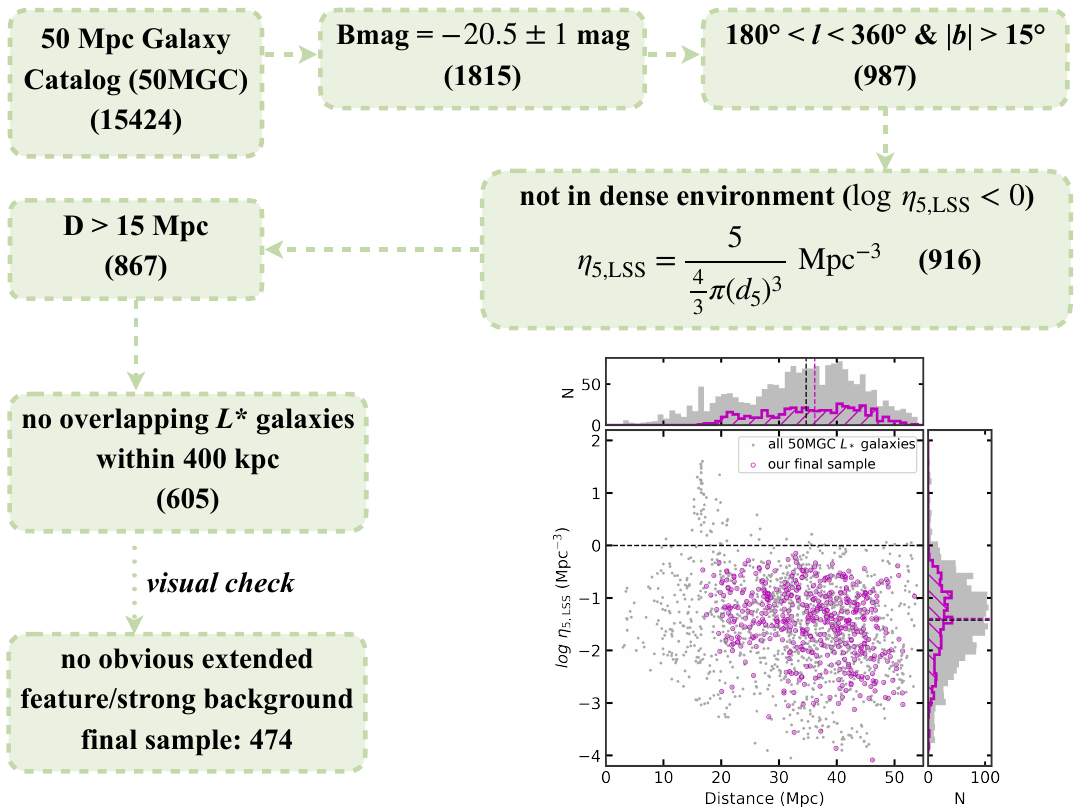}
\caption{Flow chart illustrating the sample selection process. Numbers in the bracket indicate the retained sample size at each step. The lower right panel displays the distributions of the galaxy distance and the environmental density. The Virgo cluster stands out at a distance of $\sim$16 Mpc.
}\label{fig:sampleselection}
\end{figure*}

We select $L^*$ galaxies with the following criteria: (i) $B$-band absolute magnitude $M_{\rm B}=-20.5\pm 1.0$ \citep{Tortorelli_2020}, which is readily taken from the 50MGC; 
(ii) Galactic longitude $180^{\circ} < l < 360^{\circ}$, such that  eRASS1 data is publicly available, and 
Galactic latitude $b >15^{\circ}$ or $b <-15^{\circ}$ to avoid heavy foreground absorption by the  Galactic plane;
(iii) To avoid potential contamination by a hot intra-group or intra-cluster medium, we exclude galaxies residing in dense environments (e.g., rich groups or clusters) by requiring $\rm log\eta_{5}<0$, where $\eta_5$ represents the galaxy number density within a spherical radius ($d_5$) enclosing five nearest neighbors with $M_*>10^9~M_{\odot}$ listed in the 50MGC. Member galaxies of the Virgo cluster and Fornax cluster are particularly excluded in this step;
(iv) A lower limit of distance $D > 15$ Mpc to exclude those nearest galaxies which have the largest angular extent;
(v) In cases when two $L^*$ galaxies overlap within their projected 400 kpc galactocentric radius, we only keep the closer one to avoid double counting of the halo X-ray emission; In rare cases that two $L^*$ galaxies overlap within their 200 kpc radius, we excluded them both, as the extended halo of one galaxy could boost the observed signals of the other;
(vi) Assisted with visual examination, we further exclude some galaxies that have at least one prominent diffuse feature projected within a galactocentric radius of 400 kpc, which might be caused by a background galaxy cluster/group, a luminous interloping galaxy, a foreground supernova remnant or star cluster, etc., which have been flagged by the eRASS1 source catalog. 
Alongside, we also exclude several galaxies that happen to fall on a high local background, where the background count rate significantly exceeds the typical values of the entire sample.
Extended Data Figure~\ref{fig:sampleselection} illustrates the sample selection procedure through the above steps. 
Notably, a large fraction of all (1815) $L^*$ galaxies in the 50MGC are thus dropped from our final sample, which consists of 474 $L^*$ galaxies. 
Nevertheless, having a clean sample of $L^*$ galaxies as such is necessary for a robust study of the halo X-ray emission.
Using a K-S test, we verify that our final sample is representative of the entire $L^*$ galaxies in the 50MGC, in terms of both $B$-band absolute magnitude and $g-i$ color.

\begin{figure*}[htbp]
\centering
\includegraphics[width=1.\textwidth]{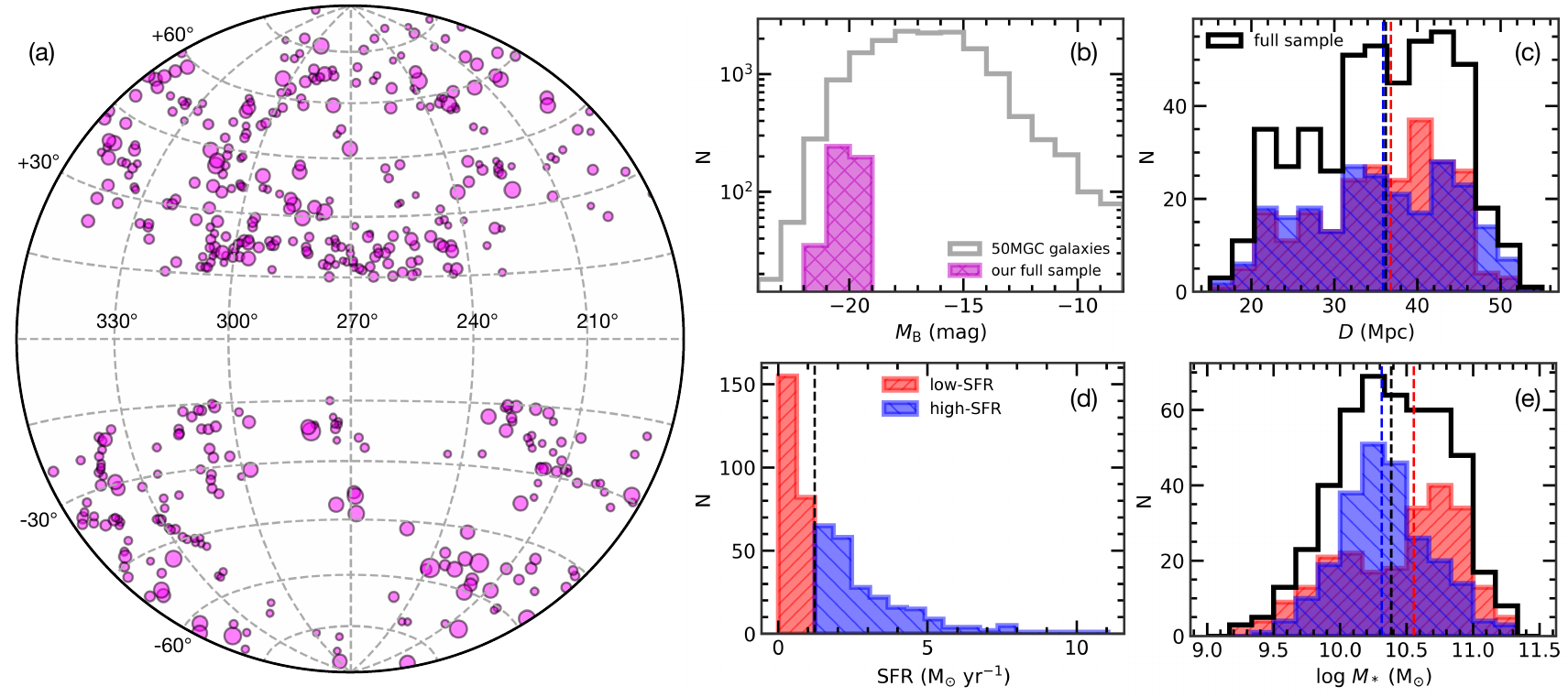}
\caption{(a): Projected sky position (in Galactic coordinates) of the 474 $L^*$ galaxies (magenta cicles), with the symbol size scaling inversely with the galaxy's distance. 
(b): Histogram of the B-band absolute magnitude of the parent sample, i.e. the 50 Mpc Galaxy Catalog (grey) and our final sample of $L^*$ galaxies (magenta). 
(c)--(e): Histograms of the distance, star formation rate and stellar mass of the sample galaxies. The high-SFR and low-SFR hosts are shown in blue and red, respectively, and the the full sample in black. The vertical dashed lines in panels (c) and (e) indicate the median values of the corresponding samples, while in panel (d) the vertical dashed line separates the high-SFR and low-SFR hosts.
}\label{fig:histo} 
\end{figure*}

Figure~\ref{fig:histo}a displays the sky locations of the selected $L^*$ galaxies, in Galactic coordinates and showing only the west hemisphere for which eRASS1 data is available. A north-south asymmetry in the sky locations is appreciable, which is an inherent property of nearby galaxies due to the MW's specific position in the local universe. Otherwise the sample galaxies are distributed relatively evenly on an angular scale of degrees. The right panels of Figure~\ref{fig:histo} display histograms of $D$, $M_*$ and SFR of our $L^*$ galaxies, where the galaxy distance $D$ is directly taken from the 50MGC, while $M_*$ and SFR are derived in a homogeneous fashion based on WISE data (see Appendix~\ref{sec:WISE}).  The majority of the sample galaxies have their $M_*$ fall between $10^{10}-10^{11} M_\odot$. It is noteworthy that the MW, with an often adopted value of $M_* \sim 6\times10^{10} M_\odot$, would find itself in the upper half of this distribution. Most sample galaxies have a moderate SFR $\lesssim 2~M_\odot\rm~yr^{-1}$, which is also similar to the MW. There are only few starburst galaxies (i.e. with SFR $\gtrsim 10~M_\odot\rm~yr^{-1}$) in this sample. Notably, galaxies with a higher SFR have an on-average lower $M_*$ (compare the blue and red groups in Figure~\ref{fig:histo}), which can be understood as a natural outcome of our sample selection based on the $B$-band luminosity, as young stars associated star-forming galaxies produce more blue light, compensating for a lower stellar mass.  

We further divide our full sample of 474 $L^*$ galaxies into different sets of subsamples. First, we separate the galaxies into three subsamples according to their distances: 15--22.5 (near), 22.5--33.75 (medium) and 33.75--50 (far) Mpc, such that the galaxy distance in each subsample varies by less than $\pm20\%$, justifying a direct stacking of the individual galaxies. This helps to eliminate the need for image rescaling among different galaxies and minimizing any potential bias thus introduced (see Section~\ref{subsec:stack}).
The near, medium and far subsamples contain 44, 144 and 286 galaxies, respectively.
Second, we separate the galaxies into two subsamples, high-mass hosts and low-mass hosts, according to their stellar mass (above or below $2.4\times10^{10}$ $M_{\odot}$), each containing 237 galaxies. 
Third, we separate the galaxies into two subsamples, high-SFR host and low-SFR host, according to their SFR (above or below 1.2 $M_{\odot}~\rm yr^{-1}$), each containing 237 galaxies. Lastly, we separate the galaxies into two subsamples, AGN host and non-AGN host, according to whether there is a nuclear X-ray source detected in the eRASS1 image (see Section~\ref{subsec:data}). The AGN and non-AGN subsamples contain 114 and 360 galaxies, respectively.

\section{data preparation and stacking method} \label{sec:method}

\subsection{eRASS1 data}
\label{subsec:data}
We utilized the Spectrum-Roentgen-Gamma/eROSITA data from its first all-sky scan (eRASS1; \citealp{Merloni_2024}). To facilitate usage, the eROSITA-DE team has processed the eRASS1 data with a standard pipeline, which are then divided into 4700 sky tiles, each with a size of $3.6^{\circ} \times 3.6^{\circ}$ \citep{Merloni_2024}. For each object in our $L^*$ galaxy sample (Section~\ref{sec:sample}), we retrieved the sky tile(s) enclosing a galactocentric radius of 400 kpc (corresponding to an angular extent of $26'-82'$), which is well exceeding the virial radius ($\sim$200 kpc) of a MW-sized galaxy.
In some cases, this radius spans over two or more neighboring tiles.
We downloaded the calibrated events files, counts images and vignetting-corrected exposure maps in the 0.2--2.3 keV (soft) and 2.3--5 keV (hard) bands for the relevant sky tiles from the eROSITA data archive. 
For those galaxies falling on multiple tiles, we reprojected the neighboring images to form a mosaic using the {\it reproject} package in Python \citep{robitaille_2018_1162674}. 
To maximize the photon counts, we decided to include data from all seven telescope modules, i.e., TM1-TM7. 
While TM5 and TM7 are known to be subject to light leak at low energies ($\lesssim$ 0.5 keV), we have done a test by including or excluding photons between 0.2--0.5 keV and found consistent X-ray intensity profiles between the two cases, with the latter having a poorer $S/N$ (see left panel of Figure~\ref{fig:testprofile}). Therefore we preserve the use of TM5 and TM7 and adopt the energy range of 0.2--2.3 keV, taking full advantage of the eROSITA data. 
\\

\subsection{Point source masking} 
The eRASS1 main source catalog contains nearly 930,000 X-ray sources, both point-like and extended, detected in the 0.2--2.3 keV band \citep{Merloni_2024}.
A small fraction of these sources fall within the footprint of our sample galaxies, due to their relative large angular sizes. 
The left panel of Figure~\ref{fig:eRASS1src} displays the 0.2--2.3 keV flux distribution of the point sources detected within a galactocentric radius of 400 kpc around the sample galaxies. The detection completeness, approximated by the peak of this distribution, is $4.5\times10^{-14}\rm~erg~s^{-1}~cm^{-2}$, or $7.1 \times10^{39}\rm~erg~s^{-1}$ at the median distance of 36.2 Mpc of our sample galaxies.
We found no significant radial variation in the azimuthally-averaged surface density profile as a function of the galactocentric radius combining all sample galaxies (middle panel of Figure~\ref{fig:eRASS1src}), except for a central peak which can be attributed to sources associated with the stellar content of the host galaxies and/or a central AGN. This ensures a statistically uniform detection and subsequent removal of the point sources (the majority being cosmic X-ray background [CXB] sources) and thus an unbiased quantification for the unresolved X-ray emission. 
We adopted different masking radius for the detected point and extended sources. For point sources defined by EXT $=$ 0, we used three times the APE\_RADIUS\_1, which equals to about 1.5 times the 90$\%$ enclosed energy radius. Here EXT and APE\_RADIUS\_1 are parameters provided by the eRASS1 main source catalog. For extended sources, EXT is defined as the best-fit core size of the {\it beta} model fitted to the source, and we used five times EXT to remove them. 
The source masking was done for both the counts map and exposure map.
We have checked the resultant cheese images to ensure that there are no significant residual counts outside the masking radius around individual bright sources. 
\\

\subsection{Identification of nuclear X-ray sources} 
To identify a putative X-ray AGN, we cross-matched the nominal positions of our sample galaxies, provided by the 50MGC, with the eRASS1 main source catalog. We adopted a matching radius of $10^{\prime\prime}$, i.e., the 99 percentile of the position error of the soft-band sources, which corresponds to a linear scale of 0.8 to 2.6 kpc at the distance of the sample galaxies. This yields 114 X-ray nuclei out of the 474 host galaxies, with 0.2-2.3 keV luminosity ($L_{0.2-2.3}$) ranging from $1.8\times 10^{39}~\rm erg~s^{-1}$ to $5.5\times 10^{41}~\rm erg~s^{-1}$. 
We also investigated X-ray sources associated with the satellite galaxies. For each sample (central) galaxy, we search for its potential satellites within a projected separation of $\lesssim$ 400 kpc from the 50MGC catalog, which are less massive than the central and have not been included in our $L^*$ galaxy sample. 
This results in a total of 1046 satellites for the 474 centrals. 
We note that some of these might not be true satellites, because their line-of-sight distance could be significantly different from the associated central $L^*$ galaxy. Nevertheless, we still designate such galaxies as satellite because their X-ray emission could be a potential contamination to the central galaxy.  
We then cross-matched the coordinates of the satellites with the eRASS1 main source catalog using the same $10^{\prime\prime}$ matching radius and found 26 X-ray counterparts, with $L_{0.2-2.3}$ ranging between $4.7\times 10^{38}-1.2\times 10^{41}\rm ~erg~s^{-1}$. These 26 satellites have a stellar mass ranging between $3.2\times 10^7~M_{\odot}$ to $1.4\times 10^{10}~M_{\odot}$, and
owing to their relatively small sizes, the X-ray counterparts are essentially spatially coincident with their galactic nuclei.
We compare the $L_{0.2-2.3}$ distribution of the nuclear sources in the central and satellite galaxies in the right panel of Figure~\ref{fig:eRASS1src}. 
Given the range of $L_{0.2-2.3}$, the nuclear sources associated with both the central and satellite galaxies are consistent with low-luminosity AGNs, although we cannot rule out the possibility that a fraction of them, especially the ones with the lowest $L_{0.2-2.3}$, are XRBs.

\subsection{Stacking eRASS1 images}
\label{subsec:stack}
Due to an expected low count rate of the halo X-ray emission in any individual galaxy, it is necessary to stack the sample galaxies to bring out the X-ray signal. We performed the stacking on the image-level data, i.e., codding the 0.2--2.3 keV counts maps and exposure maps after masking the detected sources, similar to the procedure adopted by recent studies based on eROSITA data \citep{Chadayammuri_2022,Zheng_2023,Li_2024}.
To achieve this, we made a cutout image of a uniform size of 800 kpc $\times$ 800 kpc centered on each galaxy. The cutout image,  with a natal pixel size of 4$^{\prime\prime}$, was then rescaled to a common spaxel scale according a reference distance of 50 Mpc, i.e., the maximum distance of our sample galaxies,  by an interpolation conserving the total counts and exposure.

We applied this stacking method to our full sample and various subsamples, except for three subsamples of varied distance ranges, whose imaging stacking is performed without rescaling (i.e., the pixel size does not change). The ``near", ``medium" and ``far" subsample each contains galaxies at a distance within $\pm 20\%$ deviation of its median value, hence one can safely neglect a small difference in the angular-size-to-physical-scale conversion among galaxies within a given group, which allows for a direct coadding. This ensures that the sky background, which shall not vary with the distance of the source, not be subject to a distance-dependent weighting when stacking the images. 

\section{Results}\label{sec:results}
\subsection{Stacked X-ray surface brightness profiles}
\begin{figure}[htbp]
\centering
\includegraphics[width=1.0\textwidth]{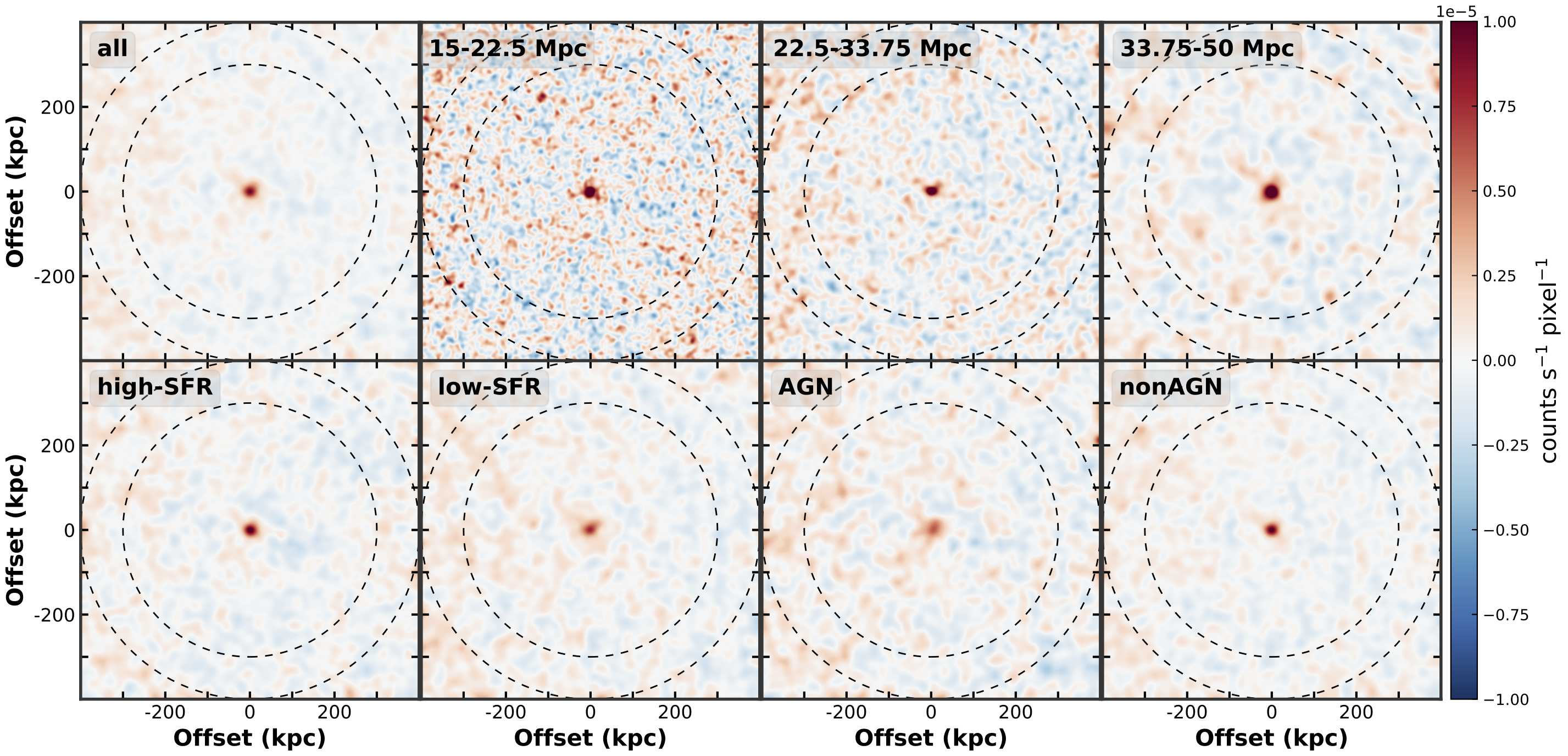}
\caption{Point source and local background-subtracted, exposure-corrected 0.2--2.3 keV surface brightness images stacking the full sample and subsamples of $L^*$ galaxies, smoothed with a Gaussian kernel size of 10 natal pixels (40$^{\prime\prime}$). {\it Top row}: from left to right is the total sample and distance-separated subsamples of 15--22.5, 22.5--33,75, and 33.75--50 Mpc bins. {\it Bottom row}: from left to right is the subsamples of high-SFR hosts, low-SFR hosts, and AGN-hosts and non-AGN hosts. Each image has a physical size of 800 kpc $\times$ 800 kpc. The dashed annulus with inner and outer radii of 300 and 400 kpc defines the local background region.}
\label{fig:stackimage}
\end{figure}

The stacked images of the full sample and various subsamples are shown in Figure~\ref{fig:stackimage}.
Significant X-ray emission are seen in all stacked images, especially near the inner region, demonstrating the effectiveness of the stacking. 
At large radii, the stacked images appear smooth and even, suggesting no contamination by unwanted large-scale features. To test the robustness of our stacking results, we also investigate the stacked images and profiles of random fields, as well as those of the the sample galaxies in hard (2.3--5 keV) X-ray band, no clear sign of extended emission beyond the central region is found in both cases (see details in Appendix~\ref{sec:test}). This strengthens our presumption that the detected diffuse soft X-ray emission originate predominantly, if not entirely, from the hot baryons pervading the halos of these nearby $L_*$ galaxies. 

\begin{figure}[htbp]
\centering
\includegraphics[width=1.\textwidth]{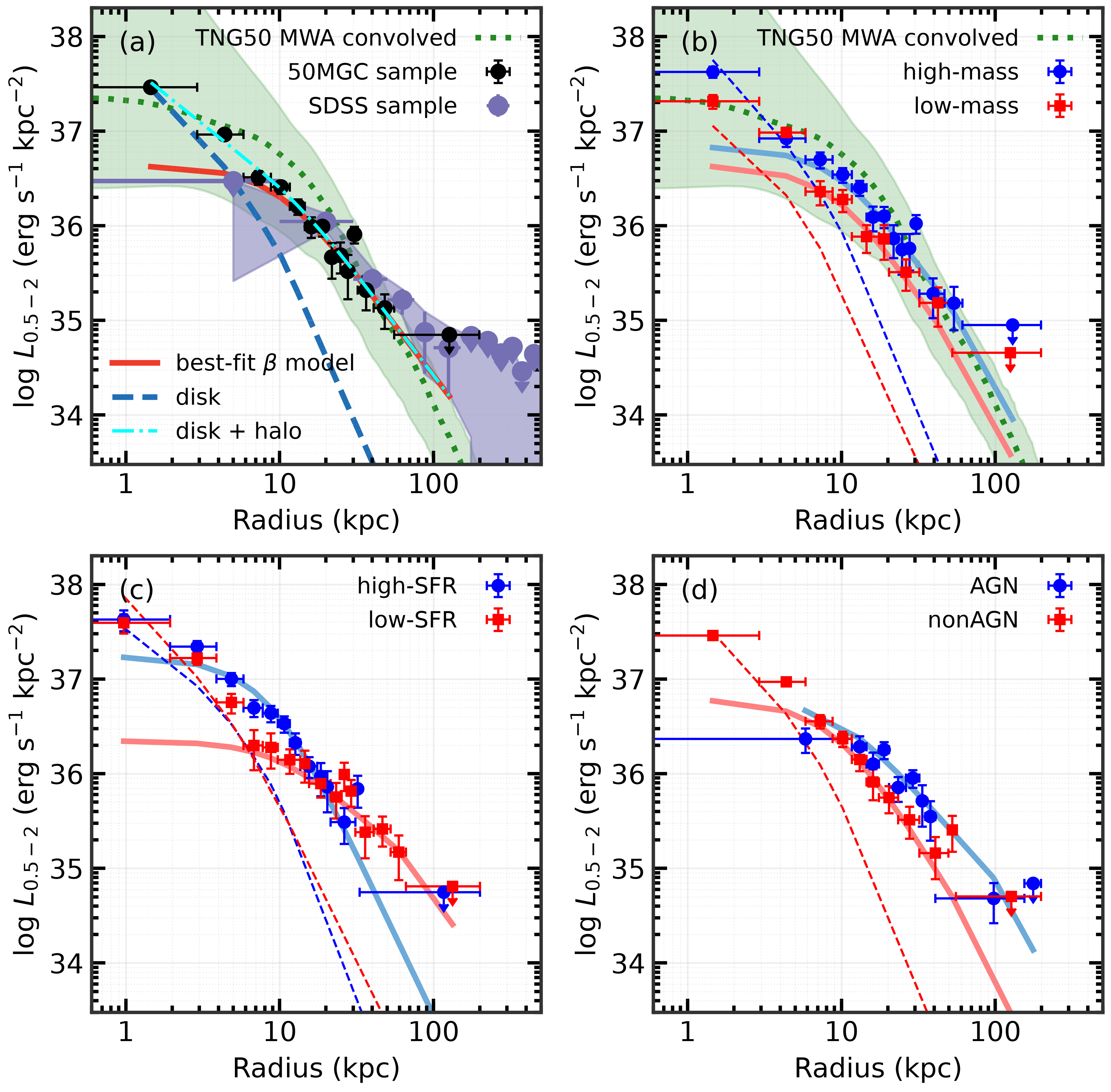}
\caption{(a): Background-subtracted 0.5--2 keV radial intensity profiles of the full sample of $L^*$ galaxies (black circles with error bars), adaptively binned to achieve a signal-to-noise ratio greater than 2 in each bin. 
The angular scale has been converted to a physical scale assuming a nominal distance of 50 Mpc,
while the observed 0.2--2.3 keV net count rate has been converted into unabsorbed energy flux using the best-fit spectral model to the stacked halo X-ray spectrum (see Section~\ref{subsec:spectrum}).
The last radial bin with the arrow denotes 2\,$\sigma$ upper limit. The error bars include only statistical uncertainties from the stacking, except for the outermost bins, where the upper limits were calculated using both the statistical uncertainty from stacking and the systematic uncertainty in the background estimates.
The blue dashed and red solid curves show
the normalized WISE near-infrared starlight profile (for the disk component; Appendix~\ref{sec:WISE}) 
and PSF-convolved $\beta$-model (for the halo component), respectively, while their sum is shown by the cyan dash-dotted curve. 
For comparison, the eROSITA 0.5--2 keV intensity profile obtained by stacking $\sim 3\times10^4$ Milky Way-sized galaxies mainly drawn from the Sloan Digital Sky Survey, taken from \cite{Zhang_2024}, is shown with purple symbols. Also shown as the green solid curve is the mean intensity profile calculated for 152 Milky Way analogs in the Illustris-TNG50 simulations, as defined by \cite{Pillepich_2024}. This simulation-predicted profile has been convolved with the eROSITA PSF, and its 1\,$\sigma$ scatter is shown by the light green strip.   
(b): Background-subtracted 0.5--2 keV intensity profiles of the high-mass (blue) and low-mass (red) galaxies. The TNG50-predicted profile in panel (a) is also overlaid, which compares better with the high-mass subgroup with stellar masses more comparable to the MW.  (c): Similar to panel (b), but for the high-SFR (blue) and low-SFR (red) galaxies. (d): Similar to panel (b), but for the AGN hosts (blue) and non-AGN hosts (red).}
\label{fig:profile}
\end{figure}

To better reveal the diffuse X-ray emission and quantify its spatial extent, we extracted the azimuthally averaged radial intensity profile out to a projected galactocentric radius of 200 kpc, of each stacked group of $L^*$ galaxies (i.e., the full sample or a given subsample). 
Figure~\ref{fig:profile}a displays the exposure-corrected, background-subtracted intensity profile of the full sample, which declines almost monotonically with radius, spanning about three orders of magnitude in the intensity.  
Significant soft X-ray emission is revealed out to at least $\sim50$ kpc, beyond which point an upper limit of the X-ray intensity is obtained. The profiles have been converted to $\rm erg~s^{-1}~kpc^{-2}$ space to match with the physical unit of radius in $x$-axis, assuming the same nominal distance and a uniform energy conversion factor (ECF).

While detected point sources associated with the host galaxies have been identified and carefully removed in the stacked images, residual X-ray emission from discrete sources, both from unresolved stellar populations and the unmasked halo of the eROSITA point-spread function (PSF) of resolved sources, could still be present, and even dominate, in the inner galactic (disk) regions. 
We approximated the residual stellar X-ray emission under the assumption that it intrinsically follows the stellar distribution of the host galaxies (see Section~\ref{subsec:starresidual} for details). 
The hence derived residual stellar X-ray intensity profile, averaged over the sample galaxies and convolved with the PSF, is shown by a blue dashed curve (labelled ``disk'') in Figure~\ref{fig:profile}a.
This stellar component dominates the observed diffuse X-ray emission in the central region, but due to a steep radial decline of the stellar distribution, becomes subdominant beyond a radius of $\sim$10 kpc, which is consistent with the typical disk scale of a MW-sized galaxy.

Next, we fitted the X-ray intensity profile, mostly over 10--50 kpc, by adding to the disk component a PSF-convolved $\beta$-model, in the form of $S_{\rm X}(R)=S_{\rm X,0}[1+(\frac{R}{R_c})^2]^{0.5-3\beta}$ as a function of the projected radius $R$, where $R_c$ is the core radius. 
To do so, we converted the PSF (in angular units) to a physical scale adopting the median distance (36.2 Mpc) of the sample galaxies, and used the Markov chain Monte Carlo (MCMC) method for the model fitting, taking the best-fit value as the median value and $1\sigma$ uncertainties as the values at $16^{\rm th}$ and $84^{\rm th}$ percentiles. We also included an additive background offset shared by all radial bins and propagated its uncertainty into the fitted halo parameters and derived quantities.
This yields $R_c = 8.7^{+3.7}_{-2.7}$ kpc and $\beta = 0.52_{-0.06}^{+0.08}$, i.e., a converging intensity profile. 
Notably, the MW's CGM is also characterized by  $\beta\approx$ 0.5, according to the modeling of a number of X-ray absorption lines \citep{Bregman_2018}.
Integrating the best-fit $\beta$-model between 10--200 kpc yields a total 0.5--2 keV luminosity of $L_{\rm 0.5-2} \approx 5\times10^{39}\rm~erg~s^{-1}$ per sample galaxy. 
If we assume that all diffuse X-ray emission approximated by the $\beta$-model originates from a hot CGM and that the gas is isothermal (both assumptions are not necessarily true; see discussions below and in Section~\ref{sec:discussion}), the radial distribution of the gas density would have the form of $n_{\rm hot}(r)=n_{0}[1+(\frac{r}{R_c})^2]^{-3\beta/2}$, where $r$ is the spherical radius. Integrating this gas density distribution over 10--200 kpc (10--50 kpc), we derive an inferred hot gas mass of $M_{\rm hot} \approx 2\times10^{10} M_\odot$ ($3\times 10^9~M_{\odot}$) per $L^*$ galaxy\setcounter{footnote}{0}\footnote{This calculation relies on the radial distribution of the hot gas and the adopted spectral model. Here we use the fitted parameters of the APEC component listed in Table~\ref{tab:spec_fit}. For a detailed discussion of how the baryon mass estimate depends critically on assumptions regarding the gas density profile, temperature, metallicity, and other parameters, we refer the readers to \citep{Zhang2026A&A...706A.102Z}.}.

\subsection{Comparison between subsamples}

The stacked images and radial surface brightness profiles of the ``near", ``medium" and ``far" subsample are shown in Figure~\ref{fig:stackimage} and Figure~\ref{fig:radprofiles} (also the right panel of Figure~\ref{fig:testprofile}), respectively, in which extended diffuse emission is present with a varied degree but is overall consistent with each other and with the full sample.
Notably, the nearest subsample shows a somewhat lower local background level than the two more distant groups. This might be understood as more X-ray sources (e.g., XRBs and/or satellite AGNs) in and around the nearest galaxies having been removed from the stacked image, due to an on-average lower limiting luminosity ($L_{0.2-2.3}\sim 2\times10^{39}\rm~erg~s^{-1}$) for point source detection.

The X-ray intensity profiles of the high-mass and low-mass subsamples (i.e., galaxies with a stellar mass above or below  $2.4\times10^{10}~{M_\odot}$) both show an overall radial trend  similar to the full sample (Figure~\ref{fig:profile}b). 
However, the profile of the high-mass subsample is almost everywhere higher than its low-mass counterpart. 
The 0.5--2 keV luminosity per high-mass galaxy between 10--200 kpc is about 2.2 times the value per low-mass galaxy, based on the respectively fitted $\beta$-model.
This suggests that the amount of halo X-ray emission may scale with the stellar mass. 
Similarly, the X-ray intensity profiles of the high-SFR and low-SFR subsamples (i.e., galaxies with a current star-formation rate above or below 1.2 ${M_\odot}\rm~yr^{-1}$) both follow the overall radial trend of the full sample (Figure~\ref{fig:profile}c). 
However, the high-SFR subsample exhibits an overall faster decline than the low-SFR subsample beyond 10 kpc, such that the former has systematically higher (lower) X-ray intensities than the latter inside (outside) a radius of $\sim$20 kpc. 
Although the high-SFR and low-SFR galaxies differ slightly in stellar mass (by $\sim 0.2$ dex, as shown in Figure~\ref{fig:histo}e), the observed difference in their radial slopes is not driven by this mass offset. We have tested that the same trend persists after controlling the $M_*$ of these SFR subsamples, which however results in coarser radial bins owing to the reduced sample size and hence is not presented in Figure~\ref{fig:profile}.
This finding is consistent with the simple expectation that the positive effect of star-forming activities, such as energy and mass injection to the CGM, is strongest in the inner halo region. On larger scales, however, this effect may turn negative, suggested by the total 0.5--2 keV luminosity per high-SFR galaxy between 10--50 kpc (10--200 kpc) being 63\% (21\%) of the value per low-SFR galaxy. This reverse trend is not straightforward to interpret, but could be related to previous episodes of galactic feedback, which have sufficient time to increase the amount of hot gas in the outer halo, and in the meantime, to reduce the amount of gas fuel for otherwise continued star formation in the disk.

Lastly, the X-ray intensity profiles of the AGN and non-AGN subsamples (i.e., with and without a detected nuclear X-ray source) also exhibit a similar shape, if one neglects the inner 10 kpc region where masking of the nuclear sources leads to an artificial inward flattening in the profile of the AGN subsample.
The smaller size of the AGN subsample (114 galaxies) results in a poorer counting statistics and hence coarser radial bins, compared to the non-AGN subsample (360 galaxies). 
Nevertheless, the intensity profile of the AGN subsample is almost everywhere higher than the non-AGN counterpart\footnote{We performed a bootstrap test: we randomly selected 114 galaxies from the non-AGN sample (i.e., matching the AGN sample size) and repeated the stacking analysis 100 times. The resulting median profiles remain essentially unchanged, suggesting the observed differences between the AGN and non-AGN subsamples are not due to sample size or other systematic effects.}, and the total 0.5--2 keV luminosity of the former ($1.3\times10^{40}\rm~erg~s^{-1}$), integrated between 10--200 kpc, is about four times higher than the latter ($3.4\times10^{39}\rm~erg~s^{-1}$).
Even after subtracting the residual contribution from the unmasked PSF halos of detected ``AGNs'' (as discussed in Section~\ref{subsec:spectrum}), which has an average luminosity of $1.2\times10^{39}\rm~erg~s^{-1}$, the difference in integrated luminosity between the AGN and non-AGN samples remains large.
This may indicate a positive effect of nuclear activity on the hot CGM, even though the majority of these ``AGNs'' have a relatively low luminosity ($L_X \lesssim 10^{41}\rm~erg~s^{-1}$) and should trace a weakly accreting supermassive black hole. Such a positive effect is also found in recent simulations \citep{Zou2026ApJ..1003..199Z}, in which AGN feedback suppresses the density while increasing the temperature and entropy of the CGM, leading to an enhanced surface brightness profile by a factor of a few compared to the case without AGN feedback.
The fitted parameters of all X-ray intensity profiles and inferred integrated properties of the hot CGM are listed in Table~\ref{tab:tab1}. We note that the tabulated luminosities and hot-gas masses rely on the assumption that the gas properties of the different subsamples are identical to those of the full sample, which is not necessarily valid. In particular, the temperature and metallicity of the halo gas can differ substantially depending on the feedback history. However, the current data do not provide sufficient constraints to determine the temperature and metallicity independently for each subsample. We therefore adopt the same spectral model solely to provide a common basis for comparing the subsamples.

\begin{deluxetable*}{ccccccccc}
\tablecaption{Characterization of the diffuse X-ray emission in various samples \label{tab:tab1}}
\tablehead{
\colhead{Sample} & \colhead{Definition} & \colhead{Size} & \colhead{SNR} & \colhead{$R_c$} & \colhead{$\beta$} & \colhead{$L_{0.5-2}^{10-50}$} & \colhead{$L_{0.5-2}^{10-200}$} & \colhead{$M_{\rm hot}^{10-200}$}
}
\colnumbers
\startdata
full & $L^*$ galaxies & 474 & 7.4 & $8.7_{-2.7}^{+3.7}$ & $0.52_{-0.06}^{+0.08}$ & $39.4_{-0.4}^{+0.3}$ & $39.7_{-0.5}^{+0.4}$ & $10.3_{-0.5}^{+0.2}$ \\
\hline
near & $15<D<22.5$ & 44 & 5.3 & $9.3_{-3.1}^{+3.8}$ & $0.53_{-0.12}^{+0.15}$ & $39.5_{-0.6}^{+0.4}$ & $39.7_{-0.8}^{+0.6}$ & $10.3_{-0.6}^{+0.4}$ \\
medium & $22.5<D<33.75$ & 144 & 2.8 & $9.4_{-3.2}^{+3.8}$ & $0.60_{-0.20}^{+0.23}$ & $39.3_{-0.8}^{+0.6}$ & $39.5_{-1.0}^{+0.9}$ & $10.1_{-0.8}
^{+0.6}$ \\
far & $33.75<D<50$ & 286 & 5.1 & $8.8_{-2.9}^{+3.9}$ & $0.63_{-0.12}^{+0.17}$ & $39.5_{-0.7}^{+0.4}$ & $39.6_{-0.7}^{+0.5}$ & $10.1_{-0.7}^{+0.5}$ \\
\hline
high-mass & log $M_* > 10.38$ & 237 & 7.0 & $9.1_{-3.0}^{+3.9}$ & $0.55_{-0.07}^{+0.10}$ & $39.6_{-0.5}^{+0.3}$ & $39.8_{-0.6}^{+0.4}$ &
$10.3_{-0.6}^{+0.3}$ \\
low-mass & log $M_* < 10.38$ & 237 & 3.5 & $8.5_{-2.7}^{+4.1}$ & $0.58_{-0.14}^{+0.19}$ & $39.3_{-0.7}^{+0.5}$ & $39.5_{-0.8}^{+0.7}$ & $10.1_{-0.8}
^{+0.5}$ \\
\hline
high-SFR & SFR $> 1.23$ & 237 & 3.3 & $7.8_{-2.2}^{+3.7}$ & $0.76_{-0.13}^{+0.13}$ & $39.3_{-0.7}^{+0.5}$ & $39.4_{-0.8}^{+0.5}$ & $9.8_{-0.7}
^{+0.5}$ \\
low-SFR & SFR $< 1.23$ & 237 & 7.2 & $9.4_{-3.2}^{+3.7}$ & $0.41_{-0.06}^{+0.07}$ & $39.5_{-0.4}^{+0.3}$ & $40.0_{-0.5}^{+0.4}$ & $10.5_{-0.5}
^{+0.1}$ \\
\hline
AGN & with X-ray nuclei & 114 & 6.1 & $9.1_{-3.0}^{+3.9}$ & $0.48_{-0.05}^{+0.07}$ & $39.8_{-0.5}^{+0.3}$ & $40.1_{-0.5}^{+0.3}$ & $10.5_{-0.4}
^{+0.2}$ \\
nonAGN & no X-ray nuclei & 360 & 4.6 & $8.3_{-2.5}^{+4.0}$ & $0.61_{-0.10}^{+0.14}$ & $39.3_{-0.6}^{+0.4}$ & $39.5_{-0.7}^{+0.5}$ & $10.1_{-0.7}
^{+0.4}$ \\
\enddata
\tablecomments{(1)$\&$(2) Name and definition of the full sample and various subsamples. The distance ($D$), stellar mass ($M_*$) and star formation rate (SFR) involved in the sample definitions are in units of Mpc, $M_{\odot}$, and $M_{\odot}~\rm yr^{-1}$, respectively; (3) Size of each $L^*$ galaxy sample; (4) Signal-to-noise ratios calculated within the radial range of 10--50 kpc, where the diffuse X-ray emission are most significant; (5)$\&$(6) Core radius in unit of kpc and the $\beta$ value of the X-ray surface brightness profile, fitted by a PSF-convolved $\beta$-model; (7) Logarithm of 0.5--2 keV luminosity (in units of erg~s$^{-1}$), calculated within a radial range of 10--50 kpc; (8) Logarithm of 0.5--2 keV luminosity (in units of erg~s$^{-1}$), calculated within a radial range of 10--200 kpc; (9) Logarithm of hot gas mass (in units of $M_{\odot}$), calculated within a radial range of 10--200 kpc. Quoted errors are at $1\sigma$ confidence level.
}
\end{deluxetable*}

\subsection{Stacking and characterizing eRASS1 spectra}
\label{subsec:spectrum}
We also stack the X-ray spectrum extracted from individual galaxies to obtain an averaged halo spectrum with a sufficient $S/N$. 
Guided by the stacked radial intensity profile (Figure~\ref{fig:profile}), the source region for spectral extraction is chosen to be an annulus with inner-to-outer radii of 10--50 kpc, over which the observed halo X-ray emission is most prominent and subject to little contamination by emission from the galactic disk. 
The background region is chosen to be an annulus with inner-to-outer radii of 300--400 kpc, i.e., same as for the radial intensity profile. 
We employ Xstack (https://github.com/AstroChensj/Xstack), a novel X-ray spectral stacking code, which is originally designed to consistently combine eROSITA rest-frame PI spectra and corresponding instrumental responses, with appropriate weighting factors to preserve spectral shapes (see details in \citealp{Chen_2025,Zxy_2024}). The inputs for this code are: individual source PI spectra, additional lists of individual source redshifts and neutral hydrogen column densities (to shift all sources to a common rest-frame and correct the Galactic absorption, which is procedurally necessary but not crucial for our sample galaxies), the background PI spectra, the ancillary response files (ARF, extracted from the source regions) and response matrix files (RMF, extracted for each source, but actually identical among all sources). Specifically, we use the following command to perform the stacking: {\it runXstack filelist.txt -\,-rsp\_weight\_method FLX -\,-extended},
which is appropriate for the spectral stacking of extended sources, weighting the instrumental responses by exposure $\times$ angular size of the extended source.

\begin{figure}[htbp]
\centering
\includegraphics[width=0.32\textwidth]{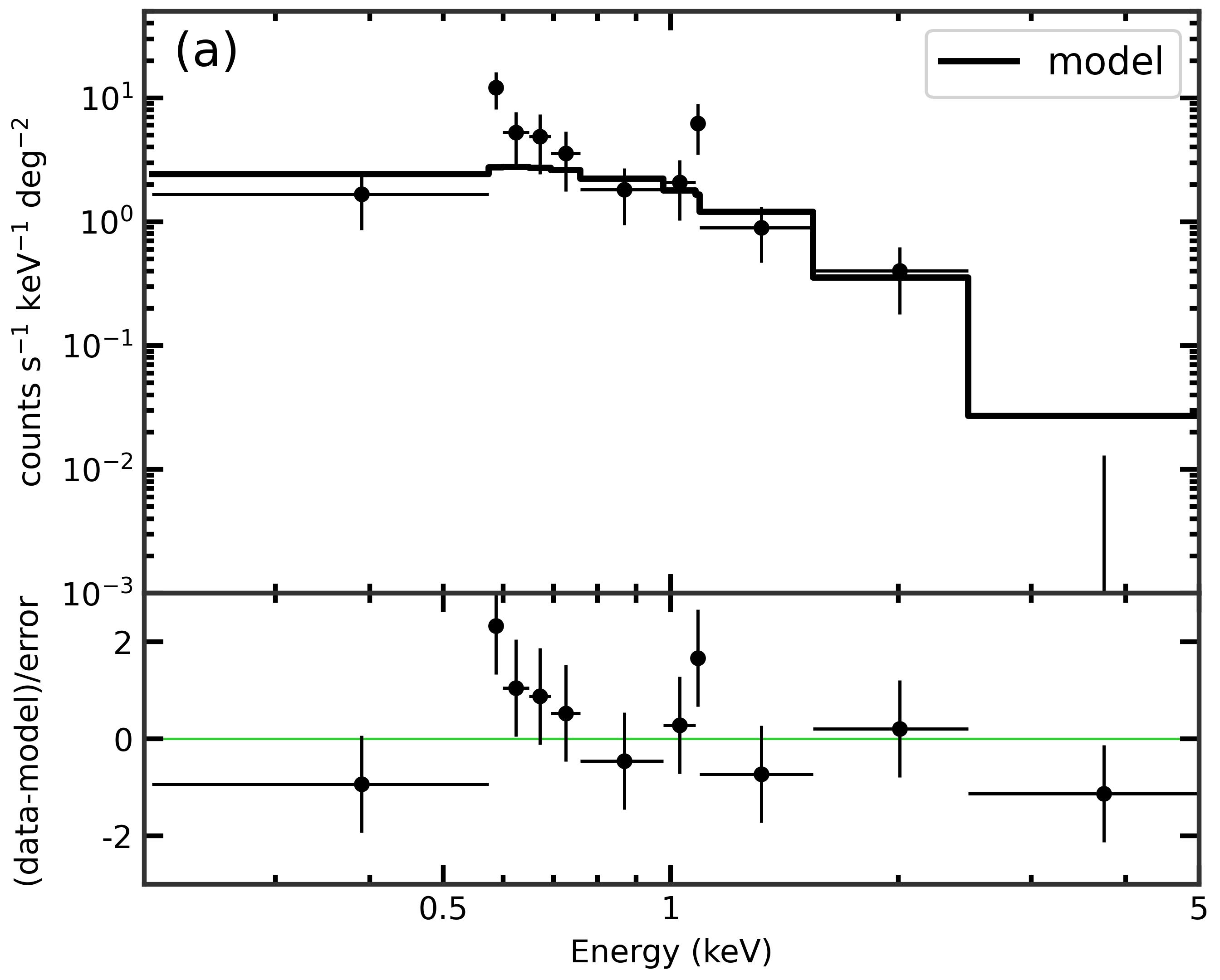}
\includegraphics[width=0.32\textwidth]{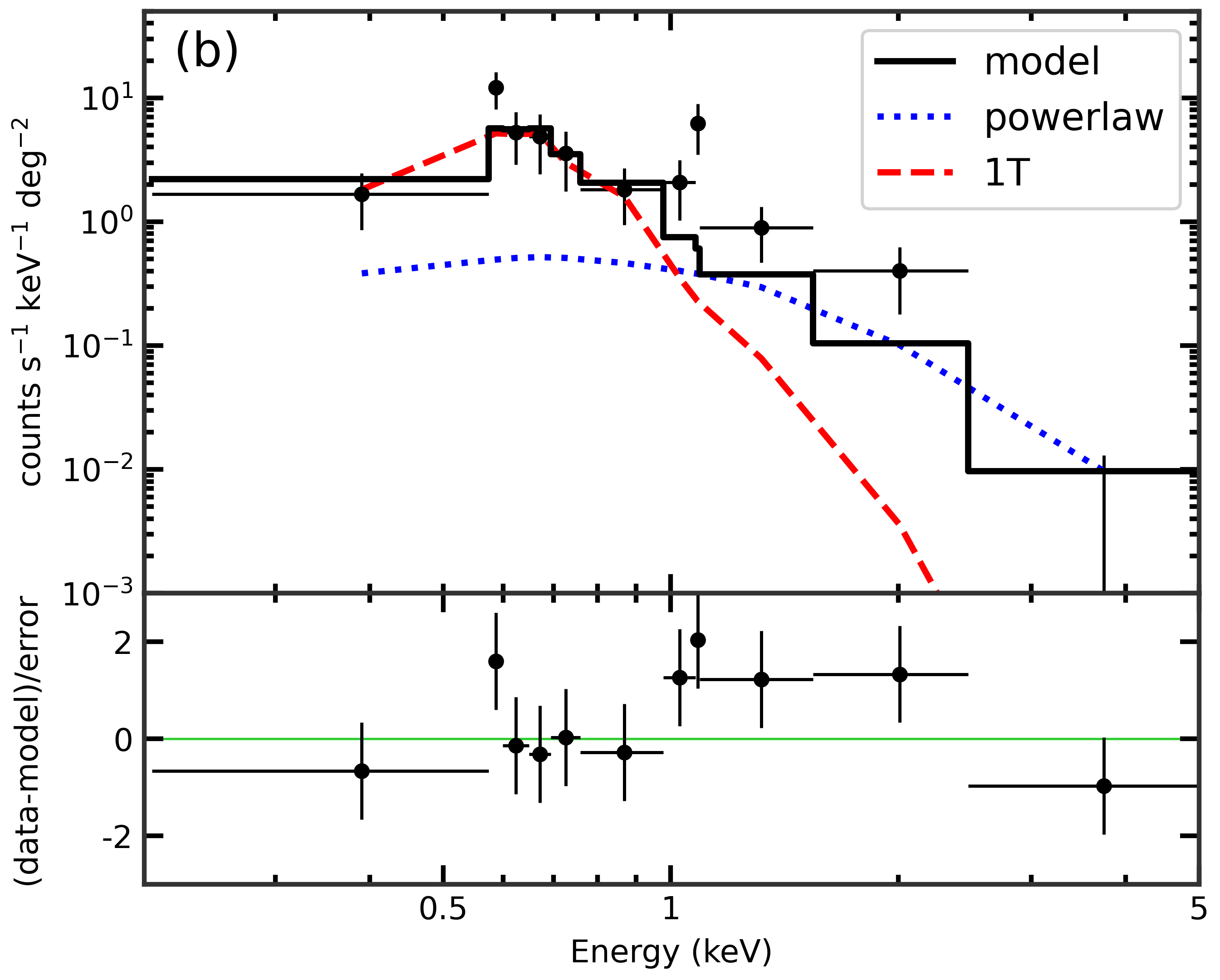}
\includegraphics[width=0.32\textwidth]{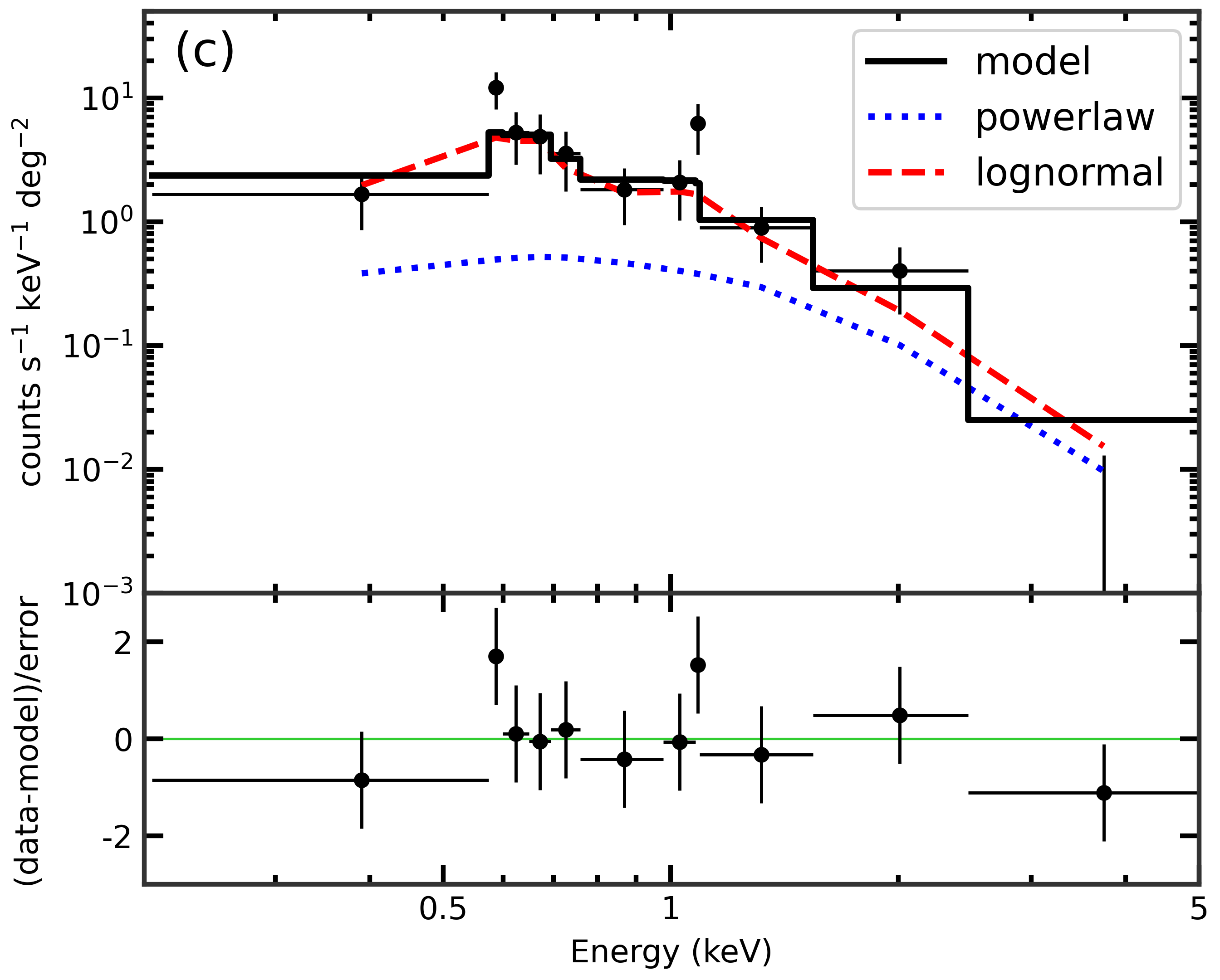}
\caption{Background-subtracted stacked eROSITA X-ray spectrum, extracted from the radial range of 10--50 kpc around the full sample of $L^*$ galaxies. The spectrum has been adaptively grouped to achieve a signal-to-noise ratio greater than 2 per bin. From left to right shows the fitted model: an absorbed power-law, an absorbed single-temperature thermal plasma plus a fixed power-law, and an absorbed thermal plasma with a log-normal temperature distribution plus a fixed power-law. In all panels, the total model is represented by a solid black curve. In panels (b) and (c), the hot plasma component and the power-law component are indicated by red dashed and blue dotted lines, respectively. See details in Section~\ref{subsec:spectrum}.}
\label{fig:spec}
\end{figure}

The stacked spectrum thus derived over 0.2--5 keV is shown in Figure \ref{fig:spec}. 
This spectrum, albeit with a moderate $S/N$, appears more than a smooth continuum.   
Indeed, when fitting the spectrum with a simple absorbed power-law model ({\it tbabs*powerlaw}) in Xspec \citep{Arnaud1996ASPC..101...17A}, the resultant $\chi^2/d.o.f. \approx 13.4/9$ suggests a poor fit, which in particular fails to match a small bump at 0.6--0.7 keV. 
We have also attempted an absorbed thermally Comptonized continuum model ({\it tbabs*nthcomp} in Xspec; \citealp{1996MNRAS.283..193Z,1999MNRAS.309..561Z}), in which blackbody seed photons from the cosmic microwave background are assumed, i.e. with a fixed 2.7 K blackbody seed photon, 
along with a high-energy cutoff electron energy fixed at $1$ GeV. We note that this latter parameter cannot be tightly constrained, due to the fitted photon energy range being orders of magnitude lower than the electron energy, but the assumed value is well consistent with the expectation of a cosmic ray-dominated halo \citep{Hopkins_2025}.
The fitted asymptotic power-law photon-index, $\Gamma=2.2\pm0.3$, is identical with the $\Gamma$ of the absorbed power-law model, and the resultant $\chi^2/d.o.f. \approx 13.4/9$ suggests these two models are essentially the same.
In both cases, the foreground absorption column density is fixed at $N_{\rm H} = 3\times10^{20}\rm~cm^{-2}$, which is typical of the Galactic foreground for our sample galaxies.

We then adopt an absorbed APEC plus power-law model ({\it tbabs}*[{\it apec+powerlaw}] in Xspec) to fit the spectrum.
The foreground absorption column density is also fixed at $N_{\rm H} = 3\times10^{20}\rm~cm^{-2}$. 
We first allow both the powerlaw and apec parameters to vary freely. This yields a statistically acceptable fit ($\chi^2/d.o.f. \approx 8.4/7$) with $\Gamma=1.23^{+0.69}_{-0.79}$ and $kT=0.19^{+0.04}_{-0.03}$~keV, and the two components contributing comparably to the total flux. We further evaluate the likely origin of this power-law component. 
As addressed in Section~\ref{subsec:starresidual}, residual emission of central X-ray point sources, which cannot be completely masked due to the relatively large PSF of eROSITA, is expected to contribute up to 15\% of the 0.5--2 keV halo emission. The wing of disk component (as seen in Figure~\ref{fig:profile}) accounts for an additional 5\%.  Hence, the maximum possible contribution from discrete sources to the total 0.5--2 keV spectrum amounts to 20\%. Motivated by this, we then considered a constrained model in which the power-law parameters are fixed. Specifically, the photon-index is fixed at 1.77, which is derived from the collective spectrum of detected sources within the galactic disk regions. The normalization of the power-law is also fixed as justified above. As shown in Figure~\ref{fig:spec}, this fixed power-law is well compatible with the observed spectrum above $\sim$2 keV.
The optically thin plasma model, APEC, accounts for the truly diffuse emission from the hot CGM. Fixing the abundance at the solar value, we find a best-fit plasma temperature of $0.23^{+0.04}_{-0.03}$~keV, 
which is consistent with the virial temperature of a MW-sized halo. The resultant $\chi^2/d.o.f. \approx 13.2/9$ is comparable to that obtained from the purely nonthermal model, but provides a better match to the 0.6--0.7 bump better, which might be attributed to OVII and OVIII lines. However, this model  underestimates the observed spectrum at energies between 1--2 keV, which might suggest the presence of higher-temperature (i.e. super-virial) gas. 
Therefore, we further replace the single-temperature APEC model with a custom optically thin plasma model (vlntd, \citealp{Ge2015}), in which the plasma temperature has a log-normal distribution and the abundance is again fixed at the solar value. This model, with one additional free parameter compared to APEC, provides an acceptable fit with $\chi^2/d.o.f. \approx 8.2/8$. Using the Akaike information criterion (AIC; \citealp{Akaike1100705}) and the Bayesian information criterion (BIC; \citealp{Schwarz1978AnSta...6..461S}), we compare this model with the previous {\it tbabs*powerlaw} model. The $\Delta \rm AIC$ and $\Delta \rm BIC$ are 3.2 and 2.8, respectively, indicating a statistically meaningful improvement. The best-fit peak temperature, $0.20^{+0.05}_{-0.03}$~keV, is also compatible with the virial temperature, while the temperature dispersion, $>$2.7 keV, is substantial but still compatible with values predicted by 3-dimensional simulations of the CGM around a MW-mass galaxy regulated by supernova-driven outflows \citep{Vijayan2022}.

We also attempted to fit the source and background spectra separately, which is in principle preferable for regions with large angular extent. With the best-fit background model fixed and appropriately rescaled, an additional APEC component yields a temperature of $kT\sim0.24$ keV, while no further component
is required by the data. However, this approach does not provide better constraints than the background-subtracted spectrum, which can be attributed to the degeneracy between the Milky Way halo and the target halo (i.e., with very similar temperature), combined with the large systematic uncertainties of a background spectrum averaged over half the sky. Given the moderate quality of the background-subtracted spectrum, which limits a meaningful examination of more sophisticated models, we adopt the log-normal model as our fiducial, physically motivated description and use it to convert the observed eROSITA count rates to an unabsorbed energy flux (Figure~\ref{fig:profile}).

\begin{deluxetable*}{cccccccccc}
\tablecaption{Fitting results of various models. Fixed parameters are marked with ($f$). All quoted uncertainties are at 1$\sigma$ level.\label{tab:spec_fit}}
\tablehead{
\colhead{Model} & \colhead{$\Gamma$} & \colhead{$kT_e$} & \colhead{$kT_{\rm bb}$} & \colhead{norm\_PL} & \colhead{$kT$} & \colhead{norm\_T} & \colhead{$T_{\rm mean}$} & \colhead{$\sigma$} & \colhead{$\chi^2$/dof}
}
\colnumbers
\startdata
powerlaw & $2.17 \pm 0.33$ & -- & -- & $2.23 \pm 0.04$ & -- & -- & -- & -- & $13.4/9$ \\
nthcomp & $2.17 \pm 0.33$ & $10^{6}\,(f)$ & $10^{-6}$\,(f) & $2.23 \pm 0.04$ & -- & -- & -- & -- & $13.4/9$ \\
powerlaw+apec & $1.23^{+0.69}_{-0.79}$ & -- & -- & $1.35^{+0.58}_{-0.55}$ & $0.19^{+0.04}_{-0.03}$ & $1.87^{+0.82}_{-0.88}$ & -- & -- & $8.4/7$ \\
powerlaw+apec & $1.77\,(f)$ & -- & -- & $0.49\,(f)$ & $0.23^{+0.04}_{-0.03}$ & $2.19^{+0.55}_{-0.51}$ & -- & -- & $13.2/9$ \\
powerlaw+vlntd & $1.77\,(f)$ & -- & -- & $0.49\,(f)$ & -- & -- & $0.20^{+0.05}_{-0.03}$ & $>2.7$ & $8.2/8$ \\
\enddata
\tablecomments{All models are multiplied by a Galactic absorption component (\textsc{tbabs}) with $N_{\rm H}$ fixed at $3\times10^{20}$~cm$^{-2}$. (1) Model name, the model components are: powerlaw = simple photon power law; apec = simple thermal plasma model; nthcomp = thermally comptonized continuum; vlntd = simple plasma model with a lognormal temperature distribution; (2) Photon index of powerlaw or asymptotic power-law photon index of nthcomp; (3) Electron temperature of nthcomp in keV; (4) Seed photon temperature of nthcomp in keV; (5) Normalization of powerlaw/nthcomp in units of $10^{-4}$ ph cm$^{-2}$ s$^{-1}$ keV$^{-1}$; (6) Plasma temperature of apec in keV; (7) Normalization of apec in units of $10^{-4}$ $\frac{10^{-14}}{4\pi[D_{A}(1+z)]^2}\int n_en_H dV$; (8) Mean temperature of the lognormal distribution (vlntd) in keV; (9) Temperature dispersion of vlntd in keV; (10) Goodness of fit, given by the ratio of chi-square and the degrees of freedom.} 
\end{deluxetable*}

\section{Discussion}\label{sec:discussion}
\subsection{Residual X-ray emission from discrete sources}
\label{subsec:starresidual}
Some residual  emission from discrete sources still exist in the apparently diffuse X-ray emission after masking the detected sources (Section~\ref{subsec:data}), which arise primarily from unresolved stellar populations and the unmasked PSF halo of resolved sources. 
The contribution from old stellar populations, in particular low-mass X-ray binaries, is approximated by employing the near-infrared starlight distribution (Appendix~\ref{sec:WISE}) normalized according to the scaling relation $L_{\rm X}/M_* = 5.0 \times 10^{28}\rm~erg~s^{-1}~M_\odot^{-1}$ (derived from the empirical relation of stellar mass and 2--10 keV luminosity from \citealp{Lehmer_2010ApJ}, assuming an intrinsic power-law spectrum with a canonical photon-index of 1.7), as shown in Figure~\ref{fig:profile}.
According to this normalized WISE starlight profile, the old stellar populations contribute to the observed diffuse emission beyond 10 kpc at a level of $\lesssim$5\%.
Star-forming activities giving rise to high-mass X-ray binaries and supernova-heated hot gas may also produce copious X-ray emission.
Employing the scaling relation between star formation rate and 0.5--2 keV luminosity (extrapolated from \citealp{Lehmer_2010ApJ}, as mentioned above), $L_{\rm X}/SFR = 8.9 \times 10^{38}\rm~erg~s^{-1}/(M_\odot~yr^{-1})$, we estimate an average luminosity  of $1.6\times 10^{39}~\rm erg~s^{-1}$ per sample galaxy. This component, however, should concentrate in the disk region, especially where star formation is active \citep{Strickland_2004}. Hence a large fraction of the star formation-related X-ray emission should have been removed when masking the detected sources, which themselves have a tendency of concentrating in the inner galactic region (see middle panel of figure~\ref{fig:eRASS1src}).
Thus we expect that direct contribution by young stellar populations to the observed halo emission is negligible. 
On the other hand, the PSF-scattered photons of the detected nuclear sources, which may trace a low-luminosity AGN, might contribute up to 15\% of the halo emission, based on the cumulative luminosity of the nuclear sources and conservatively adopting a PSF-scattered fraction of 5\%. 
Overall, we expect that the collective residual emission from discrete sources accounts for 5--20\% of the observed halo emission\footnote{The fractions quoted here refer to ratios of integrated 0.5--2 keV luminosities within the 10--50 kpc annulus,
rather than detector-count fractions.}.

\subsection{Comparison with the SDSS MW-mass sample}

The analysis and results outlined in the above represent the first systematic investigation of the hot CGM around a representative sample of $L^*$ galaxies in the local Universe (within 50 Mpc). 
It is instructive to compare our results with recent studies \citep{Zhang_2024, Chadayammuri_2022,Comparat_2022}, which are also based on stacking the eROSITA data, but for a large number of more distant MW-sized galaxies selected from wide-field galaxy surveys such as the Sloan Digital Sky Survey (SDSS).
In particular, we compare in Figure~\ref{fig:profile}a the 0.5--2 keV intensity profile (purple symbols) derived in an influential study by \cite{Zhang_2024}, which is based on stacking the eROSITA data of currently the largest sample ($\sim$$10^4$) of low-redshift (median value of 0.08) MW-sized galaxies primarily extracted from the SDSS spectroscopic galaxy catalog (hereafter referred to as the SDSS sample).
The profile of the SDSS sample effectively begins at an inner radius of $\sim$10 kpc, which is due to a combined effect of nuclear source masking and a larger equivalent physical scale for the same eROSITA PSF, and extends to $\sim$500 kpc, although most bins beyond a radius of 200 kpc are only upper limits.
We note that the smaller distances of our 50MGC sample help compensate for the much smaller sample size, resulting in a comparable, albeit somewhat lower, signal-to-noise ratio ($S/N$) to that of the SDSS sample.

As shown in Figure~\ref{fig:profile}, the X-ray intensity profile of our 50MGC sample tends to be lower and steeper than that of the SDSS sample of \cite{Zhang_2024}. The latter is higher than the former by a factor of $\sim$2 at a radius of 50 kpc, and at larger radii, the latter is still detectable whereas the former becomes an upper limit.
We consider the cause of this difference due to one or a combination of the following factors. 
First, while both profiles are subject to the same eROSITA PSF broadening, the on-average $\sim$10 times greater distance of the SDSS sample means that its profile is more flattened in terms of physical scale. We thus compare the fitted $\beta$-model for the two samples, both of which have accounted for the PSF broadening. The value of $\beta \approx 0.52$ for our full sample indicates an intrinsically steeper profile than the SDSS sample with $\beta \approx 0.43$. The unabsorbed 0.5--2 keV luminosity between 10--200 kpc is also smaller in our 50MGC sample (${\rm log} L_{0.5-2} \approx 39.7$) compared to that of the SDSS sample (${\rm log} L_{0.5-2} \approx 39.9$).
Second, the small size of the 50MGC sample, which is about 1\% of the SDSS sample, may introduce a selection bias. However, our final sample of $L^*$ galaxies is derived from a physical volume of about (80 Mpc)$^3$ equivalent, which is generally considered to be cosmologically presentative. For instance, the smallest volume simulated by the IllstriusTNG suites is about (50 Mpc)$^3$. 
Third, given the proximity of the 50MGC sample, the point-source detection limit (figure~\ref{fig:eRASS1src}), in terms of apparent luminosity, is $\lesssim$100 times lower than that achieved in the SDSS sample, which is on-average 10 times more distant. This allows to detect and remove more low-luminosity ($L_{0.5-2}\sim10^{39-41}\rm~erg~s^{-1}$) sources that are not possible for the SDSS sample. 
We estimate that the resolved sources within a radius of 200 kpc, after statistically subtracting the CXB sources, on-average account for a total $L_{0.5-2} \approx 3.2\times10^{39}\rm~erg~s^{-1}$ per $L^*$ galaxy, which is on the same order of magnitude as the diffuse X-ray emission, highlighting the importance of masking them. 
In addition, we have discarded $L^*$ galaxies against a dense environment and/or a high diffuse background. Qualitatively, this means that our 50MGC sample is likely less contaminated by the large-scale environment than the SDSS sample.

\begin{figure}[htbp]
\centering
\includegraphics[width=0.5\textwidth]{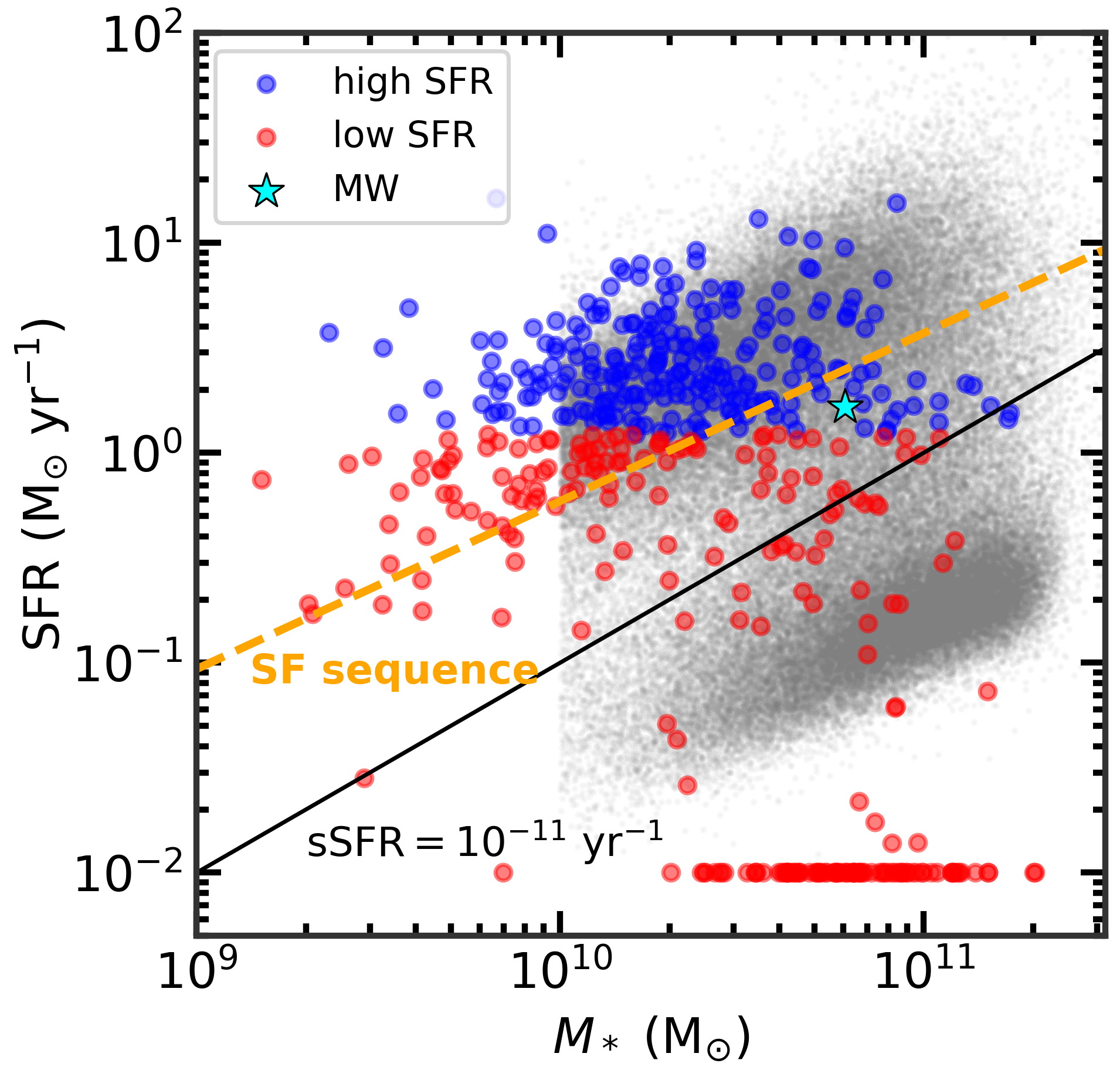}
\caption{Distribution of the 50MGC $L^*$ galaxies (large dots) and the SDSS central galaxies (small dots) from \cite{Zhang_2025} in the SFR vs. $M_*$ plane. The $M_*$ and SFR of the 50MGC $L^*$ galaxies are derived from WISE images, while those of the SDSS central galaxies are taken from  \cite{Chen2012} and \cite{Brinchmann2004}. Blue and red symbols distinguish the high-SFR and low-SFR galaxies defined in this work, respectively. 
For 50MGC galaxies whose WISE-based SFR measurements fall below $0.01$ (i.e., unreliablely low values), a nominal SFR $=0.01~M_{\odot}~\rm yr^{-1}$ is artificially assigned for illustrative purposes. The Milky Way is represented by a cyan star, with SFR and $M_*$ derived from \cite{Licquia2015ApJ...806...96L}.
The orange dashed curve indicates the main-sequence of star-forming galaxies from \cite{Chang2015} and the black line marks a specific star formation rate (sSFR) $=10^{-11}~\rm yr^{-1}$, introduced in \cite{Zhang_2025} to separate star-forming and quiescent galaxies.}
\label{fig:M-SFR}
\end{figure}

Fourth, in the SDSS sample, MW-sized galaxies are defined as galaxies with a stellar mass between $10.5 < {\rm log}(M_*/M_\odot) < 11.0$, which is in fact more massive than half of our $L^*$ galaxies. Given a {\it naive} expectation that the amount of hot CGM scales with galaxy mass, the SDSS sample with a systematically higher stellar mass should also have a systematically higher X-ray luminosity and plausibly a flatter profile. 
If we focused on the high-mass subsample of $L^*$ galaxies, which have a stellar mass of $10.4 < {\rm log} (M_*/M_\odot) < 11.3$ (i.e., more comparable to the SDSS sample), the two samples indeed have comparable $L_{0.5-2}$ between 10--200 kpc, but they still differ in $\beta$ (0.55 vs. 0.43), meaning that the 50MGC sample remains the one with a steeper intensity profile.
Moreover, as shown in figure \ref{fig:M-SFR}, the two sample differ in the fraction of quiescent galaxies. In the 50MGC sample, most galaxies lie on or near the so-called main-sequence of star-forming galaxies, which follows a tight correlation between $M_*$ and SFR \citep{Chang2015}, 
whereas in the SDSS sample more than half are truly quiescent galaxies that fall substantially below the main-sequence. It is not straightforward to directly compare the behavior of the actively star-forming galaxies in the two samples, because many of the low-SFR galaxies defined in our 50MGC sample also tend to have a specific star formation rate (i.e., SFR/$M_*$) close to the main-sequence. Still, the low-SFR subsample exhibits an on-average 0.6 (3.8) times higher $L_{0.5-2}$ within 10--50 (10--200) kpc. Notably, this is consistent with the finding of \cite{Zhang_2025} that  the stacked X-ray emission from the hot CGM of their star-forming galaxies is actually no brighter than that of the quiescent galaxies of a similar stellar mass range. 

Fifth, and lastly, the SDSS sample consists of galaxies at a slightly earlier evolutionary stage of the Universe. 
The median redshift of the SDSS sample, $\bar{z} \sim 0.08$, corresponds to a look-back time of about 1 Gyr.
On the one hand, studies  of black hole mass growth rate and star formation rate as a function of redshift \citep{Shankar2009}, based on empirical measurements, suggest a non-negligible ($\lesssim$50\%) decrease of both rates from $z \sim 0.1$ to $z = 0$, which in principle implies for a decaying energy input to the CGM by stellar feedback and/or AGN feedback.
On the other hand, given a typical halo sound crossing time of $\tau_{\rm dyn} \sim \rm 100~kpc/300~km~s^{-1} \sim 0.3~Gyr$, 
and a characteristic radiative cooling time of $\tau_{\rm cool} \sim E_{\rm hot}/L_{0.5-2} \sim (M_{\rm hot}kT)/({\mu}m_{\rm H}L_{0.5-2}) \sim$~85 Gyr, a systematic reduction of the hot CGM over the past 1 Gyr would seem rather unlikely. 

In summary, the various factors considered in the above may conspire to cause the apparent difference in the stacked X-ray intensity profiles of the two samples. Regardless, it is evident that the hot CGM around nearby $L^*$ galaxies on-average have a relatively small content (a few $10^{10}\rm~M_\odot$; Table~\ref{tab:tab1}), with indications that the hot gas mass may depend on both the star formation rate and nuclear activity. 

\subsection{Comparison with cosmological simulations}

We further compare the observed X-ray intensity profile of the $L^*$ galaxies with the prediction by numerical models of galaxy formation and evolution.
The prevalence of a tenuous, hot CGM around $L^*$ galaxies is a generic prediction of state-of-the-art cosmological simulations, such as EAGLE \citep{Schaye_2015,Crain_2015,McAlpine_2016} and IllustrisTNG \citep{Pillepich_2018,Nelson_2018}. 
A comparison with the IllustrisTNG simulation is particularly relevant, as the latter offers a high-resolution simulation suite known as the TNG50, which traces the cosmological evolution of a co-moving volume of $\sim$(50 Mpc)$^3$ and produces a significant number of low-redshift MW analogs with ${\rm log}(M_*/M_\odot) \sim 10^{10.5-11}$ \citep{Pillepich_2024}. We thus obtain the predicted X-ray intensity profile of 152 TNG50 MW analogs (see details in Appendix) and convolve it with the eROSITA PSF according to a physical scale at the median distance of the 50MGC sample (green dashed curve in Figure~\ref{fig:profile}). Evidently, the observed profile is quite consistent with the TNG50-predicted profile over the radial range of 10--50 kpc. 
On the other hand, it has been shown that the predicted amount of the hot CGM around MW-sized galaxies generally exhibits a significant variation from galaxy to galaxy within the same simulation suite and across different simulation suites, which might be attributed to the specific feedback prescriptions employed \citep{Oppenheimer2020ApJ...893L..24O,Silich_2025}. According to \citealp{Silich_2025}, simulations employing a kinetic-mode AGN feedback (e.g., TNG, Simba \citealp{Dave2019MNRAS.486.2827D}) or  employing only stellar feedback (e.g., FOGGIE \citealp{Peeples2019ApJ...873..129P}, TEMPEST \citealp{Hummels2019ApJ...882..156H}) predict large-scale outflows into the hot CGM, 
while simulations employing only thermal-mode AGN feedback (e.g., EAGLE) tend to produce more spherical hot CGM with a steeper density decline.

\subsection{Predominantly thermal origin of the diffuse X-ray emission}
A number of recent studies have also compared the observational result of \cite{Zhang_2024} with the prediction by various numerical simulations \citep{Silich_2025,Vladutescu-Zopp_2025,Zhang_2025ApJ...991..170Z}. Due to the relatively flat X-ray intensity profile of the SDSS sample, however, these comparisons often came to the conclusion that the simulations tend to predict a deficit of hot CGM around MW-sized galaxies, if the observed values of the SDSS sample were taken as nominal (this situation is now alleviated in view of the nearby $L^*$ galaxies). 
This has invoked the proposal that part, if not all, of the observed halo diffuse X-ray emission is of a non-thermal origin, where inverse Compton scattering between the cosmic microwave background (CMB) photons and relativistic electrons in a cosmic-ray (CR) prevalent halo produces the observed diffuse X-ray emission \citep{Hopkins_2025}.
In the extreme case of this non-thermal scenario, the CGM is primarily supported by the CR pressure, without the need for a substantial hot gas component, thus only little thermal X-ray emission should be present.   
To test the non-thermal scenario, 
we examine the spectrum of the diffuse X-ray emission stacking the full sample of $L^*$ galaxies over a radial range of 10--50 kpc (Figure~\ref{fig:spec}; see Section~\ref{subsec:spectrum} for details). 
Fitting various phenomenological models to this stacked spectrum reveals that neither a single non-thermal model (i.e., a power-law continuum or a Comptonized continuum) or a single thermal model (i.e., an optically thin plasma consisting of a bremsstrahlung continuum and various atomic lines) provides a satisfactory description to the observed spectrum. Instead, a composite model combining an optically thin plasma and a power-law continuum characterizes the observed spectrum well (Figure~\ref{fig:spec}c). 
In the latter case, the extra power-law component accounts for the observed photons above $\sim2$ keV and is well consistent with the estimated residual emission from discrete sources (see Section~\ref{subsec:spectrum}). The hot plasma is assumed to follow a log-normal temperature distribution (with mean $kT\sim 0.2$ keV and $\sigma >$ 2.7 keV), while a single temperature plasma ($kT\sim 0.2$ keV) with additional non-thermal contribution is also compatible with the current spectrum. In both models, the thermal emission contributes approximately half or more of the total flux in the
0.5--2 keV band, favoring a substantial thermal contribution rather than a predominantly non-thermal origin of the observed halo X-ray emission.

\subsection{Examine the azimuthal anisotropies of the hot CGM}
\begin{figure}[htbp]
\centering
\includegraphics[width=0.45\textwidth]{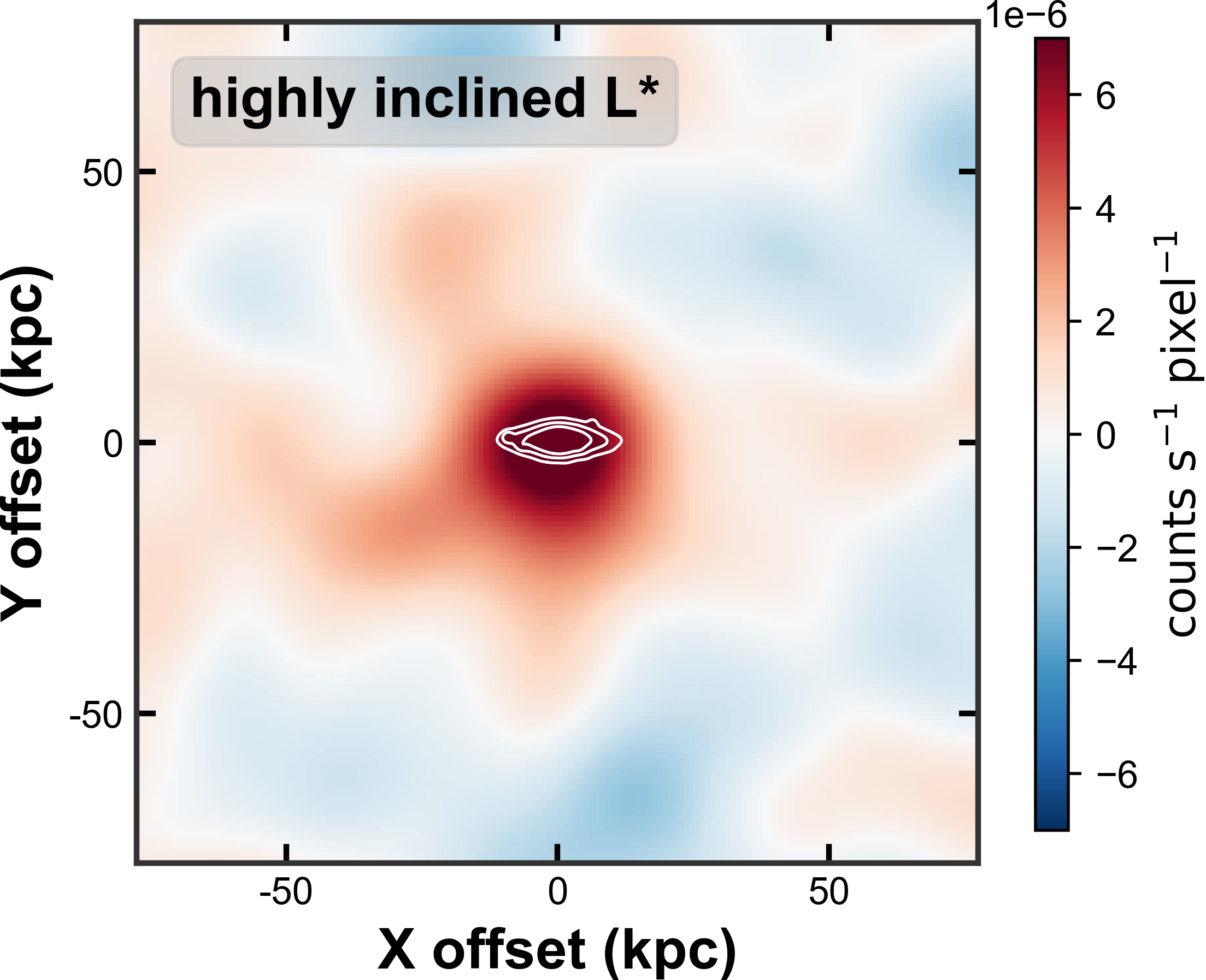}
\includegraphics[width=0.4\textwidth]{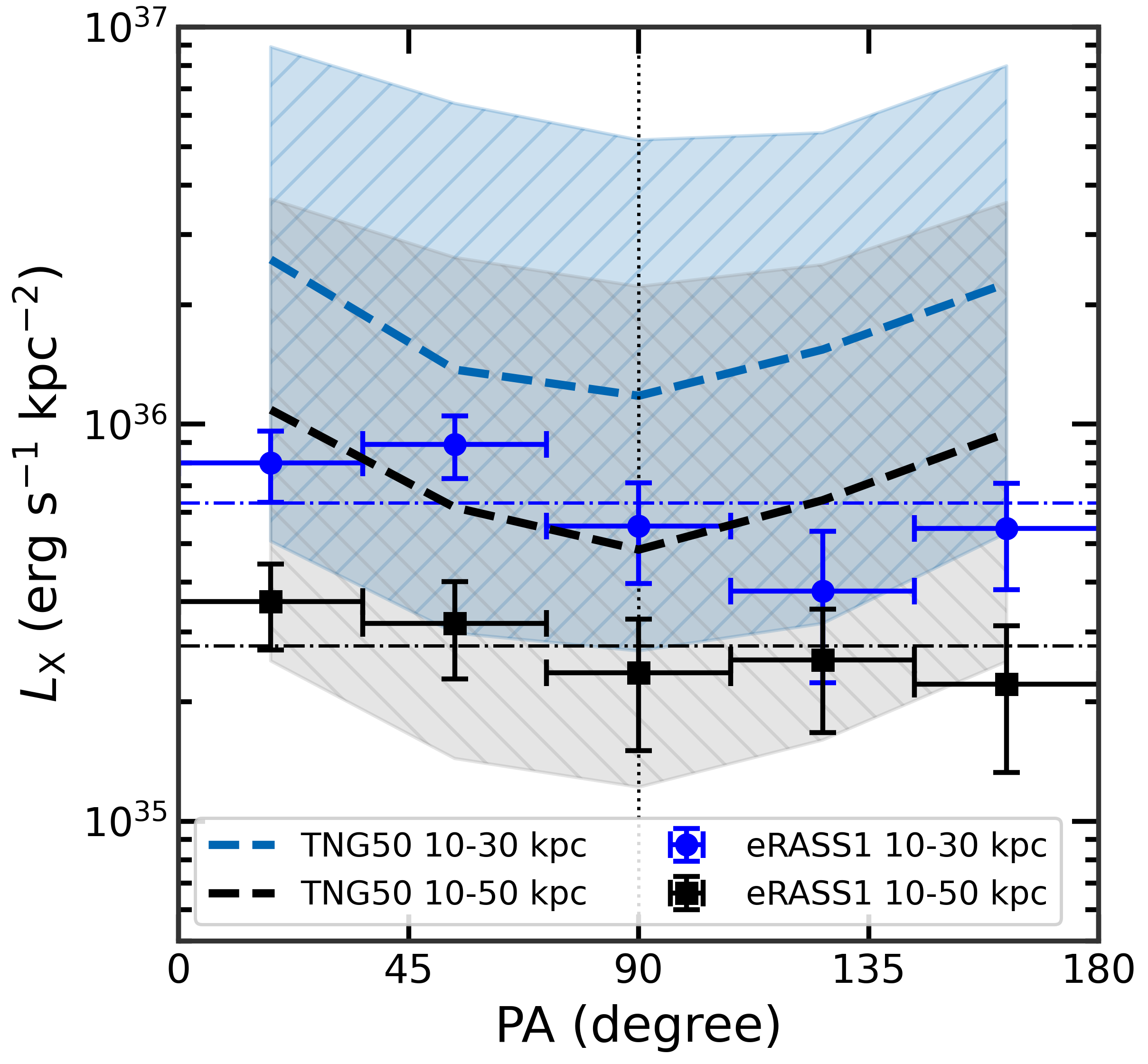}
\caption{{\it Left}: Stacked background-subtracted 0.2--2.3 keV surface brightness image of 114 galaxies with a highly inclined disk, defined as having $b/a<0.5$, where $a$ and $b$ are the semi-major and semi-minor axes of the $I_{22,W1}$ isophote derived from WISE W1 images (see Appendix \ref{sec:WISE}). The X-ray images of individual galaxies have been rotated such that their major-axis is aligned with the horizontal (X) direction. The white contours are derived from the WISE image aligned and stacked in the same way, illustrating the extent of the inclined stellar disk. {\it Right}: Azimuthal 0.2–2.3 keV intensity profiles extracted from the stacked X-ray image of the highly inclined galaxies, over a radial range of 10–30 kpc (blue) and 10--50 kpc (black), which avoids the disk region. A position angle of $0^{\circ}$ corresponds to the positive X-axis (i.e. the galactic plane) and $90^{\circ}$ indicates the minor-axis. Each bin is $36^{\circ}$ wide, with counts from regions above and below the galactic plane combined to improve the the $S/N$. The horizontal
lines represent the average intensity. Both profiles are consistent with no significant azimuthal dependence. 
Synthetic intensity profiles of the TNG50 MW analogs are plotted as the blue (10--30 kpc) and black (10--50 kpc) dashed lines, with 1$\sigma$ scatters represented by the strips.
}\label{fig:azimuthal}
\end{figure}

Both stellar feedback and/or AGN feedback are able to drive a large-scale outflow \citep{Veilleux_2005,Fabian_2012,Hopkins_2012,Cicone_2014}, either continuously or episodically, which propagates into the CGM and modifies the latter's structure by depositing mass, energy and momentum.
If the outflow has a preferred orientation, in particular along the rotation axis of the galactic disk as generally expected for bipolar outflows driven by an AGN or nuclear starburst, the hot CGM should exhibit some degree of anisotropy in response to the outflow. We test this prediction by examining the azimuthal distribution of the diffuse X-ray emission from the halos of the $L^*$ galaxies. Specifically, we align and stack the individual {\rm 0.2--2.3 keV} images of 114 highly inclined $L^*$ galaxies, defined as having a minor-to-major axis ratio less than 0.5 in the WISE W1-band isophote (roughly corresponding to a stellar disk viewed at an angle greater than 60$^\circ$). As shown in the left panel of Figure~\ref{fig:azimuthal}, no significant azimuthal variation is evident in the stacked image. This is further supported by a more quantitative view of the azimuthal X-ray intensity distribution averaged within a radial range of 10--30 kpc or 10--50 kpc (Figure~\ref{fig:azimuthal} right), which is consistent with a constant distribution. 
We also test splitting this sample into different SFR and mass subgroups and find that the results do not show significant differences within each group and are consistent with those obtained from the full sample of highly inclined galaxies. At last, we extracted the corresponding azimuthal intensity profiles from the MW analogs in the TNG50 simulation.  As shown as dashed lines in the right panel of Figure~\ref{fig:azimuthal}, the synthetic X-ray intensity profiles tend to exhibit lower surface brightness along the minor axis, qualitatively consistent with the common presence of X-ray cavities seen in TNG50 MW analogs \citep{Pillepich2021MNRAS.508.4667P}. Overall, neither our observed azimuthal profiles nor the TNG50 predictions show the sinusoidal pattern expected from bipolar outflows or bubbles, which may reflect that such outflows are launched at different epochs, and their varying evolutionary phases act to smear out any detectable anisotropy. On the other hand, a recent study by \cite{Sacchi_2025} reported an excess of extraplanar X-ray emission at $\lesssim$14 kpc along the minor axis, based on the aligned and stacked 0.5--2 keV {\it Chandra} image of $\sim$100 highly inclined galaxies located at a median distance of 130 Mpc. The apparent discrepancy can be understood if in reality the degree of the anisotropy degrades rapidly with increasing radius. This can happen if the {\it Chandra} sample in \cite{Sacchi_2025} contains systematically more galaxies with kpc-scale active bipolar outflows, which unfortunately, cannot be resolved by the eROSITA data. Alternatively, galactic-scale bipolar hot gas outflows may be intrinsically rare in the local Universe, as discussed in \cite{He_2025b}. Overall, our results suggest the hot CGM of 50MGC galaxies does not show detectable anisotropy on scales $\gtrsim$ 30 kpc. However, given the modest signal-to-noise ratio and the relatively large eROSITA PSF, the current data do not permit a definitive conclusion.

\section{Summary} \label{sec:summary}
Using the unprecedented X-ray data afforded by the SRG/eROSITA all-sky survey, we detect and characterize the long-sought hot circumgalactic medium (CGM) around a representative sample of nearby $L^*$ galaxies, based on stacking their X-ray images and spectra. Our conclusions can be summarized as follows:

(1) We obtain a firm detection ($7.4\sigma$) of diffuse X-ray emission over a radial range of 10—50 kpc, and provide a stringent upper limit for 50—200 kpc, i.e. out to the virial radius of a MW-sized galaxy. The collective contribution from discrete unresolved sources and PSF scattering accounts for 5$\%$--20$\%$ of the observed diffuse emission.

(2) The diffuse emission exhibits a relatively steep intensity profile ($\beta\sim 0.5$) and reflects a moderate amount (a few $10^{10}~M_{\odot}$) of hot gas in the halo, based on the PSF-convolved $\beta$-model fitting. The hot gas distribution and content may depend on both the star formation rate and nuclear activity.

(3) The stacked spectrum can be well characterized by a composite model combining an optically thin hot plasma with a log-normal temperature distribution and a power-law continuum. The latter is required to account for the residual emission from discrete sources. A predominantly non-thermal origin of the observed halo X-ray emission, such as an inverse Compton continuum as expected for a cosmic-ray pressure-dominating halo, is not preferred by the data. 

(4) We found no detectable anisotropy in the halos of a subset of $L^*$ galaxies with highly inclined disk and the MW analogs in the TNG50 simulation, which can be interpreted as the varying evolutionary phases of the bipolar outflows/bubbles act to smear out any detectable anisotropy in the stacking analysis.

This study provides the first systematic detection, and constraints on the statistical properties of, the hot CGM around $L^*$ galaxies in the local Universe.
The moderate amount of the hot gas inferred, along with a moderately steeply declining density distribution, 
shall not only invite a close comparison with state-of-the-art and next-generation numerical models of MW-like galaxies, but also encourage new multi-wavelength observations toward individual $L*$ systems, in particular with future X-ray  instruments (e.g., HUBS, \citealp{Cui2020JLTP..199..502C}, and NewAthena, \citealp{Cruise2025NatAs...9...36C}) featuring a large collecting area and wide field-of-view.

\section*{Acknowledgements}
The authors acknowledge support by the National Natural Science Foundation of China (grant 12225302) and the Fundamental Research Funds for the Central Universities (grant KG202502).
The authors wish to thank Shijiang Chen, Meicun Hou, Zhensong Hu, Feng Yuan, Yi Zhang, Xueying Zheng for helpful discussions on various aspects of this study. 

This work is based on data from eROSITA, the soft X-ray instrument aboard SRG, a joint Russian-German science mission supported by the Russian Space Agency (Roskosmos), in the interests of the Russian Academy of Sciences represented by its Space Research Institute (IKI), and the Deutsches Zentrum für Luft- und Raumfahrt (DLR). 
The SRG spacecraft was built by Lavochkin Association (NPOL) and its subcontractors, and is operated by NPOL with support from the Max Planck Institute for Extraterrestrial Physics (MPE). The development and construction of the eROSITA X-ray instrument was led by MPE, with contributions from the Dr. Karl Remeis Observatory Bamberg $\&$ ECAP (FAU Erlangen-Nuernberg), the University of Hamburg Observatory, the Leibniz Institute for Astrophysics Potsdam (AIP), and the Institute for Astronomy and Astrophysics of the University of Tübingen, with the support of DLR and the Max Planck Society. The Argelander Institute for Astronomy of the University of Bonn and the Ludwig Maximilians Universität Munich also participated in the science preparation for eROSITA. 
The eROSITA data shown here were processed using the eSASS software system developed by the German eROSITA consortium. 

This research made use of Photutils, an Astropy package for detection and photometry of astronomical sources \cite{bradley_2026_19636730}. The full sample of our stacked galaxies is available at https://doi.org/10.5281/zenodo.21216062.

\appendix
\setcounter{figure}{0}
\renewcommand{\thefigure}{A\arabic{figure}}

\section{Measuring galaxy and disk properties from WISE photometry}\label{sec:WISE}
It is desired to gather fundamental integrated properties, in particular the stellar mass ($M_*$) and star formation rate (SFR), of our sample galaxies.  
While $M_*$ is one of the parameters offered by the 50MGC, the source of $M_*$ is inhomogeneous and is not available for all cataloged galaxies. 
Moreover, the 50MGC provides no information about the SFR.
Therefore, to obtain a uniform measurement for $M_*$ and SFR, as well as the disk orientation and inclination angles of our sample galaxies, we make use of the Wide-field Infrared Survey Explorer (WISE)  images \citep{Wright_2010}, following the method outlined in \cite{Hu_2024}. 
Specifically, for each galaxy, 
the infrared images were downloaded from the WISE portal (https://irsa.ipac.caltech.edu/applications/wise/). 
We defined the galaxy size as the $22\ \mathrm{mag \ arcsec^{-2}}$ (Vega) isophote in the $W1$ band ($I_{22,W1}$), which were constructed with the Python tool {\tt photuitils} (https://photutils.readthedocs.io/en/stable/index.html) and corresponds to the 2.5$\sigma$ sky surface brightness level \citep{Jarrett_2019}. 
The major-axis and position angle of the $I_{22,W1}$ isophote were taken to be the length and the position angle of the putative galactic disk. 
$M_*$ was then calculated using the stellar mass-luminosity scaling relation of \cite{Jarrett_2019}, 
based on the absolute magnitude of $W1$ ($3.4\ \mathrm{\mu m}$) and $W2$ ($4.6\ \mathrm{\mu m}$) bands within the $I_{22,W1}$ isophote.
SFR was measured using the  SFR-luminosity scaling relation of \cite{Jarrett_2019}, based on the $W1$ and $W3$ ($12\ \mathrm{\mu m}$) band magnitudes within the $I_{22,W1}$ isophote.

We further quantify the $W1$ band radial surface brightness distribution averaged over the full sample of $L^*$ galaxies, in a way similar to that for the eRASS1 X-ray data.   
This was done by first stacking WISE $W1$ band images of individual galaxies, which have been aligned to their respective center and rescaled to the same spaxel scale according to a common distance of 50 Mpc.
For consistency with the stacked X-ray image, we have also applied circular masks according to the presence of nuclear X-ray sources in some galaxies (Section~\ref{subsec:data}).
We considered two versions of the stacked image, i.e., with and without convolution of the eROSITA PSF. 
The former case is relevant when unresolved stellar emission is concerned, because the eROSITA PSF is much larger than the WISE/$W1$ PSF (FWHM 30$''$ versus 6$''$).
We adopted the PSF image\footnote{We used the PSF image from the eSASS DR1 CALDB (named as tm1\_2dpsf\_190220v03.fits) for the convolution of images, which was from ground-based PANTER measurements. Here TM1 is used because the CALDB PSF models of all seven telescope modules are currently identical. For the PSF-aware $\beta$-model fitting, we use the PSF model adopted by \cite{Zhang_2024}, which is based on the in-flight calibrated PSF. We note that differences between the on-ground and in-flight PSF models have negligible effect on our results.} from the eROSITA Science Analysis Software System (eSASS; \citealp{Brunner_2022}) calibration file and convolved it to the individual $W1$ images before rescaling and coadding.
We then extracted the infrared surface brightness profile for both cases. A local sky background was estimated and subtracted from the surface brightness profile.
Fitting the profile with a S{\'e}rsic model, $I(R)=I_0~e^{-k_n R^{1/n}}$, 
we found a S{\'e}rsic index $n \sim 1.2$ and $n \sim 2.1$ with and without the eROSITA PSF convolution.
The latter value is reasonable for a typical galactic disk, validating our stacking procedure.
A third version of stacked image was also attempted, this time aligning the major-axis of the individual galaxies before direct co-adding, which is mainly to trace the two-dimensional starlight distribution as outlined by the contours shown in the left panel of Figure~\ref{fig:azimuthal}.

\section{Test image stacking}
\label{sec:test}
\begin{figure}[htbp]
\centering
\includegraphics[width=0.33\textwidth]{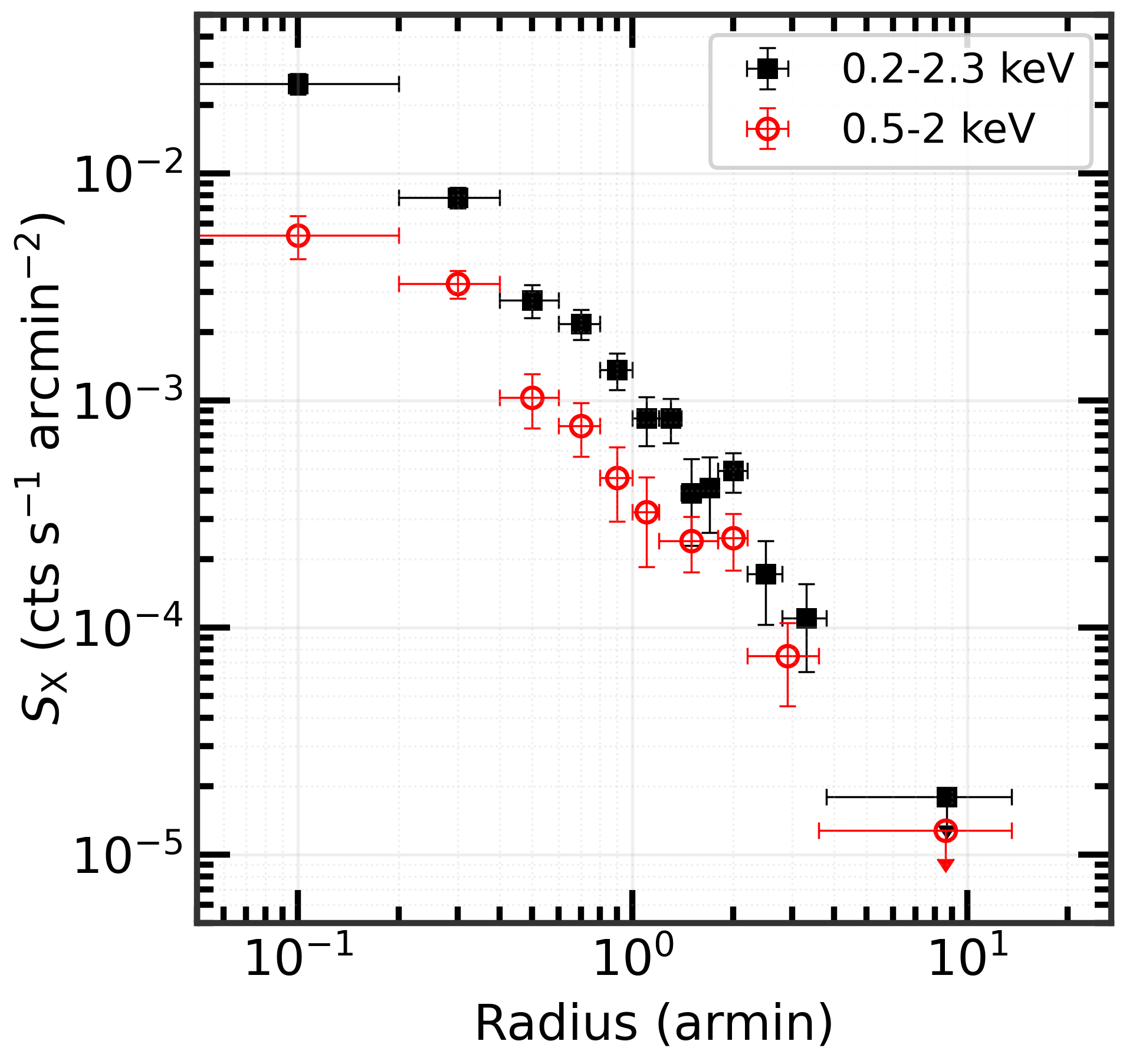}
\includegraphics[width=0.33\textwidth]{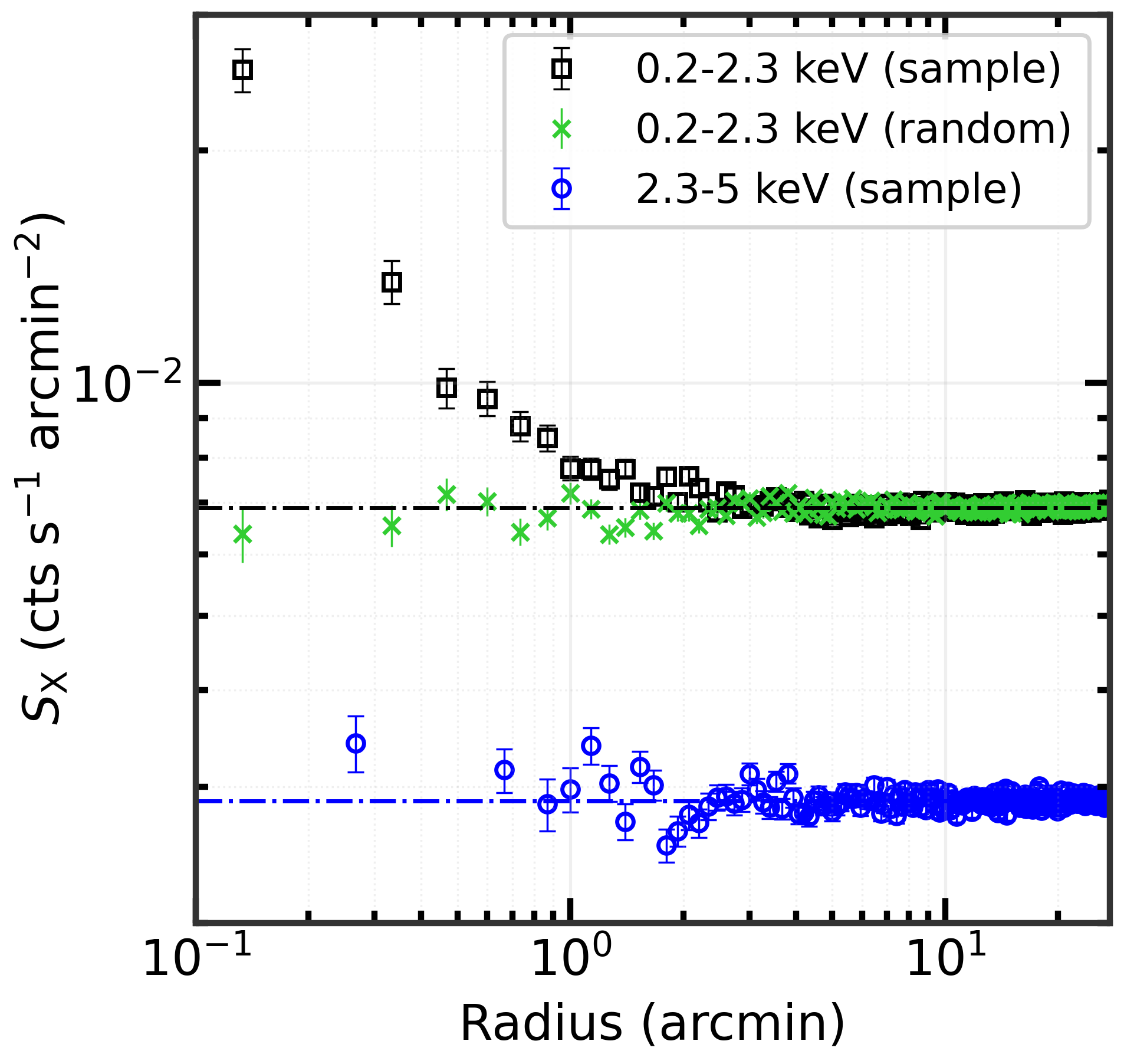}
\includegraphics[width=0.32\textwidth]{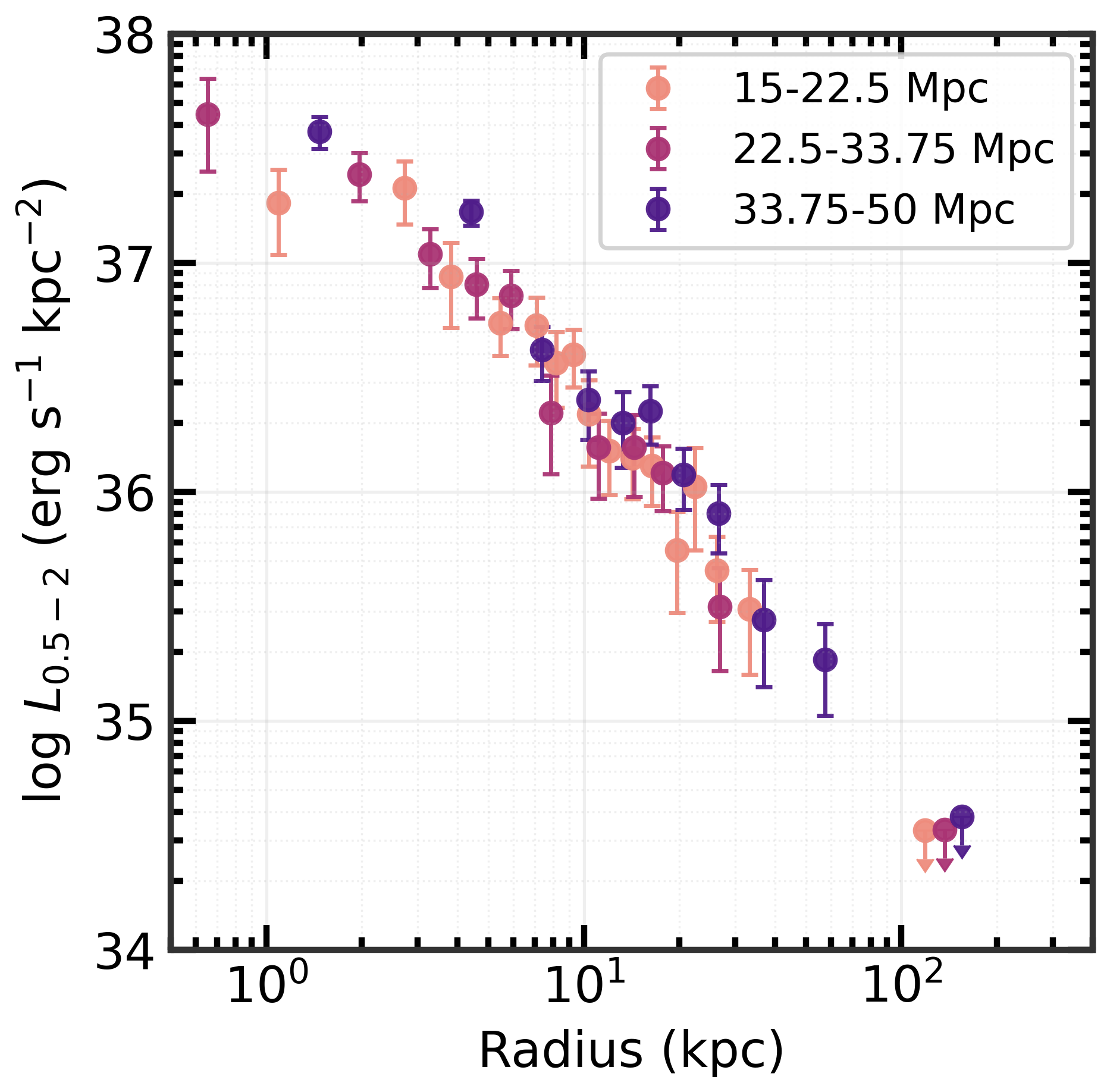}
\caption{{\it Left}: Comparison of background-subtracted surface brightness profiles in the 0.2--2.3 (black) and 0.5--2 keV (red) bands. The overall consistency between the two profiles suggests that the low-energy data is not affected by potential light leak related to the eROSITA telescope modules TM5 and TM5. 
{\it Middle}: Comparison of the stacked surface brightness profiles of our sample galaxies in the soft (0.2--2.3 keV, black) and hard (2.3--5 keV, blue) bands, as well as the 0.2--2.3 keV profile of the randomly chosen sky centroids (green). Horizontal lines indicate the background level.
{\it Right}: Comparison of background-subtracted 0.2--2.3 keV intensity profiles of galaxies within three groups of different distance range: 15–22.5 (near), 22.5–33.75 (medium), 33.75–50 (far) Mpc, with darker colors representing farther distances. For each group, the median distance of included galaxies is adopted to convert the projected angular offset into a physical radius.
}\label{fig:testprofile}
\end{figure}

\begin{figure}[htbp]
\centering
\includegraphics[width=0.45\textwidth]{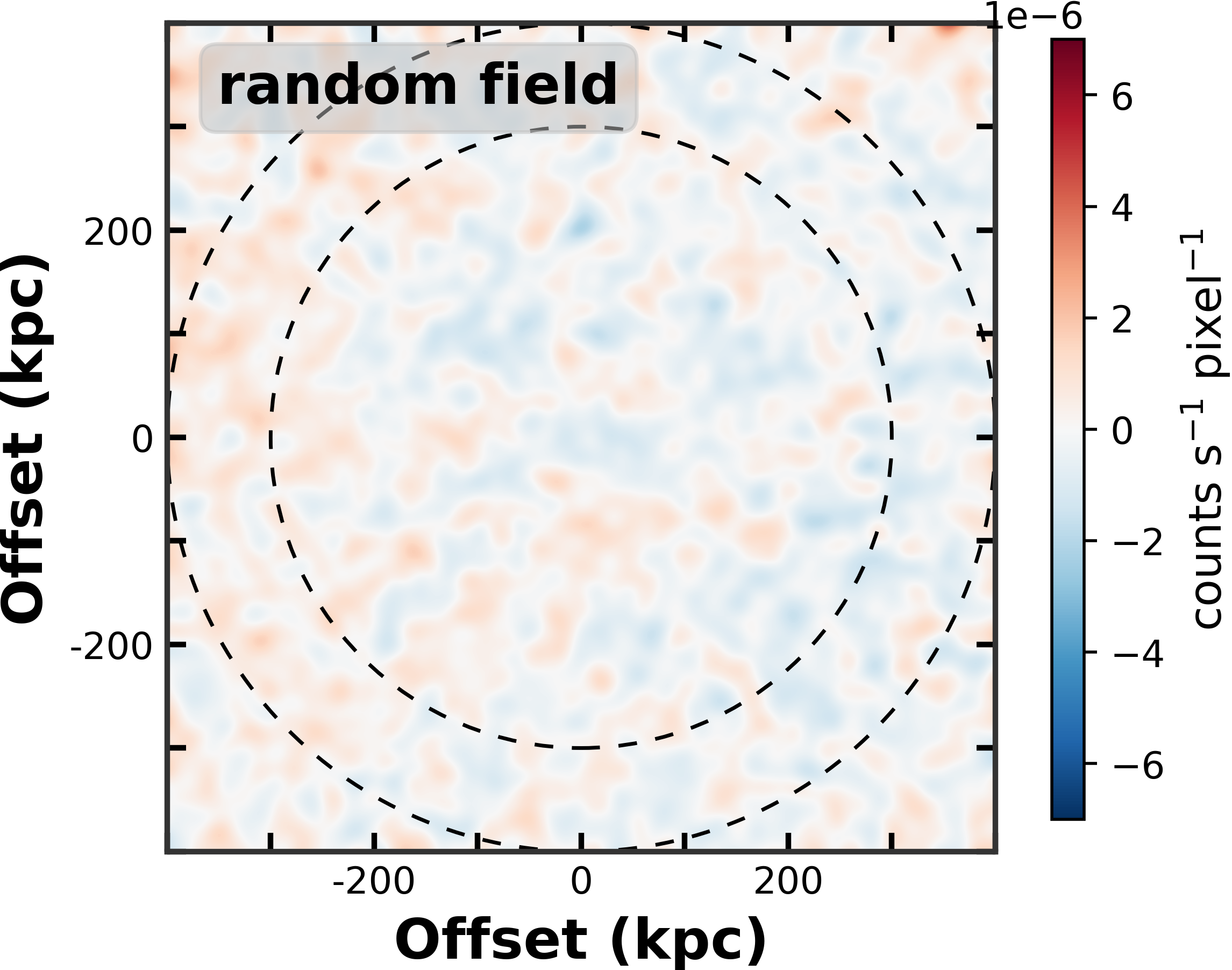}
\includegraphics[width=0.45\textwidth]{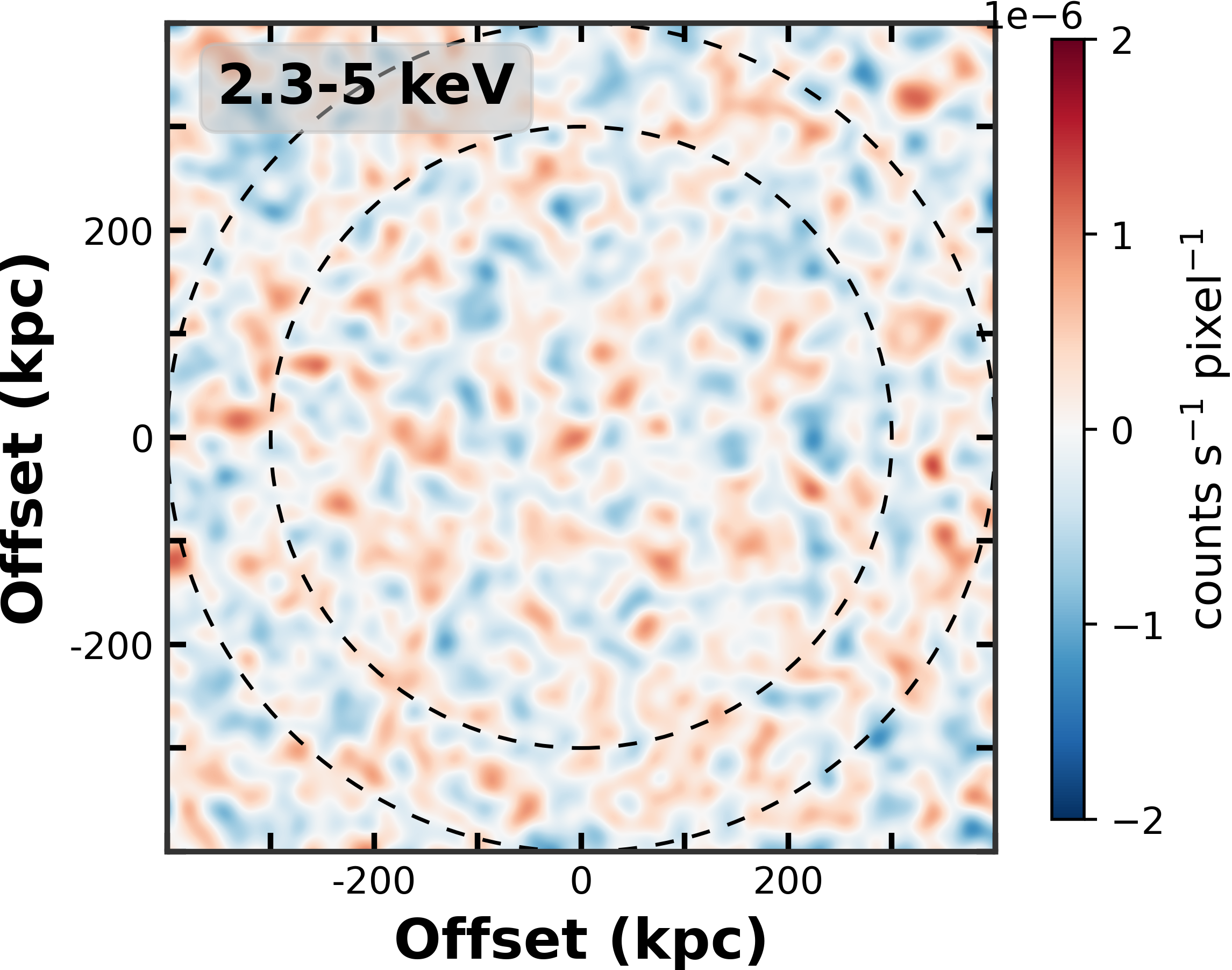}
\caption{{\it Left}: Stacked background-subtracted 0.2--2.3 keV surface brightness image of the ``random field". {\it Right}: Stacked background-subtracted 2.3--5 keV surface brightness image of the full $L^*$ sample. Both images are smoothed with a Gaussian kernel size of 10 pixels. The dashed annulus has inner-to-outer radii of 300--400 kpc, indicating the defined background region.
}\label{fig:control}
\end{figure}

The robustness of our stacking procedure is further supported by a test image stacking regions of the same angular size but around randomly chosen centroids in the sky tiles, i.e., where no prior information exists about whether or not there is excess X-ray emission. 
The stacked image of this random field (left panel of Figure \ref{fig:control}) and the corresponding radial surface brightness profile (middle panel of Figure~\ref{fig:testprofile})
exhibit no significant signals, but are rather consistent with pure statistical fluctuations, as expected. 

We also investigate the hard X-ray emission of the sample galaxies by stacking their 2.3--5 keV images from eRASS1, again after masking the detected sources.
Any hard X-ray emission associated with the host galaxy is expected to be dominated by discrete sources such as XRBs, since the characteristic temperature (sub-keV) of the hot CGM is too low to produce copious hard X-ray photons.  
Consistent with this expectation, both the stacked 2.3--5 keV image (right panel of Figure~\ref{fig:control}) and 2.3--5 keV surface brightness profile (middle panel of Figure~\ref{fig:testprofile}) exhibit no clear sign of extended emission beyond the central region ($\lesssim 2'$). This is compatible with our presumption that the detected diffuse soft X-ray emission originate predominantly, if not entirely, from the hot baryons pervading the halos of these nearby $L_*$ galaxies.  

\section{TNG50 Simulation and the Milky Way Analogs}
\label{sec:TNG}
We utilized simulation data from IllustrisTNG \citep{Nelson_2019a,Nelson_2019b,Pillepich_2019}, a suite of state-of-the-art cosmological simulations carried out using the AREPO moving-mesh code. We focused on the highest-resolution run, TNG50, which is designed to simulate a large, representative cosmological volume at a numerical resolution approaching the modern zoomed-in simulations of individual galaxies. TNG50 evolves $2160^3$ gas cells in a (51.7 cMpc)$^3$ box with a median spatial resolution of 100--140 pc and a baryonic mass resolution of $8.5\times 10^4~M_{\odot}$, facilitating a detailed view of the gas structures both within and around individual galaxies.

For direct usage and reference, TNG50 has released the Milky Way $+$ Andromeda sample, which includes 198 Milky Way-like and Andromeda-like (MW/M31-like) galaxies having a stellar disky morphology and stellar mass in the range of $10^{10.5}-10^{11.2}~M_{\odot}$, and moreover, residing within a MW-like 500 kpc-scale environment at $z=0$ \citep{Pillepich_2024}. To be more consistent with our $L^*$ galaxy sample and the MW-sized sample of \citep{Zhang_2024}, we only considered the TNG50 MW analogs with stellar mass of $10^{10.5}-10^{11.0}~M_{\odot}$, which consists of 152 simulated galaxies.  
We downloaded from the TNG data archive the snapshot cutouts at $z=0$ of the individual galaxies, and extracted information on their X-ray-emitting hot gas from the particle-level data following the method outlined in \cite{He_2025a}. Specifically, for each galaxy (or subhalo in the simulation context), we counted all bounded gas cells which may contribute to X-ray emission with a given cell temperature and metallicity. To achieve this, we first generated X-ray spectral templates using the PyAtomDB code \citep{Foster_2020}, considering a grid of $(T,Z)$, where $T$ is the cell temperature in units of keV, ranging from 0.08 to 10 keV, and $Z$ is the metallicity ranging from 0 to 2. 
These spectral templates assume collisional ionization equilibrium, which is a reasonable and commonly adopted approximation for the hot CGM.
The total X-ray luminosity in a given volume (e.g. a spherical shell) is then calculated by integrating the emission of all enclosed gas cells over the energy range of 0.5--2 keV. To obtain the radial intensity profiles, we adopt sequential concentric annuli with a width of 0.5 kpc. The predicted X-ray intensity profile from TNG50, shown in Figure~\ref{fig:profile}, is averaged from the radial profiles of all 152 MW analogs and further convolved with the eROSITA PSF, with the $1\sigma$ errors taken as the $16^{\rm th}$ and $84^{\rm th}$ percentiles.
The azimuthal intensity profiles of the TNG50 MW analogs presented in Figure~\ref{fig:azimuthal} are similarly generated.

\begin{figure}[htbp]
\centering
\includegraphics[width=0.33\textwidth]{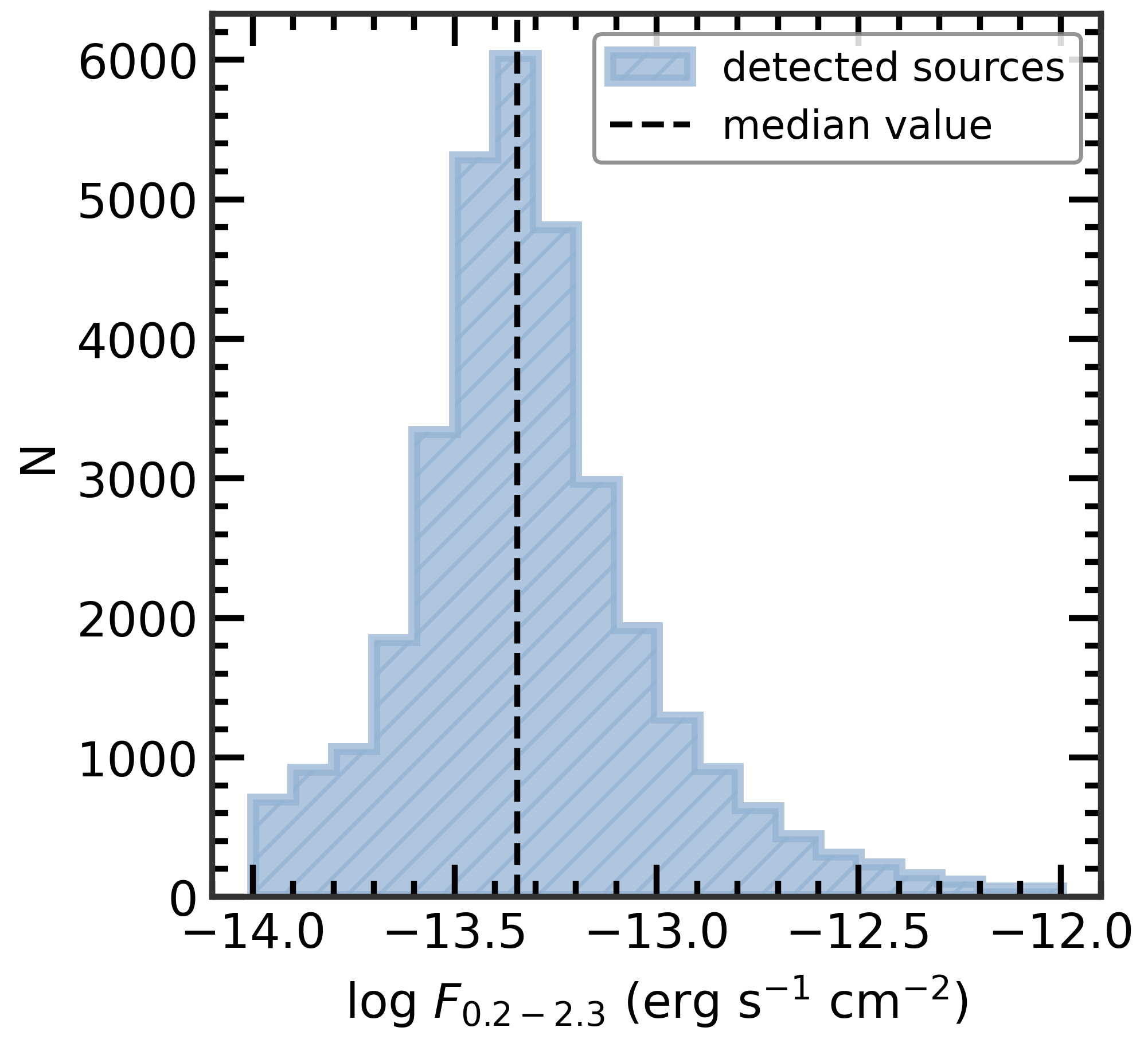}
\includegraphics[width=0.33\textwidth]{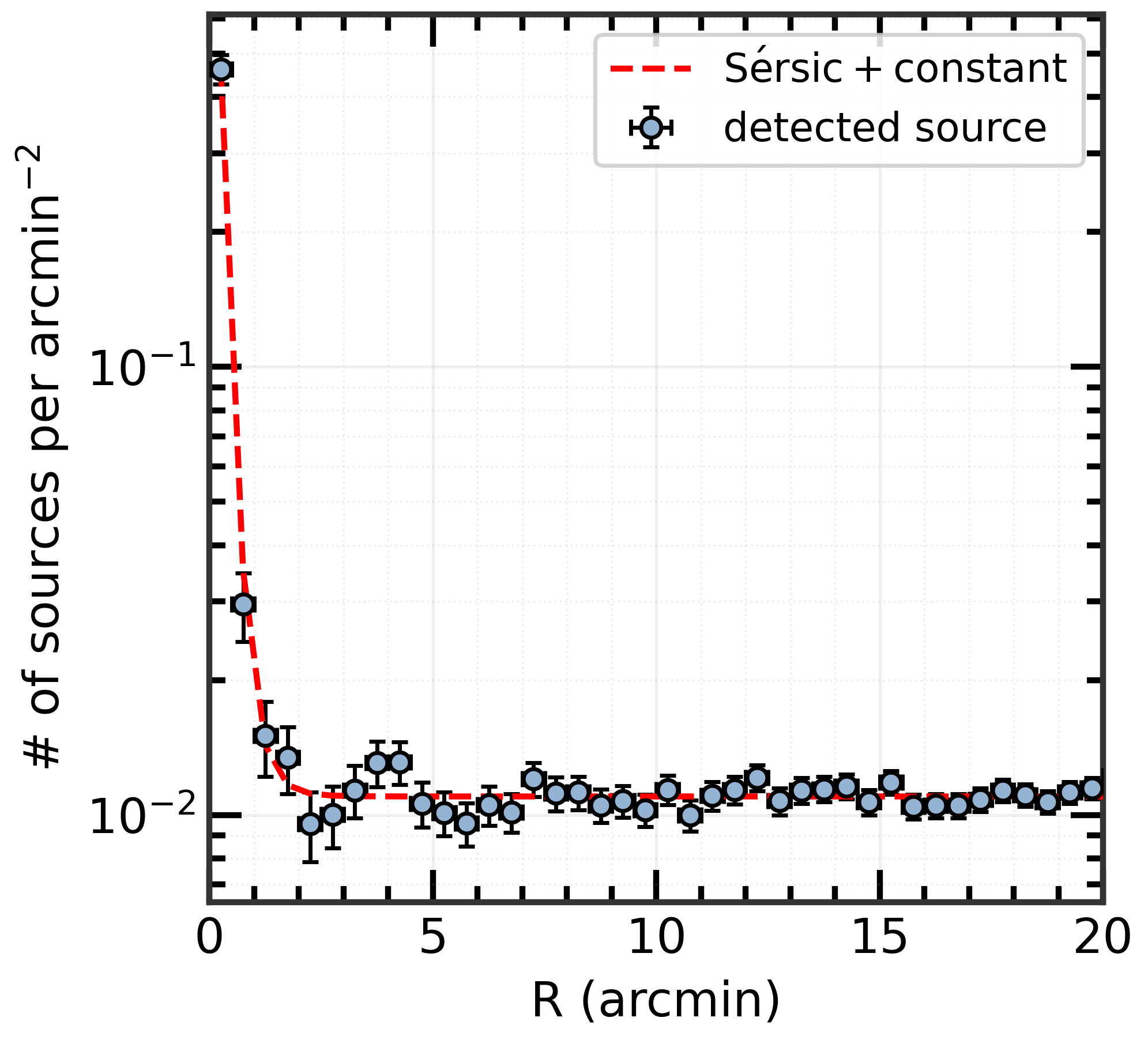}
\includegraphics[width=0.305\textwidth]{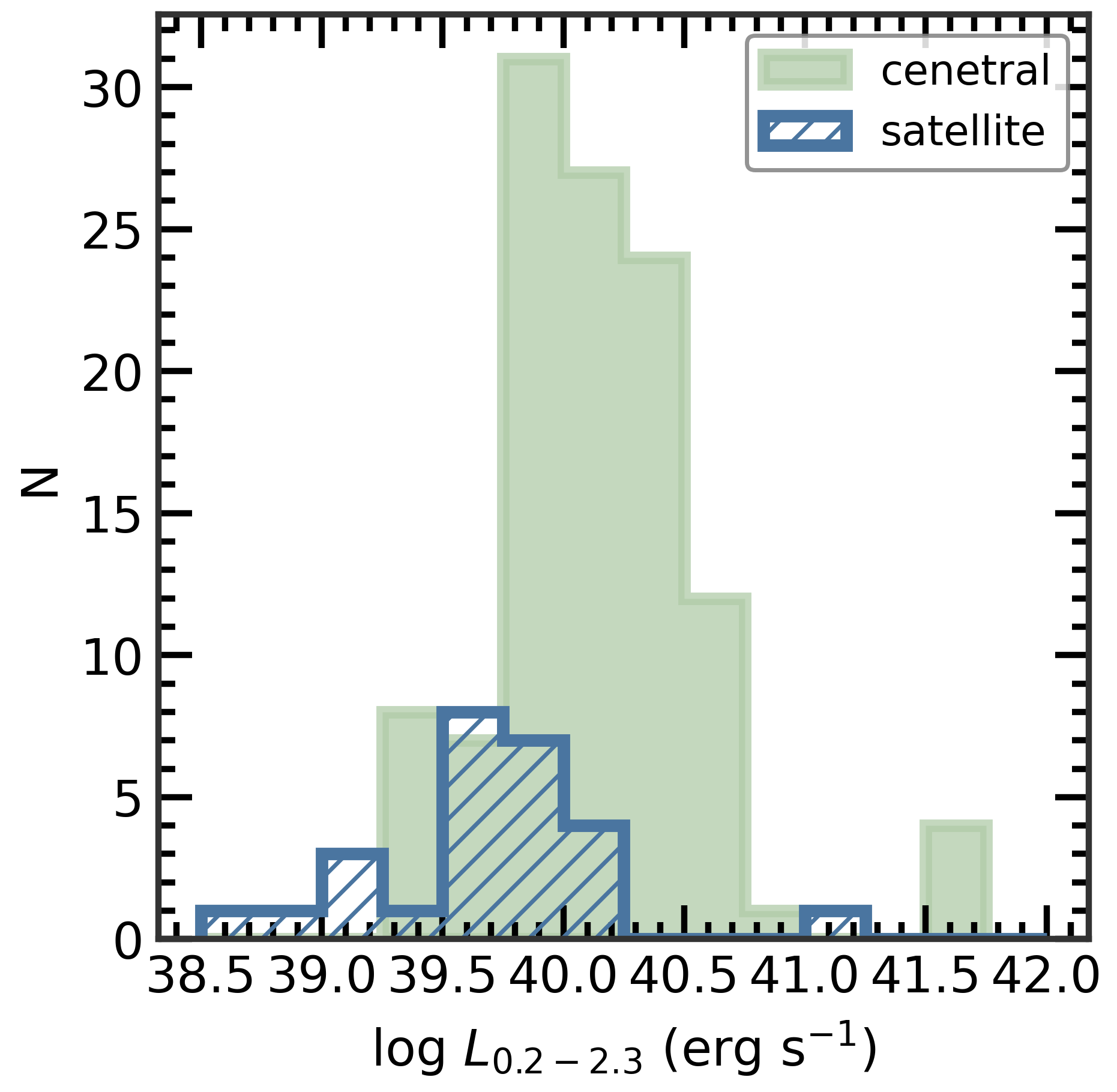}
\caption{{\it Left}: 0.2--2.3 keV flux distribution of 33695 eRASS1 point-like sources falling within the footprint of our sample galaxies. The vertical line indicates the median value. {\it Middle}: Azimuthally averaged surface density profile of the eRASS1 point-like sources as a function of the galactocentric radius. 
The profile can be roughly described by the combination of two components (red curve): a $\rm S\acute{e}rsic$ profile accounting for sources associated with the host galaxies, and a constant accounting for the cosmic X-ray background sources. 
Here the $\rm S\acute{e}rsic$ index $n$ = 2.1 is derived from the stacked WISE $W1$ image of the sample galaxies (Appendix~\ref{sec:WISE}). {\it Right}: 0.2--2.3 keV luminosity distribution of sources associated with the nuclei of central (green) and satellite (blue) galaxies.
}\label{fig:eRASS1src}
\end{figure}

\begin{figure}[htbp]
\centering
\includegraphics[width=1.0\textwidth]{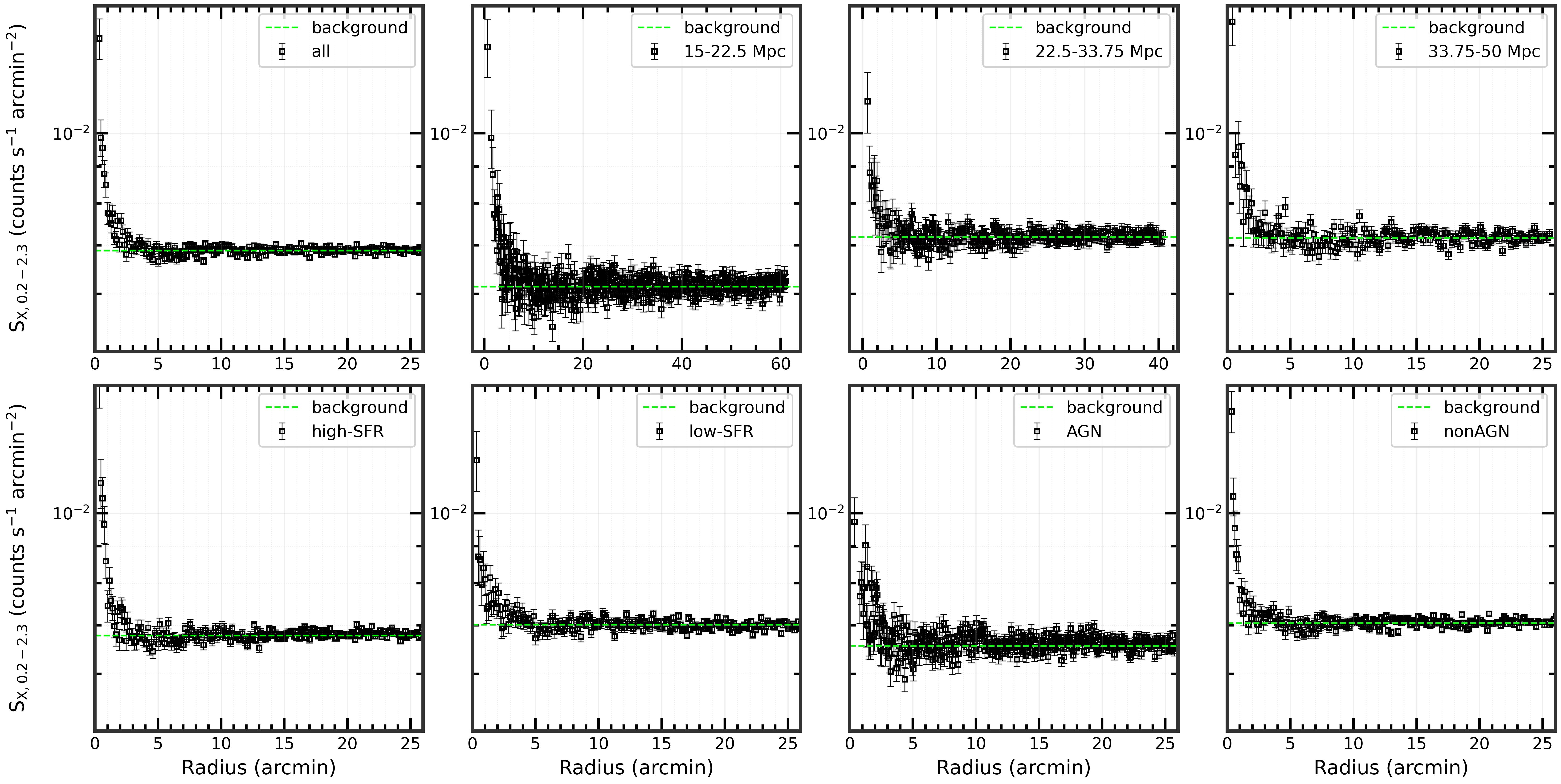}
\caption{Exposure-corrected 0.2--2.3 keV surface brightness profiles stacking the full sample and various subsamples, without subtracting the local background. In each panel, the blue dashed curve represents the best-fit local background.
}
\label{fig:radprofiles}
\end{figure}

\bibliography{ref}{}

@article{Merloni_2024,
   title={The SRG/eROSITA all-sky survey: First X-ray catalogues and data release of the western Galactic hemisphere},
   volume={682},
   ISSN={1432-0746},
   url={http://dx.doi.org/10.1051/0004-6361/202347165},
   DOI={10.1051/0004-6361/202347165},
   journal={\aap},
   publisher={EDP Sciences},
   author={Merloni, A. and Lamer, G. and Liu, T. and Ramos-Ceja, M. E. and Brunner, H. and Bulbul, E. and Dennerl, K. and Doroshenko, V. and Freyberg, M. J. and Friedrich, S. and Gatuzz, E. and Georgakakis, A. and Haberl, F. and Igo, Z. and Kreykenbohm, I. and Liu, A. and Maitra, C. and Malyali, A. and Mayer, M. G. F. and Nandra, K. and Predehl, P. and Robrade, J. and Salvato, M. and Sanders, J. S. and Stewart, I. and Tubín-Arenas, D. and Weber, P. and Wilms, J. and Arcodia, R. and Artis, E. and Aschersleben, J. and Avakyan, A. and Aydar, C. and Bahar, Y. E. and Balzer, F. and Becker, W. and Berger, K. and Boller, T. and Bornemann, W. and Brüggen, M. and Brusa, M. and Buchner, J. and Burwitz, V. and Camilloni, F. and Clerc, N. and Comparat, J. and Coutinho, D. and Czesla, S. and Dannhauer, S. M. and Dauner, L. and Dauser, T. and Dietl, J. and Dolag, K. and Dwelly, T. and Egg, K. and Ehl, E. and Freund, S. and Friedrich, P. and Gaida, R. and Garrel, C. and Ghirardini, V. and Gokus, A. and Grünwald, G. and Grandis, S. and Grotova, I. and Gruen, D. and Gueguen, A. and Hämmerich, S. and Hamaus, N. and Hasinger, G. and Haubner, K. and Homan, D. and Ider Chitham, J. and Joseph, W. M. and Joyce, A. and König, O. and Kaltenbrunner, D. M. and Khokhriakova, A. and Kink, W. and Kirsch, C. and Kluge, M. and Knies, J. and Krippendorf, S. and Krumpe, M. and Kurpas, J. and Li, P. and Liu, Z. and Locatelli, N. and Lorenz, M. and Müller, S. and Magaudda, E. and Mannes, C. and McCall, H. and Meidinger, N. and Michailidis, M. and Migkas, K. and Muñoz-Giraldo, D. and Musiimenta, B. and Nguyen-Dang, N. T. and Ni, Q. and Olechowska, A. and Ota, N. and Pacaud, F. and Pasini, T. and Perinati, E. and Pires, A. M. and Pommranz, C. and Ponti, G. and Poppenhaeger, K. and Pühlhofer, G. and Rau, A. and Reh, M. and Reiprich, T. H. and Roster, W. and Saeedi, S. and Santangelo, A. and Sasaki, M. and Schmitt, J. and Schneider, P. C. and Schrabback, T. and Schuster, N. and Schwope, A. and Seppi, R. and Serim, M. M. and Shreeram, S. and Sokolova-Lapa, E. and Starck, H. and Stelzer, B. and Stierhof, J. and Suleimanov, V. and Tenzer, C. and Traulsen, I. and Trümper, J. and Tsuge, K. and Urrutia, T. and Veronica, A. and Waddell, S. G. H. and Willer, R. and Wolf, J. and Yeung, M. C. H. and Zainab, A. and Zangrandi, F. and Zhang, X. and Zhang, Y. and Zheng, X.},
   year={2024},
   month=jan, pages={A34} }

@ARTICLE{Ohlson_2024,
       author = {{Ohlson}, David and {Seth}, Anil C. and {Gallo}, Elena and {Baldassare}, Vivienne F. and {Greene}, Jenny E.},
        title = "{The 50 Mpc Galaxy Catalog (50 MGC): Consistent and Homogeneous Masses, Distances, Colors, and Morphologies}",
      journal = {\aj},
         year = 2024,
        month = jan,
       volume = {167},
       number = {1},
          eid = {31},
        pages = {31},
          doi = {10.3847/1538-3881/acf7bc},
archivePrefix = {arXiv},
       eprint = {2309.05701},
 primaryClass = {astro-ph.GA},
       adsurl = {https://ui.adsabs.harvard.edu/abs/2024AJ....167...31O}
}

@article{Chen_2025,
   title={The average soft X-ray spectra of eROSITA active galactic nuclei},
   volume={701},
   ISSN={1432-0746},
   url={http://dx.doi.org/10.1051/0004-6361/202554737},
   DOI={10.1051/0004-6361/202554737},
   journal={\aap},
   publisher={EDP Sciences},
   author={Chen, Shi-Jiang and Buchner, Johannes and Liu, Teng and Hagen, Scott and Waddell, Sophia G. H. and Nandra, Kirpal and Salvato, Mara and Igo, Zsofi and Aydar, Catarina and Merloni, Andrea and Ni, Qingling and Kang, Jia-Lai and Cai, Zhen-Yi and Wang, Jun-Xian and Li, Ruancun and Ramos-Ceja, Miriam E. and Sanders, Jeremy and Georgakakis, Antonis and Zhang, Yi},
   year={2025},
   month=sep, pages={A144} }

@ARTICLE{Wright_2010,
       author = {{Wright}, Edward L. and {Eisenhardt}, Peter R.~M. and {Mainzer}, Amy K. and {Ressler}, Michael E. and {Cutri}, Roc M. and {Jarrett}, Thomas and {Kirkpatrick}, J. Davy and {Padgett}, Deborah and {McMillan}, Robert S. and {Skrutskie}, Michael and {Stanford}, S.~A. and {Cohen}, Martin and {Walker}, Russell G. and {Mather}, John C. and {Leisawitz}, David and {Gautier}, III, Thomas N. and {McLean}, Ian and {Benford}, Dominic and {Lonsdale}, Carol J. and {Blain}, Andrew and {Mendez}, Bryan and {Irace}, William R. and {Duval}, Valerie and {Liu}, Fengchuan and {Royer}, Don and {Heinrichsen}, Ingolf and {Howard}, Joan and {Shannon}, Mark and {Kendall}, Martha and {Walsh}, Amy L. and {Larsen}, Mark and {Cardon}, Joel G. and {Schick}, Scott and {Schwalm}, Mark and {Abid}, Mohamed and {Fabinsky}, Beth and {Naes}, Larry and {Tsai}, Chao-Wei},
        title = "{The Wide-field Infrared Survey Explorer (WISE): Mission Description and Initial On-orbit Performance}",
      journal = {\aj},
         year = 2010,
        month = dec,
       volume = {140},
       number = {6},
        pages = {1868-1881},
          doi = {10.1088/0004-6256/140/6/1868},
archivePrefix = {arXiv},
       eprint = {1008.0031},
 primaryClass = {astro-ph.IM},
       adsurl = {https://ui.adsabs.harvard.edu/abs/2010AJ....140.1868W}
}

@ARTICLE{Hu_2024,
       author = {{Hu}, Zhensong and {Hou}, Meicun and {Li}, Zhiyuan},
        title = "{Uncovering an Excess of X-Ray Point Sources in the Halos of Virgo Late-type Galaxies}",
      journal = {\apj},
         year = 2024,
        month = jun,
       volume = {968},
       number = {1},
          eid = {41},
        pages = {41},
          doi = {10.3847/1538-4357/ad429d},
archivePrefix = {arXiv},
       eprint = {2404.15057},
 primaryClass = {astro-ph.GA},
       adsurl = {https://ui.adsabs.harvard.edu/abs/2024ApJ...968...41H}
}

@ARTICLE{Jarrett_2019,
       author = {{Jarrett}, T.~H. and {Cluver}, M.~E. and {Brown}, M.~J.~I. and {Dale}, D.~A. and {Tsai}, C.~W. and {Masci}, F.},
        title = "{The WISE Extended Source Catalog (WXSC). I. The 100 Largest Galaxies}",
      journal = {\apjs},
         year = 2019,
        month = dec,
       volume = {245},
       number = {2},
          eid = {25},
        pages = {25},
          doi = {10.3847/1538-4365/ab521a},
archivePrefix = {arXiv},
       eprint = {1910.11793},
 primaryClass = {astro-ph.GA},
       adsurl = {https://ui.adsabs.harvard.edu/abs/2019ApJS..245...25J}
}

@ARTICLE{Brunner_2022,
       author = {{Brunner}, H. and {Liu}, T. and {Lamer}, G. and {Georgakakis}, A. and {Merloni}, A. and {Brusa}, M. and {Bulbul}, E. and {Dennerl}, K. and {Friedrich}, S. and {Liu}, A. and {Maitra}, C. and {Nandra}, K. and {Ramos-Ceja}, M.~E. and {Sanders}, J.~S. and {Stewart}, I.~M. and {Boller}, T. and {Buchner}, J. and {Clerc}, N. and {Comparat}, J. and {Dwelly}, T. and {Eckert}, D. and {Finoguenov}, A. and {Freyberg}, M. and {Ghirardini}, V. and {Gueguen}, A. and {Haberl}, F. and {Kreykenbohm}, I. and {Krumpe}, M. and {Osterhage}, S. and {Pacaud}, F. and {Predehl}, P. and {Reiprich}, T.~H. and {Robrade}, J. and {Salvato}, M. and {Santangelo}, A. and {Schrabback}, T. and {Schwope}, A. and {Wilms}, J.},
        title = "{The eROSITA Final Equatorial Depth Survey (eFEDS). X-ray catalogue}",
      journal = {\aap},
         year = 2022,
        month = may,
       volume = {661},
          eid = {A1},
        pages = {A1},
          doi = {10.1051/0004-6361/202141266},
archivePrefix = {arXiv},
       eprint = {2106.14517},
 primaryClass = {astro-ph.HE},
       adsurl = {https://ui.adsabs.harvard.edu/abs/2022A&A...661A...1B}
}

@article{Tortorelli_2020,
   title={Measurement of the B-band galaxy Luminosity Function with Approximate Bayesian Computation},
   volume={2020},
   ISSN={1475-7516},
   url={http://dx.doi.org/10.1088/1475-7516/2020/09/048},
   DOI={10.1088/1475-7516/2020/09/048},
   number={09},
   journal={Journal of Cosmology and Astroparticle Physics},
   publisher={IOP Publishing},
   author={Tortorelli, Luca and Fagioli, Martina and Herbel, Jörg and Amara, Adam and Kacprzak, Tomasz and Refregier, Alexandre},
   year={2020},
   month=sep, pages={048–048} }

@ARTICLE{Sunyaev_2021,
       author = {{Sunyaev}, R. and {Arefiev}, V. and {Babyshkin}, V. and {Bogomolov}, A. and {Borisov}, K. and {Buntov}, M. and {Brunner}, H. and {Burenin}, R. and {Churazov}, E. and {Coutinho}, D. and {Eder}, J. and {Eismont}, N. and {Freyberg}, M. and {Gilfanov}, M. and {Gureyev}, P. and {Hasinger}, G. and {Khabibullin}, I. and {Kolmykov}, V. and {Komovkin}, S. and {Krivonos}, R. and {Lapshov}, I. and {Levin}, V. and {Lomakin}, I. and {Lutovinov}, A. and {Medvedev}, P. and {Merloni}, A. and {Mernik}, T. and {Mikhailov}, E. and {Molodtsov}, V. and {Mzhelsky}, P. and {M{\"u}ller}, S. and {Nandra}, K. and {Nazarov}, V. and {Pavlinsky}, M. and {Poghodin}, A. and {Predehl}, P. and {Robrade}, J. and {Sazonov}, S. and {Scheuerle}, H. and {Shirshakov}, A. and {Tkachenko}, A. and {Voron}, V.},
        title = "{SRG X-ray orbital observatory. Its telescopes and first scientific results}",
      journal = {\aap},
         year = 2021,
        month = dec,
       volume = {656},
          eid = {A132},
        pages = {A132},
          doi = {10.1051/0004-6361/202141179},
archivePrefix = {arXiv},
       eprint = {2104.13267},
 primaryClass = {astro-ph.HE},
       adsurl = {https://ui.adsabs.harvard.edu/abs/2021A&A...656A.132S}
}

@ARTICLE{Strickland_2004,
       author = {{Strickland}, David K. and {Heckman}, Timothy M. and {Colbert}, Edward J.~M. and {Hoopes}, Charles G. and {Weaver}, Kimberly A.},
        title = "{A High Spatial Resolution X-Ray and H{\ensuremath{\alpha}} Study of Hot Gas in the Halos of Star-forming Disk Galaxies. II. Quantifying Supernova Feedback}",
      journal = {\apj},
         year = 2004,
        month = may,
       volume = {606},
       number = {2},
        pages = {829-852},
          doi = {10.1086/383136},
archivePrefix = {arXiv},
       eprint = {astro-ph/0306598},
 primaryClass = {astro-ph},
       adsurl = {https://ui.adsabs.harvard.edu/abs/2004ApJ...606..829S}
}

@article{Zhang_2024,
   title={The hot circumgalactic medium in the eROSITA All-Sky Survey: I. X-ray surface brightness profiles},
   volume={690},
   ISSN={1432-0746},
   url={http://dx.doi.org/10.1051/0004-6361/202449412},
   DOI={10.1051/0004-6361/202449412},
   journal={\aap},
   publisher={EDP Sciences},
   author={Zhang, Yi and Comparat, Johan and Ponti, Gabriele and Merloni, Andrea and Nandra, Kirpal and Haberl, Frank and Locatelli, Nicola and Zhang, Xiaoyuan and Sanders, Jeremy and Zheng, Xueying and Liu, Ang and Popesso, Paola and Liu, Teng and Truong, Nhut and Pillepich, Annalisa and Predehl, Peter and Salvato, Mara and Shreeram, Soumya and Yeung, Michael C. H. and Ni, Qingling},
   year={2024},
   month=oct, pages={A267} }

@ARTICLE{Pillepich_2024,
       author = {{Pillepich}, Annalisa and {Sotillo-Ramos}, Diego and {Ramesh}, Rahul and {Nelson}, Dylan and {Engler}, Christoph and {Rodriguez-Gomez}, Vicente and {Fournier}, Martin and {Donnari}, Martina and {Springel}, Volker and {Hernquist}, Lars},
        title = "{Milky Way and Andromeda analogues from the TNG50 simulation}",
      journal = {\mnras},
         year = 2024,
        month = dec,
       volume = {535},
       number = {2},
        pages = {1721-1762},
          doi = {10.1093/mnras/stae2165},
archivePrefix = {arXiv},
       eprint = {2303.16217},
 primaryClass = {astro-ph.GA},
       adsurl = {https://ui.adsabs.harvard.edu/abs/2024MNRAS.535.1721P}
}

@ARTICLE{Li_2024,
       author = {{Li}, Dawei and {Fang}, Taotao and {Ge}, Chong and {Liu}, Teng and {He}, Lin and {Li}, Zhiyuan and {Nicastro}, Fabrizio and {Yang}, Xiaohu and {Zhang}, Xiaoxia and {Zheng}, Yun-Liang},
        title = "{Robust Detection of Hot Intragroup Medium in Optically Selected, Poor Galaxy Groups by eROSITA}",
      journal = {\apjl},
         year = 2024,
        month = dec,
       volume = {977},
       number = {2},
          eid = {L40},
        pages = {L40},
          doi = {10.3847/2041-8213/ad991c},
archivePrefix = {arXiv},
       eprint = {2412.01261},
 primaryClass = {astro-ph.GA},
       adsurl = {https://ui.adsabs.harvard.edu/abs/2024ApJ...977L..40L}
}

@ARTICLE{Chadayammuri_2022,
       author = {{Chadayammuri}, Urmila and {Bogd{\'a}n}, {\'A}kos and {Oppenheimer}, Benjamin D. and {Kraft}, Ralph P. and {Forman}, William R. and {Jones}, Christine},
        title = "{Testing Galaxy Feedback Models with Resolved X-Ray Profiles of the Hot Circumgalactic Medium}",
      journal = {\apjl},
         year = 2022,
        month = sep,
       volume = {936},
       number = {1},
          eid = {L15},
        pages = {L15},
          doi = {10.3847/2041-8213/ac8936},
archivePrefix = {arXiv},
       eprint = {2203.01356},
 primaryClass = {astro-ph.GA},
       adsurl = {https://ui.adsabs.harvard.edu/abs/2022ApJ...936L..15C}
}

@ARTICLE{Mathews_1971,
       author = {{Mathews}, William G. and {Baker}, James C.},
        title = "{Galactic Winds}",
      journal = {\apj},
         year = 1971,
        month = dec,
       volume = {170},
        pages = {241},
          doi = {10.1086/151208},
       adsurl = {https://ui.adsabs.harvard.edu/abs/1971ApJ...170..241M}
}

@ARTICLE{Chevalier_1985,
       author = {{Chevalier}, R.~A. and {Clegg}, A.~W.},
        title = "{Wind from a starburst galaxy nucleus}",
      journal = {\nat},
         year = 1985,
        month = sep,
       volume = {317},
       number = {6032},
        pages = {44-45},
          doi = {10.1038/317044a0},
       adsurl = {https://ui.adsabs.harvard.edu/abs/1985Natur.317...44C}
}

@article{Choi_2020,
   title={The Impact of Outflows Driven by Active Galactic Nuclei on Metals in and around Galaxies},
   volume={904},
   ISSN={1538-4357},
   url={http://dx.doi.org/10.3847/1538-4357/abba7d},
   DOI={10.3847/1538-4357/abba7d},
   number={1},
   journal={The Astrophysical Journal},
   publisher={American Astronomical Society},
   author={Choi, Ena and Brennan, Ryan and Somerville, Rachel S. and Ostriker, Jeremiah P. and Hirschmann, Michaela and Naab, Thorsten},
   year={2020},
   month=nov, pages={8} }

@ARTICLE{White_1978,
       author = {{White}, S.~D.~M. and {Rees}, M.~J.},
        title = "{Core condensation in heavy halos: a two-stage theory for galaxy formation and clustering.}",
      journal = {\mnras},
         year = 1978,
        month = may,
       volume = {183},
        pages = {341-358},
          doi = {10.1093/mnras/183.3.341},
       adsurl = {https://ui.adsabs.harvard.edu/abs/1978MNRAS.183..341W}
}

@ARTICLE{White_1991,
       author = {{White}, Simon D.~M. and {Frenk}, Carlos S.},
        title = "{Galaxy Formation through Hierarchical Clustering}",
      journal = {\apj},
         year = 1991,
        month = sep,
       volume = {379},
        pages = {52},
          doi = {10.1086/170483},
       adsurl = {https://ui.adsabs.harvard.edu/abs/1991ApJ...379...52W}
}

@ARTICLE{Comparat_2022,
       author = {{Comparat}, Johan and {Truong}, Nhut and {Merloni}, Andrea and {Pillepich}, Annalisa and {Ponti}, Gabriele and {Driver}, Simon and {Bellstedt}, Sabine and {Liske}, Joe and {Aird}, James and {Br{\"u}ggen}, Marcus and {Bulbul}, Esra and {Davies}, Luke and {Villalba}, Justo Antonio Gonz{\'a}lez and {Georgakakis}, Antonis and {Haberl}, Frank and {Liu}, Teng and {Maitra}, Chandreyee and {Nandra}, Kirpal and {Popesso}, Paola and {Predehl}, Peter and {Robotham}, Aaron and {Salvato}, Mara and {Thorne}, Jessica E. and {Zhang}, Yi},
        title = "{The eROSITA Final Equatorial Depth Survey (eFEDS). X-ray emission around star-forming and quiescent galaxies at 0.05 < z < 0.3}",
      journal = {\aap},
         year = 2022,
        month = oct,
       volume = {666},
          eid = {A156},
        pages = {A156},
          doi = {10.1051/0004-6361/202243101},
archivePrefix = {arXiv},
       eprint = {2201.05169},
 primaryClass = {astro-ph.GA},
       adsurl = {https://ui.adsabs.harvard.edu/abs/2022A&A...666A.156C}
}

@ARTICLE{Crain_2015,
       author = {{Crain}, Robert A. and {Schaye}, Joop and {Bower}, Richard G. and {Furlong}, Michelle and {Schaller}, Matthieu and {Theuns}, Tom and {Dalla Vecchia}, Claudio and {Frenk}, Carlos S. and {McCarthy}, Ian G. and {Helly}, John C. and {Jenkins}, Adrian and {Rosas-Guevara}, Yetli M. and {White}, Simon D.~M. and {Trayford}, James W.},
        title = "{The EAGLE simulations of galaxy formation: calibration of subgrid physics and model variations}",
      journal = {\mnras},
         year = 2015,
        month = jun,
       volume = {450},
       number = {2},
        pages = {1937-1961},
          doi = {10.1093/mnras/stv725},
archivePrefix = {arXiv},
       eprint = {1501.01311},
 primaryClass = {astro-ph.GA},
       adsurl = {https://ui.adsabs.harvard.edu/abs/2015MNRAS.450.1937C}
}

@ARTICLE{Schaye_2015,
       author = {{Schaye}, Joop and {Crain}, Robert A. and {Bower}, Richard G. and {Furlong}, Michelle and {Schaller}, Matthieu and {Theuns}, Tom and {Dalla Vecchia}, Claudio and {Frenk}, Carlos S. and {McCarthy}, I.~G. and {Helly}, John C. and {Jenkins}, Adrian and {Rosas-Guevara}, Y.~M. and {White}, Simon D.~M. and {Baes}, Maarten and {Booth}, C.~M. and {Camps}, Peter and {Navarro}, Julio F. and {Qu}, Yan and {Rahmati}, Alireza and {Sawala}, Till and {Thomas}, Peter A. and {Trayford}, James},
        title = "{The EAGLE project: simulating the evolution and assembly of galaxies and their environments}",
      journal = {\mnras},
         year = 2015,
        month = jan,
       volume = {446},
       number = {1},
        pages = {521-554},
          doi = {10.1093/mnras/stu2058},
archivePrefix = {arXiv},
       eprint = {1407.7040},
 primaryClass = {astro-ph.GA},
       adsurl = {https://ui.adsabs.harvard.edu/abs/2015MNRAS.446..521S}
}

@ARTICLE{McAlpine_2016,
       author = {{McAlpine}, S. and {Helly}, J.~C. and {Schaller}, M. and {Trayford}, J.~W. and {Qu}, Y. and {Furlong}, M. and {Bower}, R.~G. and {Crain}, R.~A. and {Schaye}, J. and {Theuns}, T. and {Dalla Vecchia}, C. and {Frenk}, C.~S. and {McCarthy}, I.~G. and {Jenkins}, A. and {Rosas-Guevara}, Y. and {White}, S.~D.~M. and {Baes}, M. and {Camps}, P. and {Lemson}, G.},
        title = "{The EAGLE simulations of galaxy formation: Public release of halo and galaxy catalogues}",
      journal = {Astronomy and Computing},
         year = 2016,
        month = apr,
       volume = {15},
        pages = {72-89},
          doi = {10.1016/j.ascom.2016.02.004},
archivePrefix = {arXiv},
       eprint = {1510.01320},
 primaryClass = {astro-ph.GA},
       adsurl = {https://ui.adsabs.harvard.edu/abs/2016A&C....15...72M}
}

@ARTICLE{Nelson_2018,
       author = {{Nelson}, Dylan and {Pillepich}, Annalisa and {Springel}, Volker and {Weinberger}, Rainer and {Hernquist}, Lars and {Pakmor}, R{\"u}diger and {Genel}, Shy and {Torrey}, Paul and {Vogelsberger}, Mark and {Kauffmann}, Guinevere and {Marinacci}, Federico and {Naiman}, Jill},
        title = "{First results from the IllustrisTNG simulations: the galaxy colour bimodality}",
      journal = {\mnras},
         year = 2018,
        month = mar,
       volume = {475},
       number = {1},
        pages = {624-647},
          doi = {10.1093/mnras/stx3040},
archivePrefix = {arXiv},
       eprint = {1707.03395},
 primaryClass = {astro-ph.GA},
       adsurl = {https://ui.adsabs.harvard.edu/abs/2018MNRAS.475..624N}
}

@ARTICLE{Pillepich_2018,
       author = {{Pillepich}, Annalisa and {Springel}, Volker and {Nelson}, Dylan and {Genel}, Shy and {Naiman}, Jill and {Pakmor}, R{\"u}diger and {Hernquist}, Lars and {Torrey}, Paul and {Vogelsberger}, Mark and {Weinberger}, Rainer and {Marinacci}, Federico},
        title = "{Simulating galaxy formation with the IllustrisTNG model}",
      journal = {\mnras},
         year = 2018,
        month = jan,
       volume = {473},
       number = {3},
        pages = {4077-4106},
          doi = {10.1093/mnras/stx2656},
archivePrefix = {arXiv},
       eprint = {1703.02970},
 primaryClass = {astro-ph.GA},
       adsurl = {https://ui.adsabs.harvard.edu/abs/2018MNRAS.473.4077P}
}

@ARTICLE{Fabian_2012,
       author = {{Fabian}, A.~C.},
        title = "{Observational Evidence of Active Galactic Nuclei Feedback}",
      journal = {\araa},
         year = 2012,
        month = sep,
       volume = {50},
        pages = {455-489},
          doi = {10.1146/annurev-astro-081811-125521},
archivePrefix = {arXiv},
       eprint = {1204.4114},
 primaryClass = {astro-ph.CO},
       adsurl = {https://ui.adsabs.harvard.edu/abs/2012ARA&A..50..455F}
}

@ARTICLE{Veilleux_2005,
       author = {{Veilleux}, Sylvain and {Cecil}, Gerald and {Bland-Hawthorn}, Joss},
        title = "{Galactic Winds}",
      journal = {\araa},
         year = 2005,
        month = sep,
       volume = {43},
       number = {1},
        pages = {769-826},
          doi = {10.1146/annurev.astro.43.072103.150610},
archivePrefix = {arXiv},
       eprint = {astro-ph/0504435},
 primaryClass = {astro-ph},
       adsurl = {https://ui.adsabs.harvard.edu/abs/2005ARA&A..43..769V}
}

@ARTICLE{Cicone_2014,
       author = {{Cicone}, C. and {Maiolino}, R. and {Sturm}, E. and {Graci{\'a}-Carpio}, J. and {Feruglio}, C. and {Neri}, R. and {Aalto}, S. and {Davies}, R. and {Fiore}, F. and {Fischer}, J. and {Garc{\'\i}a-Burillo}, S. and {Gonz{\'a}lez-Alfonso}, E. and {Hailey-Dunsheath}, S. and {Piconcelli}, E. and {Veilleux}, S.},
        title = "{Massive molecular outflows and evidence for AGN feedback from CO observations}",
      journal = {\aap},
         year = 2014,
        month = feb,
       volume = {562},
          eid = {A21},
        pages = {A21},
          doi = {10.1051/0004-6361/201322464},
archivePrefix = {arXiv},
       eprint = {1311.2595},
 primaryClass = {astro-ph.CO},
       adsurl = {https://ui.adsabs.harvard.edu/abs/2014A&A...562A..21C}
}

@ARTICLE{Hopkins_2012,
       author = {{Hopkins}, Philip F. and {Quataert}, Eliot and {Murray}, Norman},
        title = "{Stellar feedback in galaxies and the origin of galaxy-scale winds}",
      journal = {\mnras},
         year = 2012,
        month = apr,
       volume = {421},
       number = {4},
        pages = {3522-3537},
          doi = {10.1111/j.1365-2966.2012.20593.x},
archivePrefix = {arXiv},
       eprint = {1110.4638},
 primaryClass = {astro-ph.CO},
       adsurl = {https://ui.adsabs.harvard.edu/abs/2012MNRAS.421.3522H}
}

@ARTICLE{Hopkins_2025,
       author = {{Hopkins}, Philip F. and {Quataert}, Eliot and {Ponnada}, Sam B. and {Silich}, Emily},
        title = "{Cosmic Rays Masquerading as Hot CGM Gas: An Inverse-Compton Origin for Diffuse X-ray Emission in the Circumgalactic Medium}",
      journal = {The Open Journal of Astrophysics},
         year = 2025,
        month = jun,
       volume = {8},
          eid = {78},
        pages = {78},
          doi = {10.33232/001c.141293},
archivePrefix = {arXiv},
       eprint = {2501.18696},
 primaryClass = {astro-ph.HE},
       adsurl = {https://ui.adsabs.harvard.edu/abs/2025OJAp....8E..78H}
}

@ARTICLE{Zheng_2023,
       author = {{Zheng}, Yun-Liang and {Yang}, Xiaohu and {He}, Min and {Shen}, Shi-Yin and {Li}, Qingyang and {Li}, Xuejie},
        title = "{Measuring the X-ray luminosities of DESI groups from eROSITA Final Equatorial-Depth Survey - I. X-ray luminosity-halo mass scaling relation}",
      journal = {\mnras},
         year = 2023,
        month = aug,
       volume = {523},
       number = {4},
        pages = {4909-4922},
          doi = {10.1093/mnras/stad1684},
archivePrefix = {arXiv},
       eprint = {2306.02594},
 primaryClass = {astro-ph.GA},
       adsurl = {https://ui.adsabs.harvard.edu/abs/2023MNRAS.523.4909Z}
}

@article{Zxy_2024,
   title={The SRG/eROSITA all-sky survey: X-ray emission from the warm-hot phase gas in long cosmic filaments},
   volume={691},
   ISSN={1432-0746},
   url={http://dx.doi.org/10.1051/0004-6361/202450933},
   DOI={10.1051/0004-6361/202450933},
   journal={\aap},
   publisher={EDP Sciences},
   author={Zhang, X. and Bulbul, E. and Malavasi, N. and Ghirardini, V. and Comparat, J. and Kluge, M. and Liu, A. and Merloni, A. and Zhang, Y. and Bahar, Y. E. and Artis, E. and Sanders, J. S. and Garrel, C. and Balzer, F. and Brüggen, M. and Freyberg, M. and Gatuzz, E. and Grandis, S. and Krippendorf, S. and Nandra, K. and Ponti, G. and Ramos-Ceja, M. and Predehl, P. and Reiprich, T. H. and Veronica, A. and Yeung, M. C. H. and Zelmer, S.},
   year={2024},
   month=nov, pages={A234} }

@ARTICLE{Silich_2025,
       author = {{Silich}, Emily M. and {ZuHone}, John and {Bellomi}, Elena and {Hummels}, Cameron and {Oppenheimer}, Benjamin and {Hopkins}, Philip F. and {Lochhaas}, Cassandra and {Ponnada}, Sam B. and {Vikhlinin}, Alexey},
        title = "{X-Ray Emission Signatures of Galactic Feedback in the Hot Circumgalactic Medium: Predictions from Cosmological Hydrodynamical Simulations}",
      journal = {\apj},
         year = 2025,
        month = nov,
       volume = {993},
       number = {1},
          eid = {125},
        pages = {125},
          doi = {10.3847/1538-4357/ae08a3},
archivePrefix = {arXiv},
       eprint = {2506.17440},
 primaryClass = {astro-ph.GA},
       adsurl = {https://ui.adsabs.harvard.edu/abs/2025ApJ...993..125S}
}

@ARTICLE{Zhang_2025ApJ...991..170Z,
       author = {{Zhang}, Zhijie and {Zhang}, Xiaoxia and {Li}, Hui and {Fang}, Taotao and {Luo}, Yang and {Marinacci}, Federico and {Sales}, Laura V. and {Torrey}, Paul and {Vogelsberger}, Mark and {Yu}, Qingzheng and {Yuan}, Feng},
        title = "{Tracing the Origins of Hot Halo Gas in Milky Way─type Galaxies with SMUGGLE}",
      journal = {\apj},
         year = 2025,
        month = oct,
       volume = {991},
       number = {2},
          eid = {170},
        pages = {170},
          doi = {10.3847/1538-4357/ae019f},
archivePrefix = {arXiv},
       eprint = {2508.21576},
 primaryClass = {astro-ph.GA},
       adsurl = {https://ui.adsabs.harvard.edu/abs/2025ApJ...991..170Z}
}

@ARTICLE{Pillepich_2019,
       author = {{Pillepich}, Annalisa and {Nelson}, Dylan and {Springel}, Volker and {Pakmor}, R{\"u}diger and {Torrey}, Paul and {Weinberger}, Rainer and {Vogelsberger}, Mark and {Marinacci}, Federico and {Genel}, Shy and {van der Wel}, Arjen and {Hernquist}, Lars},
        title = "{First results from the TNG50 simulation: the evolution of stellar and gaseous discs across cosmic time}",
      journal = {\mnras},
         year = 2019,
        month = dec,
       volume = {490},
       number = {3},
        pages = {3196-3233},
          doi = {10.1093/mnras/stz2338},
archivePrefix = {arXiv},
       eprint = {1902.05553},
 primaryClass = {astro-ph.GA},
       adsurl = {https://ui.adsabs.harvard.edu/abs/2019MNRAS.490.3196P}
}

@ARTICLE{Nelson_2019b,
       author = {{Nelson}, Dylan and {Pillepich}, Annalisa and {Springel}, Volker and {Pakmor}, R{\"u}diger and {Weinberger}, Rainer and {Genel}, Shy and {Torrey}, Paul and {Vogelsberger}, Mark and {Marinacci}, Federico and {Hernquist}, Lars},
        title = "{First results from the TNG50 simulation: galactic outflows driven by supernovae and black hole feedback}",
      journal = {\mnras},
         year = 2019,
        month = dec,
       volume = {490},
       number = {3},
        pages = {3234-3261},
          doi = {10.1093/mnras/stz2306},
archivePrefix = {arXiv},
       eprint = {1902.05554},
 primaryClass = {astro-ph.GA},
       adsurl = {https://ui.adsabs.harvard.edu/abs/2019MNRAS.490.3234N}
}

@ARTICLE{Nelson_2019a,
       author = {{Nelson}, Dylan and {Springel}, Volker and {Pillepich}, Annalisa and {Rodriguez-Gomez}, Vicente and {Torrey}, Paul and {Genel}, Shy and {Vogelsberger}, Mark and {Pakmor}, Ruediger and {Marinacci}, Federico and {Weinberger}, Rainer and {Kelley}, Luke and {Lovell}, Mark and {Diemer}, Benedikt and {Hernquist}, Lars},
        title = "{The IllustrisTNG simulations: public data release}",
      journal = {Computational Astrophysics and Cosmology},
         year = 2019,
        month = may,
       volume = {6},
       number = {1},
          eid = {2},
        pages = {2},
          doi = {10.1186/s40668-019-0028-x},
archivePrefix = {arXiv},
       eprint = {1812.05609},
 primaryClass = {astro-ph.GA},
       adsurl = {https://ui.adsabs.harvard.edu/abs/2019ComAC...6....2N}
}

@ARTICLE{He_2025a,
       author = {{He}, Lin and {Li}, Zhiyuan and {Hou}, Meicun and {Du}, Min and {Fang}, Taotao and {Cui}, Wei},
        title = "{Probing the Hot Gaseous Halo of the Low-mass Disk Galaxy NGC 7793 with eROSITA and Chandra}",
      journal = {\apj},
         year = 2025,
        month = apr,
       volume = {983},
       number = {2},
          eid = {149},
        pages = {149},
          doi = {10.3847/1538-4357/adc0a0},
archivePrefix = {arXiv},
       eprint = {2503.10087},
 primaryClass = {astro-ph.GA},
       adsurl = {https://ui.adsabs.harvard.edu/abs/2025ApJ...983..149H}
}

@ARTICLE{Foster_2020,
       author = {{Foster}, Adam R. and {Heuer}, Keri},
        title = "{PyAtomDB: Extending the AtomDB Atomic Database to Model New Plasma Processes and Uncertainties}",
      journal = {Atoms},
         year = 2020,
        month = aug,
       volume = {8},
       number = {3},
        pages = {49},
          doi = {10.3390/atoms8030049},
       adsurl = {https://ui.adsabs.harvard.edu/abs/2020Atoms...8...49F}
}

@ARTICLE{Putman_2012,
       author = {{Putman}, M.~E. and {Peek}, J.~E.~G. and {Joung}, M.~R.},
        title = "{Gaseous Galaxy Halos}",
      journal = {\araa},
         year = 2012,
        month = sep,
       volume = {50},
        pages = {491-529},
          doi = {10.1146/annurev-astro-081811-125612},
archivePrefix = {arXiv},
       eprint = {1207.4837},
 primaryClass = {astro-ph.GA},
       adsurl = {https://ui.adsabs.harvard.edu/abs/2012ARA&A..50..491P}
}

@article{Tumlinson_2017,
   title={The Circumgalactic Medium},
   volume={55},
   ISSN={1545-4282},
   url={http://dx.doi.org/10.1146/annurev-astro-091916-055240},
   DOI={10.1146/annurev-astro-091916-055240},
   number={1},
   journal={Annual Review of Astronomy and Astrophysics},
   publisher={Annual Reviews},
   author={Tumlinson, Jason and Peeples, Molly S. and Werk, Jessica K.},
   year={2017},
   month=aug, pages={389–432} }

@ARTICLE{FG_2023,
       author = {{Faucher-Gigu{\`e}re}, Claude-Andr{\'e} and {Oh}, S. Peng},
        title = "{Key Physical Processes in the Circumgalactic Medium}",
      journal = {\araa},
         year = 2023,
        month = aug,
       volume = {61},
        pages = {131-195},
          doi = {10.1146/annurev-astro-052920-125203},
archivePrefix = {arXiv},
       eprint = {2301.10253},
 primaryClass = {astro-ph.GA},
       adsurl = {https://ui.adsabs.harvard.edu/abs/2023ARA&A..61..131F}
}

@ARTICLE{Chen_2024,
       author = {{Chen}, Hsiao-Wen and {Zahedy}, Fakhri S.},
        title = "{The Circumgalactic Medium}",
      journal = {arXiv e-prints},
         year = 2024,
        month = dec,
          eid = {arXiv:2412.10579},
        pages = {arXiv:2412.10579},
          doi = {10.48550/arXiv.2412.10579},
archivePrefix = {arXiv},
       eprint = {2412.10579},
 primaryClass = {astro-ph.GA},
       adsurl = {https://ui.adsabs.harvard.edu/abs/2024arXiv241210579C}
}

@ARTICLE{Fumagalli_2024,
       author = {{Fumagalli}, Michele},
        title = "{The multiphase circumgalactic medium and its relation to galaxies: an observational perspective}",
      journal = {arXiv e-prints},
         year = 2024,
        month = aug,
          eid = {arXiv:2409.00174},
        pages = {arXiv:2409.00174},
          doi = {10.48550/arXiv.2409.00174},
archivePrefix = {arXiv},
       eprint = {2409.00174},
 primaryClass = {astro-ph.GA},
       adsurl = {https://ui.adsabs.harvard.edu/abs/2024arXiv240900174F}
}

@ARTICLE{Anderson_2016,
       author = {{Anderson}, Michael E. and {Churazov}, Eugene and {Bregman}, Joel N.},
        title = "{A deep XMM-Newton study of the hot gaseous halo around NGC 1961}",
      journal = {\mnras},
         year = 2016,
        month = jan,
       volume = {455},
       number = {1},
        pages = {227-243},
          doi = {10.1093/mnras/stv2314},
archivePrefix = {arXiv},
       eprint = {1508.01514},
 primaryClass = {astro-ph.GA},
       adsurl = {https://ui.adsabs.harvard.edu/abs/2016MNRAS.455..227A}
}

@ARTICLE{LJT_2013a,
       author = {{Li}, Jiang-Tao and {Wang}, Q. Daniel},
        title = "{Chandra survey of nearby highly inclined disc galaxies - I. X-ray measurements of galactic coronae}",
      journal = {\mnras},
         year = 2013,
        month = jan,
       volume = {428},
       number = {3},
        pages = {2085-2108},
          doi = {10.1093/mnras/sts183},
archivePrefix = {arXiv},
       eprint = {1210.2997},
 primaryClass = {astro-ph.CO},
       adsurl = {https://ui.adsabs.harvard.edu/abs/2013MNRAS.428.2085L}
}

@ARTICLE{Bogdan_2017,
       author = {{Bogd{\'a}n}, {\'A}kos and {Bourdin}, Herv{\'e} and {Forman}, William R. and {Kraft}, Ralph P. and {Vogelsberger}, Mark and {Hernquist}, Lars and {Springel}, Volker},
        title = "{Probing the Hot X-Ray Corona around the Massive Spiral Galaxy, NGC 6753, Using Deep XMM-Newton Observations}",
      journal = {\apj},
         year = 2017,
        month = nov,
       volume = {850},
       number = {1},
          eid = {98},
        pages = {98},
          doi = {10.3847/1538-4357/aa9523},
archivePrefix = {arXiv},
       eprint = {1710.07286},
 primaryClass = {astro-ph.GA},
       adsurl = {https://ui.adsabs.harvard.edu/abs/2017ApJ...850...98B}
}

@ARTICLE{Bogdan_2013,
       author = {{Bogd{\'a}n}, {\'A}kos and {Forman}, William R. and {Kraft}, Ralph P. and {Jones}, Christine},
        title = "{Detection of a Luminous Hot X-Ray Corona around the Massive Spiral Galaxy NGC 266}",
      journal = {\apj},
         year = 2013,
        month = aug,
       volume = {772},
       number = {2},
          eid = {98},
        pages = {98},
          doi = {10.1088/0004-637X/772/2/98},
archivePrefix = {arXiv},
       eprint = {1306.0643},
 primaryClass = {astro-ph.CO},
       adsurl = {https://ui.adsabs.harvard.edu/abs/2013ApJ...772...98B}
}

@ARTICLE{Li_2007b,
       author = {{Li}, Zhiyuan and {Wang}, Q. Daniel},
        title = "{Chandra Detection of Diffuse Hot Gas in and around the M31 Bulge}",
      journal = {\apjl},
         year = 2007,
        month = oct,
       volume = {668},
       number = {1},
        pages = {L39-L42},
          doi = {10.1086/522674},
archivePrefix = {arXiv},
       eprint = {0708.3077},
 primaryClass = {astro-ph},
       adsurl = {https://ui.adsabs.harvard.edu/abs/2007ApJ...668L..39L}
}

@INPROCEEDINGS{LJT_2016,
       author = {{Li}, Jiang-Tao and {Bregman}, Joel N. and {Wang}, Daniel and {Crain}, Robert A. and {Anderson}, Michael E.},
        title = "{The Circum-Galactic Medium of MASsive Spirals (CGM-MASS) I: Introduction to the XMM-Newton Large Project and a Case Study of NGC 5908}",
    booktitle = {AAS/High Energy Astrophysics Division \#15},
         year = 2016,
       series = {AAS/High Energy Astrophysics Division},
       volume = {15},
        month = apr,
          eid = {402.04},
        pages = {402.04},
       adsurl = {https://ui.adsabs.harvard.edu/abs/2016HEAD...1540204L}
}

@ARTICLE{Sacchi_2025,
       author = {{Sacchi}, Andrea and {Bogd{\'a}n}, {\'A}kos and {Truong}, Nhut},
        title = "{Detection of Anisotropies in the Circumgalactic Medium of Disk Galaxies: Supermassive Black Hole Activity or Star Formation-driven Outflows?}",
      journal = {\apj},
         year = 2025,
        month = apr,
       volume = {983},
       number = {2},
          eid = {178},
        pages = {178},
          doi = {10.3847/1538-4357/adc38b},
archivePrefix = {arXiv},
       eprint = {2501.18681},
 primaryClass = {astro-ph.GA},
       adsurl = {https://ui.adsabs.harvard.edu/abs/2025ApJ...983..178S}
}

@ARTICLE{Lehmer_2010ApJ,
       author = {{Lehmer}, B.~D. and {Alexander}, D.~M. and {Bauer}, F.~E. and {Brandt}, W.~N. and {Goulding}, A.~D. and {Jenkins}, L.~P. and {Ptak}, A. and {Roberts}, T.~P.},
        title = "{A Chandra Perspective on Galaxy-wide X-ray Binary Emission and its Correlation with Star Formation Rate and Stellar Mass: New Results from Luminous Infrared Galaxies}",
      journal = {\apj},
         year = 2010,
        month = nov,
       volume = {724},
       number = {1},
        pages = {559-571},
          doi = {10.1088/0004-637X/724/1/559},
archivePrefix = {arXiv},
       eprint = {1009.3943},
 primaryClass = {astro-ph.CO},
       adsurl = {https://ui.adsabs.harvard.edu/abs/2010ApJ...724..559L}
}

@ARTICLE{He_2025b,
       author = {{He}, Lin and {Li}, Zhiyuan and {Li}, Zongnan and {Garc{\'\i}a-Benito}, Rub{\'e}n and {Liu}, Yuanqi and {Hou}, Meicun},
        title = "{Discovery of Giant Bubbles in the Hot Gaseous Halo of the Massive Disk Galaxy NGC 6286}",
      journal = {\apj},
         year = 2025,
        month = oct,
       volume = {992},
       number = {1},
          eid = {86},
        pages = {86},
          doi = {10.3847/1538-4357/ae0722},
archivePrefix = {arXiv},
       eprint = {2509.06470},
 primaryClass = {astro-ph.GA},
       adsurl = {https://ui.adsabs.harvard.edu/abs/2025ApJ...992...86H}
}

@ARTICLE{Vladutescu-Zopp_2025,
       author = {{Vladutescu-Zopp}, S. and {Biffi}, V. and {Dolag}, K.},
        title = "{Radial X-ray profiles of simulated galaxies: Contributions from hot gas and X-ray binaries}",
      journal = {\aap},
         year = 2025,
        month = mar,
       volume = {695},
          eid = {A2},
        pages = {A2},
          doi = {10.1051/0004-6361/202450989},
archivePrefix = {arXiv},
       eprint = {2406.02686},
 primaryClass = {astro-ph.GA},
       adsurl = {https://ui.adsabs.harvard.edu/abs/2025A&A...695A...2V}
}

@ARTICLE{Ge2015,
       author = {{Ge}, Chong and {Li}, Zhiyuan and {Xu}, Xiaojie and {Gu}, Qiusheng and {Wang}, Q. Daniel and {Roberts}, Shawn and {Kraft}, Ralph P. and {Jones}, Christine and {Forman}, William R.},
        title = "{X-Ray Emissivity of Old Stellar Populations: A Local Group Census}",
      journal = {\apj},
         year = 2015,
        month = oct,
       volume = {812},
       number = {2},
          eid = {130},
        pages = {130},
          doi = {10.1088/0004-637X/812/2/130},
archivePrefix = {arXiv},
       eprint = {1509.05863},
 primaryClass = {astro-ph.GA},
       adsurl = {https://ui.adsabs.harvard.edu/abs/2015ApJ...812..130G}
}

@ARTICLE{Vijayan2022,
       author = {{Vijayan}, Aditi and {Li}, Miao},
        title = "{X-ray spectra of circumgalactic medium around star-forming galaxies: connecting simulations to observations}",
      journal = {\mnras},
         year = 2022,
        month = feb,
       volume = {510},
       number = {1},
        pages = {568-580},
          doi = {10.1093/mnras/stab3413},
archivePrefix = {arXiv},
       eprint = {2102.11510},
 primaryClass = {astro-ph.GA},
       adsurl = {https://ui.adsabs.harvard.edu/abs/2022MNRAS.510..568V}
}

@ARTICLE{Shankar2009,
       author = {{Shankar}, Francesco and {Weinberg}, David H. and {Miralda-Escud{\'e}}, Jordi},
        title = "{Self-Consistent Models of the AGN and Black Hole Populations: Duty Cycles, Accretion Rates, and the Mean Radiative Efficiency}",
      journal = {\apj},
         year = 2009,
        month = jan,
       volume = {690},
       number = {1},
        pages = {20-41},
          doi = {10.1088/0004-637X/690/1/20},
archivePrefix = {arXiv},
       eprint = {0710.4488},
 primaryClass = {astro-ph},
       adsurl = {https://ui.adsabs.harvard.edu/abs/2009ApJ...690...20S}
}

@ARTICLE{Zhang_2025,
       author = {{Zhang}, Yi and {Comparat}, Johan and {Ponti}, Gabriele and {Merloni}, Andrea and {Nandra}, Kirpal and {Haberl}, Frank and {Truong}, Nhut and {Pillepich}, Annalisa and {Popesso}, Paola and {Locatelli}, Nicola and {Zhang}, Xiaoyuan and {Sanders}, Jeremy and {Zheng}, Xueying and {Liu}, Ang and {Liu}, Teng and {Predehl}, Peter and {Salvato}, Mara and {Bruggen}, Marcus and {Shreeram}, Soumya and {Yeung}, Michael C.~H.},
        title = "{The hot circumgalactic medium in the eROSITA All-Sky Survey: III. Star-forming and quiescent galaxies}",
      journal = {\aap},
         year = 2025,
        month = jan,
       volume = {693},
          eid = {A197},
        pages = {A197},
          doi = {10.1051/0004-6361/202452273},
archivePrefix = {arXiv},
       eprint = {2411.19945},
 primaryClass = {astro-ph.GA},
       adsurl = {https://ui.adsabs.harvard.edu/abs/2025A&A...693A.197Z}
}

@ARTICLE{Brinchmann2004,
       author = {{Brinchmann}, J. and {Charlot}, S. and {White}, S.~D.~M. and {Tremonti}, C. and {Kauffmann}, G. and {Heckman}, T. and {Brinkmann}, J.},
        title = "{The physical properties of star-forming galaxies in the low-redshift Universe}",
      journal = {\mnras},
         year = 2004,
        month = jul,
       volume = {351},
       number = {4},
        pages = {1151-1179},
          doi = {10.1111/j.1365-2966.2004.07881.x},
archivePrefix = {arXiv},
       eprint = {astro-ph/0311060},
 primaryClass = {astro-ph},
       adsurl = {https://ui.adsabs.harvard.edu/abs/2004MNRAS.351.1151B}
}

@ARTICLE{Chen2012,
       author = {{Chen}, Yan-Mei and {Kauffmann}, Guinevere and {Tremonti}, Christy A. and {White}, Simon and {Heckman}, Timothy M. and {Kova{\v{c}}}, Katarina and {Bundy}, Kevin and {Chisholm}, John and {Maraston}, Claudia and {Schneider}, Donald P. and {Bolton}, Adam S. and {Weaver}, Benjamin A. and {Brinkmann}, Jon},
        title = "{Evolution of the most massive galaxies to z= 0.6 - I. A new method for physical parameter estimation}",
      journal = {\mnras},
         year = 2012,
        month = mar,
       volume = {421},
       number = {1},
        pages = {314-332},
          doi = {10.1111/j.1365-2966.2011.20306.x},
archivePrefix = {arXiv},
       eprint = {1108.4719},
 primaryClass = {astro-ph.GA},
       adsurl = {https://ui.adsabs.harvard.edu/abs/2012MNRAS.421..314C}
}

@ARTICLE{Chang2015,
       author = {{Chang}, Yu-Yen and {van der Wel}, Arjen and {da Cunha}, Elisabete and {Rix}, Hans-Walter},
        title = "{Stellar Masses and Star Formation Rates for 1M Galaxies from SDSS+WISE}",
      journal = {\apjs},
         year = 2015,
        month = jul,
       volume = {219},
       number = {1},
          eid = {8},
        pages = {8},
          doi = {10.1088/0067-0049/219/1/8},
archivePrefix = {arXiv},
       eprint = {1506.00648},
 primaryClass = {astro-ph.GA},
       adsurl = {https://ui.adsabs.harvard.edu/abs/2015ApJS..219....8C}
}

@ARTICLE{Tullmann2006A&A...448...43T,
       author = {{T{\"u}llmann}, R. and {Pietsch}, W. and {Rossa}, J. and {Breitschwerdt}, D. and {Dettmar}, R.-J.},
        title = "{The multi-phase gaseous halos of star forming late-type galaxies. I. XMM-Newton observations of the hot ionized medium}",
      journal = {\aap},
         year = 2006,
        month = mar,
       volume = {448},
       number = {1},
        pages = {43-75},
          doi = {10.1051/0004-6361:20052936},
archivePrefix = {arXiv},
       eprint = {astro-ph/0510079},
 primaryClass = {astro-ph},
       adsurl = {https://ui.adsabs.harvard.edu/abs/2006A&A...448...43T}
}

@INPROCEEDINGS{Arnaud1996ASPC..101...17A,
       author = {{Arnaud}, K.~A.},
        title = "{XSPEC: The First Ten Years}",
    booktitle = {Astronomical Data Analysis Software and Systems V},
         year = 1996,
       editor = {{Jacoby}, George H. and {Barnes}, Jeannette},
       series = {Astronomical Society of the Pacific Conference Series},
       volume = {101},
        month = jan,
        pages = {17},
       adsurl = {https://ui.adsabs.harvard.edu/abs/1996ASPC..101...17A}
}

@ARTICLE{1999MNRAS.309..561Z,
       author = {{{\.Z}ycki}, Piotr T. and {Done}, Chris and {Smith}, David A.},
        title = "{The 1989 May outburst of the soft X-ray transient GS 2023+338 (V404 Cyg)}",
      journal = {\mnras},
         year = 1999,
        month = nov,
       volume = {309},
       number = {3},
        pages = {561-575},
          doi = {10.1046/j.1365-8711.1999.02885.x},
archivePrefix = {arXiv},
       eprint = {astro-ph/9904304},
 primaryClass = {astro-ph},
       adsurl = {https://ui.adsabs.harvard.edu/abs/1999MNRAS.309..561Z}
}

@ARTICLE{1996MNRAS.283..193Z,
       author = {{Zdziarski}, A.~A. and {Johnson}, W.~N. and {Magdziarz}, P.},
        title = "{Broad-band {\ensuremath{\gamma}}-ray and X-ray spectra of NGC 4151 and their implications for physical processes and geometry.}",
      journal = {\mnras},
         year = 1996,
        month = nov,
       volume = {283},
       number = {1},
        pages = {193-206},
          doi = {10.1093/mnras/283.1.193},
archivePrefix = {arXiv},
       eprint = {astro-ph/9607015},
 primaryClass = {astro-ph},
       adsurl = {https://ui.adsabs.harvard.edu/abs/1996MNRAS.283..193Z}
}

@ARTICLE{Oppenheimer2020ApJ...893L..24O,
       author = {{Oppenheimer}, Benjamin D. and {Bogd{\'a}n}, {\'A}kos and {Crain}, Robert A. and {ZuHone}, John A. and {Forman}, William R. and {Schaye}, Joop and {Wijers}, Nastasha A. and {Davies}, Jonathan J. and {Jones}, Christine and {Kraft}, Ralph P. and {Ghirardini}, Vittorio},
        title = "{EAGLE and Illustris-TNG Predictions for Resolved eROSITA X-Ray Observations of the Circumgalactic Medium around Normal Galaxies}",
      journal = {\apjl},
         year = 2020,
        month = apr,
       volume = {893},
       number = {1},
          eid = {L24},
        pages = {L24},
          doi = {10.3847/2041-8213/ab846f},
archivePrefix = {arXiv},
       eprint = {2003.13889},
 primaryClass = {astro-ph.GA},
       adsurl = {https://ui.adsabs.harvard.edu/abs/2020ApJ...893L..24O}
}

@ARTICLE{Licquia2015ApJ...806...96L,
       author = {{Licquia}, Timothy C. and {Newman}, Jeffrey A.},
        title = "{Improved Estimates of the Milky Way's Stellar Mass and Star Formation Rate from Hierarchical Bayesian Meta-Analysis}",
      journal = {\apj},
         year = 2015,
        month = jun,
       volume = {806},
       number = {1},
          eid = {96},
        pages = {96},
          doi = {10.1088/0004-637X/806/1/96},
archivePrefix = {arXiv},
       eprint = {1407.1078},
 primaryClass = {astro-ph.GA},
       adsurl = {https://ui.adsabs.harvard.edu/abs/2015ApJ...806...96L}
}

@ARTICLE{Dave2019MNRAS.486.2827D,
       author = {{Dav{\'e}}, Romeel and {Angl{\'e}s-Alc{\'a}zar}, Daniel and {Narayanan}, Desika and {Li}, Qi and {Rafieferantsoa}, Mika H. and {Appleby}, Sarah},
        title = "{SIMBA: Cosmological simulations with black hole growth and feedback}",
      journal = {\mnras},
         year = 2019,
        month = jun,
       volume = {486},
       number = {2},
        pages = {2827-2849},
          doi = {10.1093/mnras/stz937},
archivePrefix = {arXiv},
       eprint = {1901.10203},
 primaryClass = {astro-ph.GA},
       adsurl = {https://ui.adsabs.harvard.edu/abs/2019MNRAS.486.2827D}
}

@ARTICLE{Hummels2019ApJ...882..156H,
       author = {{Hummels}, Cameron B. and {Smith}, Britton D. and {Hopkins}, Philip F. and {O'Shea}, Brian W. and {Silvia}, Devin W. and {Werk}, Jessica K. and {Lehner}, Nicolas and {Wise}, John H. and {Collins}, David C. and {Butsky}, Iryna S.},
        title = "{The Impact of Enhanced Halo Resolution on the Simulated Circumgalactic Medium}",
      journal = {\apj},
         year = 2019,
        month = sep,
       volume = {882},
       number = {2},
          eid = {156},
        pages = {156},
          doi = {10.3847/1538-4357/ab378f},
archivePrefix = {arXiv},
       eprint = {1811.12410},
 primaryClass = {astro-ph.GA},
       adsurl = {https://ui.adsabs.harvard.edu/abs/2019ApJ...882..156H}
}

@ARTICLE{Peeples2019ApJ...873..129P,
       author = {{Peeples}, Molly S. and {Corlies}, Lauren and {Tumlinson}, Jason and {O'Shea}, Brian W. and {Lehner}, Nicolas and {O'Meara}, John M. and {Howk}, J. Christopher and {Earl}, Nicholas and {Smith}, Britton D. and {Wise}, John H. and {Hummels}, Cameron B.},
        title = "{Figuring Out Gas \& Galaxies in Enzo (FOGGIE). I. Resolving Simulated Circumgalactic Absorption at 2 {\ensuremath{\leq}} z {\ensuremath{\leq}} 2.5}",
      journal = {\apj},
         year = 2019,
        month = mar,
       volume = {873},
       number = {2},
          eid = {129},
        pages = {129},
          doi = {10.3847/1538-4357/ab0654},
archivePrefix = {arXiv},
       eprint = {1810.06566},
 primaryClass = {astro-ph.GA},
       adsurl = {https://ui.adsabs.harvard.edu/abs/2019ApJ...873..129P}
}

@ARTICLE{Hou_2024,
       author = {{Hou}, Meicun and {He}, Lin and {Hu}, Zhensong and {Li}, Zhiyuan and {Jones}, Christine and {Forman}, William and {Su}, Yuanyuan and {Wang}, Jing and {Ho}, Luis C.},
        title = "{X-Ray Constraints on the Hot Gaseous Corona of Edge-on Late-type Galaxies in Virgo}",
      journal = {\apj},
         year = 2024,
        month = feb,
       volume = {961},
       number = {2},
          eid = {249},
        pages = {249},
          doi = {10.3847/1538-4357/ad138a},
archivePrefix = {arXiv},
       eprint = {2312.04050},
 primaryClass = {astro-ph.GA},
       adsurl = {https://ui.adsabs.harvard.edu/abs/2024ApJ...961..249H}
}

@ARTICLE{Benson2010,
       author = {{Benson}, Andrew J.},
        title = "{Galaxy formation theory}",
      journal = {\physrep},
         year = 2010,
        month = oct,
       volume = {495},
       number = {2-3},
        pages = {33-86},
          doi = {10.1016/j.physrep.2010.06.001},
archivePrefix = {arXiv},
       eprint = {1006.5394},
 primaryClass = {astro-ph.CO},
       adsurl = {https://ui.adsabs.harvard.edu/abs/2010PhR...495...33B}
}

@ARTICLE{Pillepich2021MNRAS.508.4667P,
       author = {{Pillepich}, Annalisa and {Nelson}, Dylan and {Truong}, Nhut and {Weinberger}, Rainer and {Martin-Navarro}, Ignacio and {Springel}, Volker and {Faber}, Sandy M. and {Hernquist}, Lars},
        title = "{X-ray bubbles in the circumgalactic medium of TNG50 Milky Way- and M31-like galaxies: signposts of supermassive black hole activity}",
      journal = {\mnras},
         year = 2021,
        month = dec,
       volume = {508},
       number = {4},
        pages = {4667-4695},
          doi = {10.1093/mnras/stab2779},
archivePrefix = {arXiv},
       eprint = {2105.08062},
 primaryClass = {astro-ph.GA},
       adsurl = {https://ui.adsabs.harvard.edu/abs/2021MNRAS.508.4667P}
}

@ARTICLE{Bregman_2018,
       author = {{Bregman}, Joel N. and {Anderson}, Michael E. and {Miller}, Matthew J. and {Hodges-Kluck}, Edmund and {Dai}, Xinyu and {Li}, Jiang-Tao and {Li}, Yunyang and {Qu}, Zhijie},
        title = "{The Extended Distribution of Baryons around Galaxies}",
      journal = {\apj},
         year = 2018,
        month = jul,
       volume = {862},
       number = {1},
          eid = {3},
        pages = {3},
          doi = {10.3847/1538-4357/aacafe},
archivePrefix = {arXiv},
       eprint = {1803.08963},
 primaryClass = {astro-ph.GA},
       adsurl = {https://ui.adsabs.harvard.edu/abs/2018ApJ...862....3B}
}

@ARTICLE{Cruise2025NatAs...9...36C,
       author = {{Cruise}, Mike and {Guainazzi}, Matteo and {Aird}, James and {Carrera}, Francisco J. and {Costantini}, Elisa and {Corrales}, Lia and {Dauser}, Thomas and {Eckert}, Dominique and {Gastaldello}, Fabio and {Matsumoto}, Hironori and {Osten}, Rachel and {Petrucci}, Pierre-Olivier and {Porquet}, Delphine and {Pratt}, Gabriel W. and {Rea}, Nanda and {Reiprich}, Thomas H. and {Simionescu}, Aurora and {Spiga}, Daniele and {Troja}, Eleonora},
        title = "{The NewAthena mission concept in the context of the next decade of X-ray astronomy}",
      journal = {Nature Astronomy},
         year = 2025,
        month = jan,
       volume = {9},
        pages = {36-44},
          doi = {10.1038/s41550-024-02416-3},
archivePrefix = {arXiv},
       eprint = {2501.03100},
 primaryClass = {astro-ph.IM},
       adsurl = {https://ui.adsabs.harvard.edu/abs/2025NatAs...9...36C}
}

@ARTICLE{Cui2020JLTP..199..502C,
       author = {{Cui}, W. and {Chen}, L.-B. and {Gao}, B. and {Guo}, F.-L. and {Jin}, H. and {Wang}, G.-L. and {Wang}, L. and {Wang}, J.-J. and {Wang}, W. and {Wang}, Z.-S. and {Wang}, Z. and {Yuan}, F. and {Zhang}, W.},
        title = "{HUBS: Hot Universe Baryon Surveyor}",
      journal = {Journal of Low Temperature Physics},
         year = 2020,
        month = jan,
       volume = {199},
       number = {1-2},
        pages = {502-509},
          doi = {10.1007/s10909-019-02279-3},
       adsurl = {https://ui.adsabs.harvard.edu/abs/2020JLTP..199..502C}
}

@ARTICLE{Boroson2011ApJ...729...12B,
       author = {{Boroson}, Bram and {Kim}, Dong-Woo and {Fabbiano}, Giuseppina},
        title = "{Revisiting with Chandra the Scaling Relations of the X-ray Emission Components (Binaries, Nuclei, and Hot Gas) of Early-type Galaxies}",
      journal = {\apj},
         year = 2011,
        month = mar,
       volume = {729},
       number = {1},
          eid = {12},
        pages = {12},
          doi = {10.1088/0004-637X/729/1/12},
archivePrefix = {arXiv},
       eprint = {1011.2529},
 primaryClass = {astro-ph.HE},
       adsurl = {https://ui.adsabs.harvard.edu/abs/2011ApJ...729...12B}
}

@ARTICLE{Shreeram_2025a,
       author = {{Shreeram}, Soumya and {Comparat}, Johan and {Merloni}, Andrea and {Ponti}, Gabriele and {Popesso}, Paola and {Zhang}, Yi and {Nandra}, Kirpal and {Salvato}, Mara and {Marini}, Ilaria and {Buchner}, Johannes and {Locatelli}, Nicola and {Igo}, Zsofi},
        title = "{Retrieving the hot circumgalactic medium physics from the X-ray radial profile from eROSITA with an IlustrisTNG-based forward model}",
      journal = {\aap},
         year = 2025,
        month = nov,
       volume = {703},
          eid = {A137},
        pages = {A137},
          doi = {10.1051/0004-6361/202554508},
archivePrefix = {arXiv},
       eprint = {2504.03840},
 primaryClass = {astro-ph.CO},
       adsurl = {https://ui.adsabs.harvard.edu/abs/2025A&A...703A.137S}
}

@ARTICLE{Shreeram_2025b,
       author = {{Shreeram}, Soumya and {Comparat}, Johan and {Merloni}, Andrea and {Zhang}, Yi and {Ponti}, Gabriele and {Nandra}, Kirpal and {ZuHone}, John and {Marini}, Ilaria and {Vladutescu-Zopp}, Stephan and {Popesso}, Paola and {Pakmor}, Ruediger and {Seppi}, Riccardo and {Peroux}, Celine and {Sorini}, Daniele},
        title = "{Quantifying observational projection effects with a simulation-based hot CGM model}",
      journal = {\aap},
         year = 2025,
        month = may,
       volume = {697},
          eid = {A22},
        pages = {A22},
          doi = {10.1051/0004-6361/202452271},
archivePrefix = {arXiv},
       eprint = {2409.10397},
 primaryClass = {astro-ph.GA},
       adsurl = {https://ui.adsabs.harvard.edu/abs/2025A&A...697A..22S}
}

@ARTICLE{Anderson_2013ApJ...762..106A,
       author = {{Anderson}, Michael E. and {Bregman}, Joel N. and {Dai}, Xinyu},
        title = "{Extended Hot Halos around Isolated Galaxies Observed in the ROSAT All-Sky Survey}",
      journal = {\apj},
         year = 2013,
        month = jan,
       volume = {762},
       number = {2},
          eid = {106},
        pages = {106},
          doi = {10.1088/0004-637X/762/2/106},
archivePrefix = {arXiv},
       eprint = {1211.5140},
 primaryClass = {astro-ph.CO},
       adsurl = {https://ui.adsabs.harvard.edu/abs/2013ApJ...762..106A}
}

@software{bradley_2026_19636730,
  author       = {Bradley, Larry and
                  Sipőcz, Brigitta M. and
                  Robitaille, T. P. and
                  Tollerud, E. J. and
                  Vinícius, Zé and
                  Deil, Christoph and
                  Barbary, Kyle and
                  Wilson, Tom J. and
                  Busko, Ivo and
                  Donath, Axel and
                  Günther, Hans Moritz and
                  Cara, Mihai and
                  Lim, P. L. and
                  Meßlinger, Sebastian and
                  Conseil, Simon and
                  Droettboom, Michael and
                  Bostroem, K. Azalee and
                  Bray, E. M. and
                  Bratholm, Lars Andersen and
                  Burnett, Zach and
                  Jamieson, William and
                  Ginsburg, Adam and
                  Taranu, Dan and
                  Barentsen, Geert and
                  Craig, Matthew W. and
                  Morris, Brett M. and
                  Perrin, Marshall and
                  Rathi, Shivangee},
  title        = {Photutils},
  month        = apr,
  year         = 2026,
  publisher    = {Zenodo},
  version      = {3.0.0},
  doi          = {10.5281/zenodo.19636730},
  url          = {https://doi.org/10.5281/zenodo.19636730},
  swhid        = {swh:1:dir:5ea90432cd987336a26af201c1b50343b12e2daa
                   ;origin=https://doi.org/10.5281/zenodo.596036;visi
                   t=swh:1:snp:25dd84d95353ec3813baa8785c248017da7192
                   33;anchor=swh:1:rel:253cbde27398f9aea9d7368f386661
                   8c00fe16f9;path=astropy-photutils-3322558
                  },
}

@software{robitaille_2018_1162674,
  author       = {Robitaille, Thomas},
  title        = {reproject: astronomical image reprojection in
                   Python
                  },
  month        = jan,
  year         = 2018,
  publisher    = {Zenodo},
  version      = {v0.4},
  doi          = {10.5281/zenodo.1162674},
  url          = {https://doi.org/10.5281/zenodo.1162674},
}

@ARTICLE{Zou2026ApJ..1003..199Z,
       author = {{Zou}, Yuxuan and {Yuan}, Feng and {Ji}, Suoqing and {He}, Lin and {Li}, Zhiyuan and {Zhang}, Yi and {Comparat}, Johan and {Qu}, Zhijie and {Fang}, Taotao},
        title = "{A Systematic Study of AGN Feedback in a Disk Galaxy Using MACER. II. Predictions of X-Ray Surface Brightness Profiles and Comparison with eROSITA Observations}",
      journal = {\apj},
         year = 2026,
        month = jun,
       volume = {1003},
       number = {2},
          eid = {199},
        pages = {199},
          doi = {10.3847/1538-4357/ae6778},
archivePrefix = {arXiv},
       eprint = {2603.22412},
 primaryClass = {astro-ph.GA},
       adsurl = {https://ui.adsabs.harvard.edu/abs/2026ApJ..1003..199Z}
}

@ARTICLE{Zhang2026A&A...706A.102Z,
       author = {{Zhang}, Yi and {Shreeram}, Soumya and {Ponti}, Gabriele and {Comparat}, Johan and {Merloni}, Andrea and {Qu}, Zhijie and {Li}, Jiangtao and {Bregman}, Joel N. and {Fang}, Taotao},
        title = "{On the baryon budget in the X-ray-emitting circumgalactic medium of Milky Way-mass galaxies}",
      journal = {\aap},
         year = 2026,
        month = feb,
       volume = {706},
          eid = {A102},
        pages = {A102},
          doi = {10.1051/0004-6361/202556835},
archivePrefix = {arXiv},
       eprint = {2511.17313},
 primaryClass = {astro-ph.GA},
       adsurl = {https://ui.adsabs.harvard.edu/abs/2026A&A...706A.102Z}
}

@ARTICLE{Akaike1100705,
  author={Akaike, H.},
  journal={IEEE Transactions on Automatic Control}, 
  title={A new look at the statistical model identification}, 
  year={1974},
  volume={19},
  number={6},
  pages={716-723},
  doi={10.1109/TAC.1974.1100705}}

@ARTICLE{Schwarz1978AnSta...6..461S,
       author = {{Schwarz}, Gideon},
        title = "{Estimating the Dimension of a Model}",
      journal = {Annals of Statistics},
         year = 1978,
        month = jul,
       volume = {6},
       number = {2},
        pages = {461-464},
       adsurl = {https://ui.adsabs.harvard.edu/abs/1978AnSta...6..461S}
}
\bibliographystyle{aasjournal}

\end{document}